\documentclass[rmp,aps,amsmath,amsfonts,noshowkeys,noshowpacs,floatfix,fix,twocolumn]{revtex4}
\usepackage{color,bm}
\usepackage{fancybox,amsmath,amssymb,rotating,pifont}
\usepackage{epic,eepic,calc,array,shadow}
\usepackage{npsfont}
\usepackage{graphicx}
\usepackage{dcolumn}  
\usepackage{subfigure}
\usepackage{multirow}

\usepackage{mdwlist}

\makeatletter
\newlength{\earraycolsep}
\def\eqnarray{\stepcounter{equation}\let\@currentlabel%
\theequation \global\@eqnswtrue\m@th
\global\@eqcnt\z@\tabskip\@centering\let\\\@eqncr
$$\halign to\displaywidth\bgroup\@eqnsel\hskip\@centering
$\displaystyle\tabskip\z@{##}$&\global\@eqcnt\@ne \hskip
2\earraycolsep \hfil$\displaystyle{##}$\hfil &\global\@eqcnt\tw@
\hskip 2\earraycolsep $\displaystyle\tabskip\z@{##}$\hfil
\tabskip\@centering&\llap{##}\tabskip\z@\cr} \makeatother

\DeclareMathAlphabet{\mathbbm}{U}{bbm}{m}{n}
\SetMathAlphabet\mathbbm{bold}{U}{bbm}{bx}{n} 
\usepackage{hyperref}
\everymath={\displaystyle}
\usepackage{young}
\usepackage{rotating}
\usepackage{longtable}

\def\1{\mbox{l\hspace{-0.53em}1}}
\newcommand{\fr}{\frac}
\usepackage{multirow}

\newlength{\AccoHaut}

\begin{document}
\title{Baryon resonances in large $N_c$  QCD}

\author{N. Matagne\footnote{e-mail address: Nicolas.Matagne@umons.ac.be}}

\author{Fl. Stancu\footnote{e-mail address: fstancu@ulg.ac.be
}}
\affiliation{
$^a$ Service de Physique Nucl\'eaire et Subnucl\'eaire, University of Mons, Place du Parc, B-7000 Mons, Belgium \\
$^b$ University of Li\`ege, Institute of Physics B5, Sart Tilman,
B-4000 Li\`ege 1, Belgium}

\date{\today}

\begin{abstract}

We review the current status and present open challenges of large $N_c$ QCD baryon spectroscopy. After introducing the $1/N_c$
expansion method we first shortly revisit the latest achievements for the ground state properties. Next we discuss the
applicability of this method to excited states, presenting two different approaches with
their advantages and disadvantages. Selected results for the spectrum and strong and electromagnetic decays are described. 
We also present further developments  for the applicability of the method to excited states, based on the qualitative compatibility 
between the quark excitation picture and the meson-nucleon scattering picture. We show that  a quantitative comparison between results obtained from 
the mass formula of the $1/N_c$ expansion method and quark models brings convincing support to quark
models and shortly discuss the implications of different large $N_c$ limits. We stress that the SU(6) spin-flavor structure of large $N_c$ baryon
allows a convenient classification of highly excited resonances into SU(3) multiplets and predicts mass ranges for the missing partners. 
\end{abstract}

\maketitle

\clearpage

\tableofcontents

\section{Introduction}

Understanding the baryon structure directly from Quantum Chromodynamics (QCD), the theory of strong interactions,
is a basic problem of hadronic physics.
In 1974 two new papers heralded a new era in low energy QCD. One was the paper by 't Hooft \cite{HOOFT}
who proposed a perturbative expansion in  QCD, in powers of $1/N_c$,
where $N_c$ is the number of colors. The other was Wilson's paper \cite{WILSON} who discretized the continuum
Euclidean space on a grid, laying the foundation of lattice calculations.

Tremendous progress has been achieved since 1974
in lattice QCD which has reproduced the ground state baryon masses at a few percent level 
and lattice results of several groups are in agreement. However the extraction 
of resonant states remains a very difficult problem. There are large statistical and systematic errors.
Traditionally all these states are treated as stable states but exploratory steps 
have been made in the direction of resonant states.  For a recent review see, for example, 
Refs. \cite{Lang:2013eca} and \cite{Mohler:2012nh}.
Also most studies are restricted to the first positive  and negative parity resonances of total angular momentum $J = 1/2$,
namely the Roper $N(1440)1/2^+$ and the $N(1535)1/2^-$ resonance respectively (see for example Ref. \cite{Alexandrou:2013fsu}).
However, it was at least possible to show that the number of each spin and flavor states in the lowest energy bands is in agreement
with the expectations based on a weakly broken SU(6) $\times$ O(3) symmetry \cite{Edwards:2012fx}, used in quark models and in the 
treatment of excited states in large $N_c$ QCD,  as presented in this paper.

On the other hand, the $1/N_c$ expansion of QCD, proposed by 't Hooft, which has been extended by Witten \cite{WITTEN0} and applied to  
baryons  \cite{WITTEN}, has a clear phenomenological success. It offers the possibility of studying various baryon properties in a more direct way.
Presently it is considered to be a model independent, powerful and  systematic tool for
baryon spectroscopy. This  method is based on the discovery
that, for $N_f$ flavors, the ground state baryons
display an exact contracted SU($2N_f$) spin-flavor symmetry in the
large $N_c$ limit of QCD \cite{Gervais:1983wq,Dashen:1993as}. 
Such a symmetry follows from consistency conditions on meson-baryon 
scattering amplitude which must be satisfied for the theory to be unitary.  
As a consequence,
at $N_c \rightarrow \infty$ the ground state baryons are degenerate.
At large, but finite $N_c$, the spin-flavor symmetry
is broken and the mass splitting  starts at order $1/N_c$. As shown by Dashen, Jenkins and Manohar
\cite{Dashen:1993as,Jenk1,Dashen:1993jt,Dashen:1994qi},
the consistency conditions restrict the form of subleading $1/N_c$ corrections, such as 
definite predictions can be made. An operator reduction rule simplifies the  $1/N_c$ expansion.

The $1/N_c$ expansion method is closer to QCD than the quark models so that it provides a deeper 
understanding of the success of various quark models. This means that many results obtained in the
nonrelativistic quark model, the bag model or the Skyrme model can be proven in large $N_c$ QCD to order 
$1/N_c$ or $1/N_c^2$, as we shall discuss. 
Being based on group theory it allows to classify baryonic states and make predictions for the not yet discovered members of SU(6)
multiplets and study their properties.

The lattice QCD and the $1/N_c$ expansion can be combined together.
Lattice simulations with a varying number of colors are extremely useful for confirming the validity 
of the $1/N_c$ expansion. So far, one was able 
to demonstrate that the results of the real world where  $N_c$ = 3 are already ''close'' to $N_c = \infty$
\cite{Teper:1998te}.  
A summary of such recent lattice studies and the 
extrapolation to the 't Hooft limit  can be found in a recent comprehensive review
paper \cite{Lucini:2012gg}.

In addition, the existing lattice simulations at $N_c$ = 3 for ground state baryons were 
able to test important features  of the $1/N_c$ expansion results,
in particular the baryon mass relations.
Lattice data display both the $1/N_c$ expansion and SU(3) flavor-symmetry breaking 
hierarchies  \cite{Jenkins:2009wv}.

Since 1974 large $N_c$ QCD played an important role in phenomenology as well as in a number of theoretical
developments in gauge theories as for example the fundamental problems of confinement and spontaneous symmetry breaking. 
The status of large $N_c$ QCD thirty years later after its introduction by 't Hooft can be found in Ref. \cite{TRENTO2004}. 
For pure fundamental aspects, as
for example, AdS/CFT duality, gravity and string theory approaches to flavor physics, phenomenology of quark-gluon plasma, etc.
one can consult the proceedings of an workshop held in 2011 \cite{FLORENCE2011}.

Presently there are several excellent reviews on  large $N$  where one can see that the SU($N$) field theories simplify
when  $N$ becomes large and  the solutions to these theories possess an expansion in $1/N$. We could refer the reader, for example,  to Manohar's
lectures \cite{Manohar:1998xv}, partly based on the treatment proposed by Coleman  \cite{COLEMAN} with examples of theories with fields 
which transform either according to the vector representation (one index representation) or to the adjoint representation (a two-index representation),
which can be used in the case of QCD. Manohar's lectures also rely on Witten's papers, directly related to QCD \cite{WITTEN0,WITTEN}.
Also, several properties of large $N_c$ QCD were shortly described by Bhaduri \cite{BHADURI} from general arguments, where contact was made with the 
Skyrme model of the baryon, mentioned again in Sec. \ref{MN}, as equivalent to the non-relativistic quark model in the large $N_c$ limit.

The adjoint representation carries two indices, the upper one labels the basis vectors of the fundamental representation,  like for quarks
and lower index, corresponds to its complex conjugate, like for antiquarks, as described in Sec. \ref{differentlimits}.
In this representation the gluon field therefore has two indices. 
This inspired 't Hooft to introduce the double line notation for gluons (Fig. 1) which provides a simple way to keep 
track of the color index contraction and find the combinatoric 
factors in a Feynman diagram and the $N_c$-counting rules.

\begin{figure}[t]
\begin{center}
\includegraphics[width=4cm,keepaspectratio]{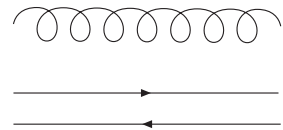}
\end{center}
\caption{A gluon in the traditional and in the double line notation.}
\label{double}
\end{figure}

\begin{figure*}[pt]
\begin{center}
\includegraphics[width=15cm,keepaspectratio]{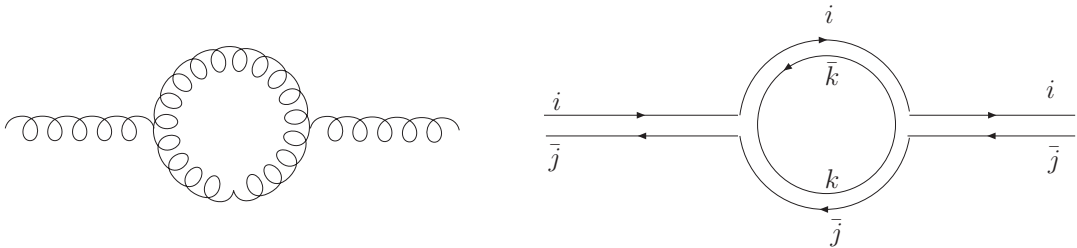}
\end{center}
\caption{The gluon vacuum polarization diagram in the standard (left) and the double line notation of 't Hooft (right).}
\label{glueball}
\end{figure*}

For example, let us  consider the one-loop gluon vacuum polarization diagram (Fig. \ref{glueball}). From the right part of Fig. \ref{glueball} it is easy to determine its combinatoric factor depending on $N_c$. Indeed, the color quantum numbers of the initial and final states are specified but not the inner index $k$ which leads to a combinatoric factor equal to $N_c$ for this Feynman diagram. At $N_c \to \infty$ the contribution of this diagram would be 
infinite.

To obtain a finite limit  for this process, one can renormalize the theory by introducing a new coupling constant $g/\sqrt{N_c}$ instead of $g$. Then 
\begin{equation}
\frac{g}{\sqrt{N_c}}\to 0\ \mathrm{when}\ N_c\to \infty,
 \label{renormalisation}
\end{equation}
where $g$ is fixed when $N_c$ becomes large. 
 In the one-loop gluon vacuum polarization we have two vertices and one combinatoric factor  $N_c$. With this renormalization, the order of the Feynman diagram in Fig. \ref{glueball} becomes 
\begin{equation}
 \left(\frac{g}{\sqrt{N_c}}\right)^2 N_c = g^2
\end{equation}
independent of $N_c$ as expected, and thus finite. For the combinatoric factors of more complex Feynman diagrams the reader 
is referred to Witten's paper \cite{WITTEN}. The main conclusion is that the leading Feynman diagrams are planar and contain 
a minimum number of quark loops.

As mentioned, the application of the  $1/N_c$ expansion method to baryon spectroscopy combines Witten's developments  \cite{WITTEN}
and the discovery that for $N_f$ flavors, the ground state baryons
display an exact contracted SU($2N_f$) spin-flavor symmetry in the
large $N_c$ limit of QCD \cite{Gervais:1983wq,Dashen:1993as}. Therefore the large $N_c$ analysis for baryons is
quite subtle, more subtle than that for mesons. The counting rules were studied by Witten \cite{WITTEN}.
It has been first used to study the ground state baryon masses, described
by the symmetric representation $\bf 56$ of SU(6)  
and other properties as, for example,  axial vector couplings,  magnetic moments, heavy quark baryons, etc. 
\cite{Dashen:1993as,Jenk1,Dashen:1993jt,Dashen:1994qi,CGO94,Luty:1993fu,Jenkins:1995td,DDJM96}.

The success of the $1/N_c$ expansion method in describing ground state baryons raised
the question whether excited baryons can be described as well by the same method. It
is not obvious that the consistency condition used for the ground state is applicable to
excited baryons because excited baryons are not generically stable asymptotic states
even at large $N_c$. Witten has shown that the characteristic width of an excited baryon is
$N^0_c$ \cite{WITTEN}. Based on the argument that for some class of states the width goes
like $N^{-1}_c$, Pirjol and Yan \cite{PY} have shown that a contracted SU($4$) 
symmetry also exists for those excited states. From requiring the pion-excited baryon scattering 
amplitude to satisfy Witten's large   $N_c$   counting rules,   they derived   consistency conditions  
analogous to those obtained by Dashen, Jenkins and Manohar \cite{Dashen:1993as,Jenk1,Dashen:1993jt,Dashen:1994qi}
for $s$-waves baryons. Pirjol and Yan have also shown that the solutions to large $N_c$
consistency conditions coincide with the predictions of the nonrelativistic quark model for
excited states. 

The legitimacy of the procedure used by Pirjol and Yan has later been questioned by 
Cohen and Lebed \cite{COLEB1}, inasmuch as the characteristic width of an excited baryon is 
$N^0_c$ according to Witten \cite{WITTEN}. Cohen and Lebed have tried to support the applicability of 
the $1/N_c$ expansion method by studying the compatibility between the scattering picture and the 
quark model picture, using quark operators as defined in Refs. \cite{Dashen:1993as,Jenk1,Dashen:1993jt,Dashen:1994qi}. 
They have shown that the two pictures share the same pattern of degeneracy, giving rise to degenerate sets of resonances,
identical in their quantum numbers $J$ and $I$ in both pictures at fixed grand spin $K$, as discussed in Sec. \ref{MN}.
From there they concluded that the two pictures are generically compatible.

In practice the extension of the $1/N_c$ expansion method
to excited states is also based on the observation that these states can approximately be classified  
as  SU($2N_f$) multiplets, and that the resonances can be grouped into 
excitation bands, $N $ = 1, 2, ..., as in quark models,
each band containing a number of SU(6) $\times$ O(3) multiplets. 

The symmetric multiplets of these bands are similar to the ground  state from group theory
point of view. Therefore they were analyzed by analogy 
to the ground state. In this case the mass splitting starts at order $1/N_c$ as well.

That is why this review  is largely be devoted to mixed symmetric states for which, by contrast to symmetric   
states, the splitting starts at order $N^0_c$. The problem is in fact that 
these states are technically more difficult to be studied and two distinct procedures have been proposed so far.
The first procedure was based on the Hartree approximation, in the spirit 
Witten's arguments \cite{Goi97}. In this procedure the system of $N_c$
quarks is split into a ground state core which creates a mean field and an excited quark moving in this field.
Then the Pauli principle is fulfilled by the core wave function only, but not by the total wave function \cite{CCGL}. 
As we shall see, the numerous applications of this procedure were  mostly restricted to the $N$ = 1 band, but concerned 
both the study of the baryon spectra and their electromagnetic and strong decays. We understand that the technical advantage of this method
was that the matrix elements of the  SU($2N_f$) generators, needed in the calculations of spectra and decays,
  were known at that time for symmetric states  only, but not for 
mixed symmetric spin-flavor states. A disadvantage is that the number
of terms included in an operator describing an observable becomes generally large and it is difficult to select 
the dominant ones.

Later an alternative procedure, based on the identity of all quarks in the system, has been proposed in Ref. \cite{Matagne:2006dj}.
There is no physical reason to  
separate the excited quark from the rest of the system.  The total
wave function is completely antisymmetric, the orbital-spin-flavor part is symmetric, being combined with an antisymmetric color part. The orbital and 
the spin-flavor 
parts have the same mixed symmetry. The method can straightforwardly be applied to all bands, including more than one excited quark. 
The analytic form of the matrix elements of the SU($2N_f$) generators for the necessary mixed symmetric spin-flavor states have 
been obtained as described in Ref. \cite{Matagne:2008kb}.
A quantitative analysis was  performed for a number of  SU($2N_f$)  $\times$ O(3) multiplets
for which data exist. They correspond to the excitation bands with $N $ = 1, 2 or 3.
In the nonstrange sector it covers a resonance mass region up to about 2.5 GeV. So far only the spectrum has been
analyzed. There is an expectation that the decays will also be considered within this approach.

One important goal of the $1/N_c$ expansion method was to understand whether the success of the nonrelativistic
quark model has a natural explanation in large $N_c$ QCD. Various studies presented here prove the compatibility
between quark models and the $1/N_c$ mass formula. An interesting outcome is the similarity between the
Regge trajectories resulting from both the $1/N_c$ expansion method and the quark models.

The paper is organized as follows. In the next  section we introduce the definition of large $N_c$ baryons 
according to 't Hooft and Witten.  In Sec. \ref{sfsymmetry} we
sketch the derivation of the contracted SU($2N_f)$ spin-flavor symmetry and recall the resulting su$( N_f)_c$ algebra.
The baryon operator expansion method is described in Sec. \ref{barexp}.
The latest results on the ground state baryons, as for example, the magnetic moments,  make the subject of Sec. \ref{groundstate}.
After introducing the formalism of the $1/N_c$ expansion, Sections \ref{exsym} - \ref{specmixed} are devoted to the 
study of excited states, with a 
special emphasis to the two distinct approaches to treating mixed symmetric spin-flavor states.
The extension to heavy baryon masses is considered in Sec. \ref{heavy}. 
Section \ref{versus} contains considerations about 
the compatibility between the  $1/N_c$ expansion and a quark model mass formula. Important qualitative support to the 
$1/N_c$ expansion  method applied to excited states is brought in Sec. \ref{MN} by a comparison between the quark excitation picture 
to order $N^0_c$ and the meson-nucleon scattering picture. 
The combined $1/N_c$ and chiral expansions are updated in Sec. \ref{combchiral}.
The present status of the strong and electromagnetic decays is summarized in Secs. \ref{strongdecays} and \ref{photo}.
A short discussion of various large $N_c$ limits, including that of 't Hooft is given in Sec. \ref{differentlimits}. 
Some of the appendices are devoted to the extended Wigner-Eckart theorem and the derivation of isoscalar factors of SU(6) generators
needed in this work. General analytic expressions  are reproduced. They could perhaps be applied to other fields, 
in particular to systems where the hypercharge is a good quantum number.


\section{Large $N_c$ baryons}\label{Ncbaryons}

According to Witten, large $N_c$ baryons are colorless bound states composed of $N_c$
valence quarks described by a completely antisymmetric color wave function of the form
\begin{equation}\label{CA}
C^A =  \varepsilon_{i_1i_2i_3\ldots i_{N_c}}q^{i_1}q^{i_2}q^{i_3}\ldots q^{i_{N_c}}.
\end{equation}

Then the total wave function of such a system can be obtained by combining $C^A$ with 
the orbital part $\psi_{\ell m}$, the spin part $\chi$ and the flavor part $\phi$ 
by using Clebsch-Gordan coefficients  of  the permutation group $S_{N_c}$ \cite{Stancu:1991rc}
to obtain a totally antisymmetric 
wave function written symbolically as
\begin{equation}
\Psi = \psi_{\ell m}~ \chi~ \phi~ C^A. 
\end{equation}
Because  $C^A$ is antisymmetric the product $\psi_{\ell m} \chi \phi$ must be symmetric.
For the ground state $\psi_{\ell m}$ is symmetric, inasmuch as all identical quarks are in an $s$-state. 
Therefore the product  $\chi \phi$ must be symmetric which makes the study of the ground state 
rather easy. For excited states described by mixed symmetric orbital states the product 
$\chi \phi$ must have the same mixed symmetry, as discussed in Sec. \ref{exactwavefunction}.

The number of quarks inside a large  $N_c$ baryon grows as $N_c$. 
Witten has proposed to describe such a system by a Hartree approximation where each quark experiences the
same average potential.  In this approximation one has 
\begin{equation}\label{Hartree}
M_{baryon} \sim \mathcal{O}(N_c).
\end{equation}
On the other hand the size of the baryon is governed by the confinement scale $\Lambda^{-1}_{QCD} \simeq $ 1fm which is fixed.
Thus the quark density must increase with $N_c$. Corrections to the Hartree approximation follow 
from the spin-flavor structure of baryons discussed below.
\section{Spin-flavor symmetry}\label{sfsymmetry}

\begin{figure}[t]
\begin{center}
\includegraphics[width=6cm,keepaspectratio]{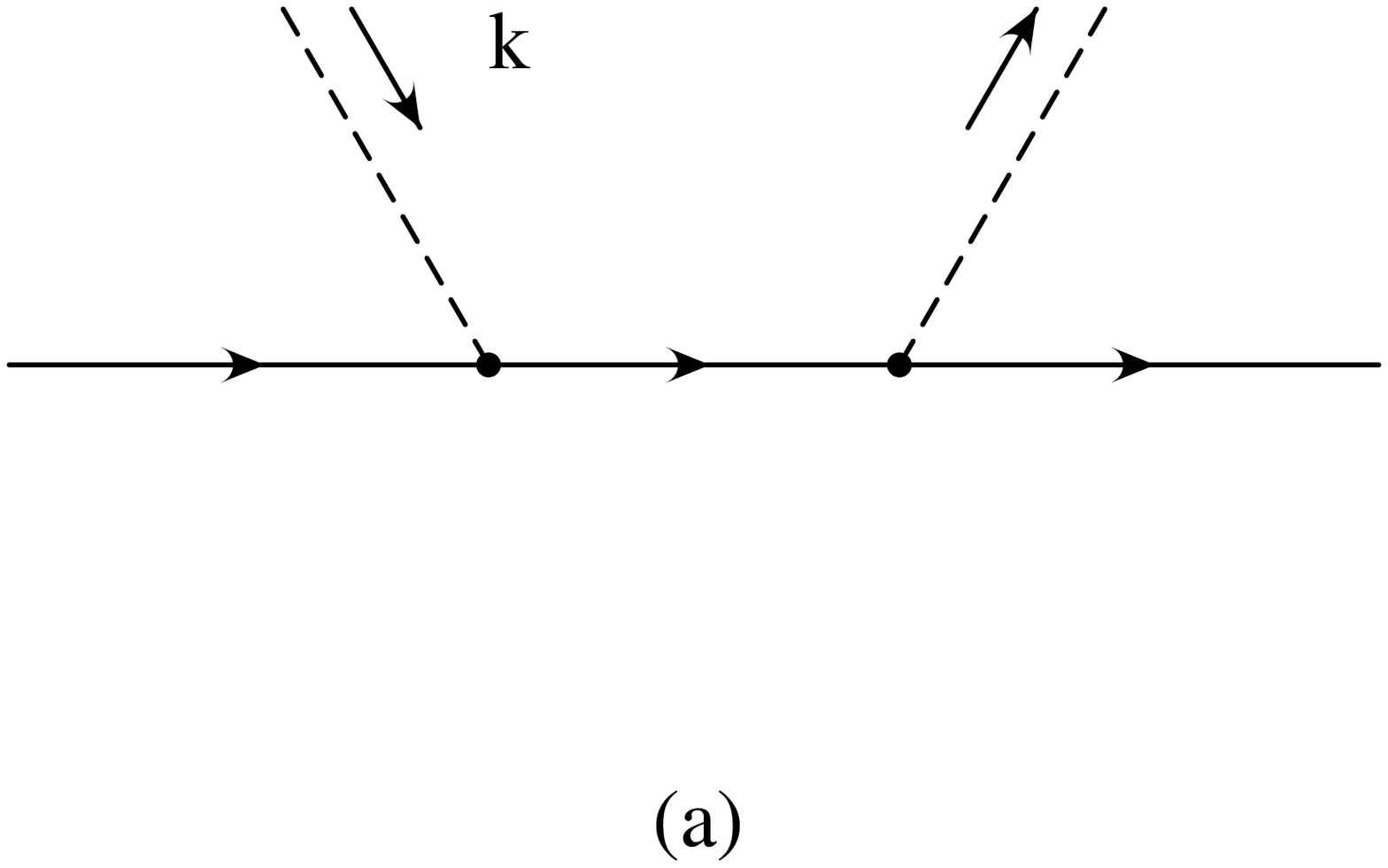} 
\hspace{2cm} 
\includegraphics[width=6cm,keepaspectratio]{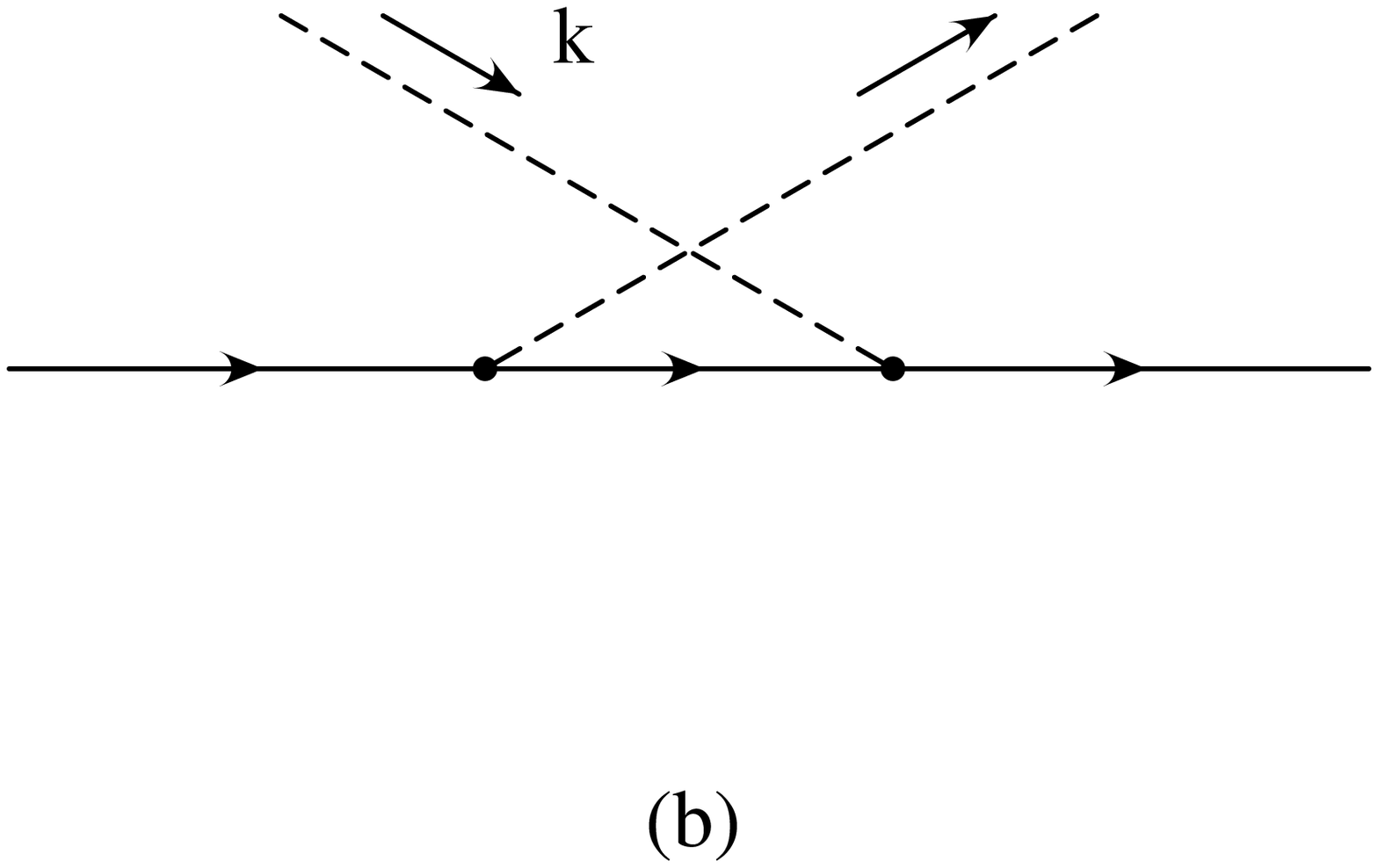}
\end{center}
\caption{Leading-order diagrams for the scattering $B+\pi \to B+\pi$.}
\label{pionbarver}
\end{figure}

In Refs. \cite{Gervais:1983wq,Dashen:1993as} a set of consistency conditions were derived for the
pion-nucleon coupling constants in the large $N_c$ limit of QCD. The arguments were based on
the large $N_c$ counting rules for meson-baryon scattering, analyzed by Witten \cite{WITTEN}, who
showed that the baryon mass, Eq. (\ref{Hartree}), and
the axial vector coupling constant $g_A$ are  $\mathcal{O}(N_c)$ and that the pion decay 
constant $f_{\pi}$ is  $\mathcal{O}({N_c}^{1/2})$. Then the pion-nucleon vertex $g_A \vec{q}/f_{\pi}$,
where $\vec{q}$ is the pion momentum, grows as ${N_c}^{1/2}$ at fixed pion energy.

In the large $N_c$ limit, as the baryon is infinitely heavy compared with the pion, the time component of the baryon-pion coupling  
vanishes. Then, the space components of the axial vector current matrix element can be written as
 \begin{equation}
  \langle B |\bar{q}\gamma^i\gamma_5T^aq|B\rangle = g N_c \langle B|X^{ia}|B \rangle,
 \end{equation}
with the coupling constant $g$ factored out so that  $g$ 
and $\langle B|X^{ia}|B \rangle$ are of order  $\mathcal{O}(N_c^0)$, which means that in this definition $X^{ia}$ is an operator 
defined on nucleon states which has a finite large $N_c$ limit.

Considering the direct + crossed diagrams, Fig.~\ref{pionbarver},
the pion-baryon scattering amplitude becomes
 \begin{equation}
 \mathcal{A(\pi B \rightarrow \pi B}) \propto
-i \frac{N_c^2 g^2}{f_\pi^2}\frac{q^iq'^j}{q^0}\left[X^{ia},X^{jb}\right],
 \end{equation}
where the initial and final baryons are on-shell and 
$X^{ia}$ is the baryon-meson vertex operator 
acting on the spin-flavor nucleon states $\mathcal{B}$. 
When $N_f$ = 2, for example, in the product of $X'$s
the sum runs over all the possible values of the spin and isospin 
intermediate baryon states. According to Refs. \cite{Gervais:1983wq,Dashen:1993as} there must be other states that cancel the
order $N_c$ of the amplitude above so that the total amplitude is order one and consistent with unitarity. These states form 
an infinite tower of degenerate baryon states which are the solutions of the following consistency condition 
\begin{equation}\label{constraint}
 N_c\left[X^{ia},X^{jb}\right]\leq \mathcal{O}(1).
\end{equation}
If one makes the expansion 
\begin{equation}\label{expan}
 X^{ia}=X^{ia}_0+\frac{1}{N_c}X^{ia}_1+\frac{1}{N_c^2}X^{ia}_2+ \cdots.
\end{equation}
the constraint (\ref{constraint}) requires
\begin{equation}
 \left[X^{ia}_0,X^{jb}_0\right]=0,
 \label{consist1}
\end{equation}
when $N_c \to \infty$. 
In this limit, the spin operators $S^i$ ($i$ = 1,2,3), the flavor operators $T^a$ ($a$ = 1,2,...,$N_f$), together with the spin-flavor operators $X^{ia}$ 
can be identified with the generators of a contracted spin-flavor group  SU($2N_f)_c$, where $N_f$ is the number of flavors.
Its algebra is 
\begin{eqnarray}
 \left[S^i,T^a\right]& = & 0, \nonumber \\
 \left[S^i,S^j\right]& = & i\varepsilon^{ijk}S^k, \ \ \left[T^a,T^b\right] = if^{abc}T^c, \nonumber \\
 \left[S^i,X_0^{ja}\right] & = & i\varepsilon^{ijk}X_0^{ka}, \ \ \left[T^a,X_0^{ib}\right]=if^{abc}X_0^{ic}, \nonumber \\
 \left[X_0^{ia},X_0^{jb}\right]& = & 0. 
\end{eqnarray}
On the other hand the su(2$N_f$) algebra reads
\begin{widetext}
\begin{eqnarray}\label{ALGEBRA}
&[S^i,S^j]  =  i \varepsilon^{ijk} S^k,
~~~~~[T^a,T^b]  =  i f^{abc} T^c,   ~~~~~~ [S^i,T^a] = 0,   \nonumber \\
&[S^i,G^{ja}]  =  i \varepsilon^{ijk} G^{ka},
~~~~~[T^a,G^{jb}]  =  i f^{abc} G^{jc}, \nonumber \\
&[G^{ia},G^{jb}] = \fr{i}{4} \delta^{ij} f^{abc} T^c
+\frac{i}{2} \varepsilon^{ijk}\left(\frac{1}{N_f}\delta^{ab} S^k 
+d^{abc} G^{kc}\right),
\end{eqnarray}
\end{widetext}
Thus the  contracted algebra su(2$N_f$)$_c$ is obtained from the commutation relations (\ref{ALGEBRA})  by taking the limit
\begin{equation}
\label{wignerlimit}
X_0^{ia}=\lim_{N_c\to \infty}\frac{G^{ia}}{N_c}.
\end{equation}

The first formal notion of the operation of group or algebra contraction was given by Segal in 1951 \cite{SEGAL51}
who has considered a sequence of Lie groups. His approach is more general than that of 
In\"on\"u and Wigner \cite{INONUWIGNER53} who have introduced the definition from the physicist's point of view in 1953. 
In their paper they investigate, in some generality, in which sense groups can be limiting cases of other groups.
The observation was that the classical mechanics 
is a limiting case of relativistic mechanics when the velocity of light becomes infinite. 
In\"on\"u and Wigner called  ${\it contraction}$ the operation of obtaining a new group by a new singular 
transformation of the infinitesimal elements of the old group. 
The contracted infinitesimal elements form an abelian invariant subgroup of the contracted group.
In this way the  Lorentz
group contracts to the Galilei group. 
In field theory the limit can be related to coupling constants \cite{HERMANN} as in  Eq. (\ref{renormalisation}). In the present case
the parameter of the singular transformation should be $\frac{1}{{\sqrt{N_c}}}$. 
In the limit $N_c \rightarrow \infty$ one obtains the operators (\ref{wignerlimit}), as elements of an algebra nonisomorphic 
to the original one. 

Dashen and Manohar \cite{Dashen:1993as}
have solved Eq. (\ref{consist1}) for $N_f$ = 2 finding in this way the simplest irreducible representations
of SU(4)$_c$. 
Dashen, Jenkins and Manohar  \cite{Dashen:1993jt,Dashen:1994qi} classified all possible representations of the
contracted spin-flavor algebra using the theory of induced representations \cite{Mackey}.
This theory gives a complete
classification of all irreducible representations of a semidirect product $\mathcal{G}  \wedge \mathcal{A}$
of a compact Lie group $\mathcal{G}$ and an abelian invariant subgroup $\mathcal{A}$.
In  large $N_c$ QCD $\mathcal{G}$ is the direct product $SU(2) \otimes SU(N_f)$ and the group $\mathcal{A}$ generates
an abelian invariant subalgebra, the elements of which are $X^{ia}_0$ defined in
Eq. (\ref{consist1}), as first pointed out by Gervais and Sakita \cite{Gervais:1983wq}.

The basis vectors of the induced representations form  infinite towers of degenerate $(S,I)$ baryon states,
each tower corresponding to a value of the grand spin $\vec{K} = \vec{I} + \vec{S}$, like in the Skyrme model
or the non-relativistic quark model in the large $N_c$ limit. The identification with physical states
can be made by assuming that $K$ = $n_s/2$ where $n_s$ is the number of strange quarks in a baryon.
The simplest irreducible representation for two flavors is a tower of states with $K$ = 0 and $S$ = $I$ = 1/2,~3/2,~5/2,....

Dashen and Manohar \cite{Dashen:1993as} and Jenkins \cite{Jenk1} found that subleading $1/N_c$ corrections to 
the Hartree approximation, Eq. (\ref{Hartree}), are constrained by additional large $N_c$ conditions and they proposed a 
systematic expansion in powers of $1/N_c$ for ground state baryons.

A diagrammatic and group-theoretical analysis of baryons in the $1/N_c$ expansion had also been used by Carone, Georgi and Osofsky
\cite{CGO94} and independently by Luty and March-Russell \cite{Luty:1993fu}
shortly before similar results were obtained in Ref. \cite{Dashen:1994qi}. The authors of Refs. \cite{CGO94,Luty:1993fu}
have used the quark representation approach which is closely tied to the intuitive picture of baryons as quark bound states.

The connection to the nonrelativistic quark model can be hinted from the work of Gervais and Sakita \cite{Gervais:1983wq} 
in the following way. To obtain the quark representations of the group $\mathcal{G}$ defined above one has to consider the symmetric state 
of $N_c$ quarks, reduce it to the representation of the direct product  $SU(2) \otimes SU(N_f)$ and take the limit   
$N_c\to \infty$. 

\section{Baryon operator expansion}\label{barexp}

According to the previous section 
there are two kinds of representations of the spin-flavor algebra of baryons in large $N_c$.  
One is the Skyrme model representation as obtained by 
Dashen, Jenkins and Manohar  \cite{Dashen:1993jt,Dashen:1994qi} using the theory of induced representations.
At $N_c \rightarrow \infty$ the spectrum consists of infinite towers of states of given spin $S$ and isospin $I$ 
combined to a fixed value of the grand spin $K$.  
The other is the quark representation proposed in Refs. \cite{CGO94,Luty:1993fu} which is closely related to
the non-relativistic quark model, and is convenient to study decays as well. 

The  Skyrme and the non-relativistic quark model representations of the large $N_c$ spin-flavor algebra for baryons 
are identical in the $N_c \rightarrow \infty$ limit. At finite $N_c$ they differ in their organization of $1/N_c$
corrections but give equivalent results at a given order in $1/N_c$. The Skyrme representations of the 
contracted algebra are infinite dimensional. The quark representations uses the algebra (\ref{ALGEBRA}) so that the 
representations are finite. For the ground state it consists of a tower of states that terminates at spin $S = N_c/2$. 
The connection between the Skyrme and the quark representations is discussed in Ref. \cite{Dashen:1994qi}.

Below we summarize the analysis of Ref. \cite{Dashen:1994qi} of the baryon operator expansion realized in the quark representation. 
Quark operators can be classified as  $n$-body quark operators $\mathcal{O}^{(n)}$. By definition each $\mathcal{O}^{(n)}$  acts on $n$ quarks, where
$0\leq n \leq N_c$. The   SU(2$N_f$) generators are one-body operators. One can construct $\mathcal{O}^{(n)}$ for any $n$
starting from the generators of SU(2$N_f$). 
For doing this, it is convenient to express them in terms of the representations of
the direct product SU(2) $\times$ SU($N_f$). One has
\begin{eqnarray}\label{sunf}
  S^i & = & \sum_{j=1}^{N_c}
q^\dagger_j\left(S^i \times \mathbbm{1}\right)q_j \ \ \ (3,1), \nonumber \\
  T^a & = & \sum_{j=1}^{N_c}
q^\dagger_j\left(\mathbbm{1} \times T^a\right)q_j \ \ \ (1,N_f^2-1), \\
  G^{ia} & = & \sum_{j=1}^{N_c}
q^\dagger_j\left(S^i \times T^a\right)q_j \ \ \ (3,N_f^2-1), \nonumber
\end{eqnarray}
where $q^\dagger_j$ and $q_j$ are creation and annihilation quark operators with
$j$ the quark line number. The operators  $q^\dagger_j$, $q_j$ obey the Bose statistics because the ground state baryons of $N_c$
quarks is in a completely 
symmetric spin-flavor state, as mentioned in Sec. \ref{Ncbaryons}.
The brackets on the right-hand side of the definitions (\ref{sunf})
denote the (SU(2), SU($N_f$)) dimensional notation of the spin $S^i$, 
flavor $T^a$ and spin-flavor generators   $G^{ia}$ of  SU(2$N_f$) in the decomposition SU(2) $\times$ SU($N_f$). 
Thus one can see that the SU(2)  baryon spin generator $S^i$ is equal to the sum of spin generators of the 
$N_c$ quarks forming the baryon. A similar remark holds for the isospin and $G^{ia}$.
Note that, for a finite $N_c$, in the
quark representation,  $G^{ia}$ recovers the form $S^i T^a$ operator of the axial current.

The latter remark requires some useful comments.  We follow closely Ref. \cite{Jenkins:1998wy}.
From Eqs. (\ref{expan}) and (\ref{wignerlimit})
one can see that the operators  $X^{ia}_0$ and $G^{ia}/N_c$ differ at the subleading order $1/N_c$.
The ambiguity in the choice of the spin-flavor generator arises from the fact that the contracted
spin-flavor algebra for baryons is exact only in the large $N_c$ limit, which means that matrix
elements of the spin-flavor generators $G^{ia}$ 
are known to leading order in $1/N_c$ up to a normalization factor. When $X^{ia}_0$ is used,
the operator basis of the $1/N_c$ expansion is the same as the operator basis of the large $N_c$
Skyrme model, whereas the operator basis constructed from the algebra (\ref{ALGEBRA}) is the 
same as the operator basis of the large $N_c$ nonrelativistic quark model. Both operator bases 
parametrize the same large $N_c$ pĥysics encoded in coefficients entering the expansion formula 
as introduced below.

Here we deal with the quark model representation. In such a case any QCD operator which transforms as an 
irreducible representation of SU(2) $\times$ SU($N_f$) can be written 
as an expansion in $n$-body quark operators $\mathcal{O}^{(n)}$, 
which transform under the same irreducible representation. For the ground state baryons one has
 \cite{Dashen:1993jt,Dashen:1994qi,Jenk1,CGO94,Luty:1993fu}
\begin{equation}
 \mathcal{O}_{\mathrm{QCD}}=\sum_{n}c^{(n)} \frac{1}{N_c^{n-1}}\mathcal{O}^{(n)},
 \label{qcdoperatorexpansion}
\end{equation}
where $c^{(n)}$ are  unknown dynamical coefficients. Each $\mathcal{O}^{(n)}$ operator is accompanied 
by a factor of $1/N_c^{n-1}$, which comes from the fact that one needs at least $n-1$ gluon exchanges 
at the quark level to generate $n$-body effective operators in the $1/N_c$ expansion out of one-body QCD operators. 
A generic $n$-body operator can be written as a homogeneous
 $n$-th degree polynomial in the generators $S^i$, $T^a$ and $G^{ia}$ (up to an occasionally zero-degree rescaling term). As we shall see, the matrix elements of 
the baryon observables can be calculated in terms of the matrix elements of the baryon spin-flavor generators $S^i$, $T^a$ and $G^{ia}$.
Quark operator identities can be used to construct a linearly independent complete operator basis of $n$-body operators.
Operator reduction rules to simplify the 
$1/N_c$ expansion have also been derived \cite{Dashen:1994qi}.

The extension to strange baryons, where SU(3)
is broken, will be presented in the following,  as well as the extension to excited states. 

The method has been applied to baryon masses, axial currents, magnetic moments, etc.
Comprehensive reviews can be found, for example,  in Refs. \cite{Jenkins:1998wy,Jenkins:2001it}.
Therefore we shall not give details of these achievements, we shall only recall a few basic results and 
mention the latest developments. 

\section{Ground state observables}\label{groundstate}

It was later noticed that SU(3) flavor breaking cannot be neglected relative to $1/N_c$ corrections.
The $1/N_c$ operator expansion can be generalized to include SU(3) breaking as shown below.

\subsection{Masses}

As far as the masses are concerned,
the original $1/N_c$ expansion (\ref{qcdoperatorexpansion}) has been combined with a perturbative flavor breaking by Jenkins and Lebed
\cite{Jenkins:1995td}. The SU(3) symmetry breaking was implemented to $\mathcal{O}(\epsilon)$ where the parameter $\epsilon$ 
represents the quark mass difference divided by the chiral symmetry breaking scale, which is of order of 1 GeV. 
This lead to a generalized form of the mass operator used 
currently in calculating the spectra of both non-strange and strange baryons. The generalized mass operator is 
\begin{equation}
\label{massoperator}
M = \sum_{i}c_i O_i + \sum_{i}d_i B_i .
\end{equation} 
The first sum contains the operators  $O_i$ which are SU($N_f$) invariants and the operators $B_i$ in the 
second sum break SU(3) explicitly and have zero expectation values for nonstrange baryons.

In the formula (\ref{massoperator}) $O_1$ is the leading spin-flavor (SF) singlet operator
proportional to $N_c$ and $O_i$ with  $i > 1$ bring $1/N_c$ corrections which estimate the amount of SF symmetry breaking. 

For further purposes, 
the operators $O_i$ of Eq.  (\ref{qcdoperatorexpansion}), with  $i > 1$ are here defined such as to be applied to orbitally excited baryons as well,
besides the ground state baryons.  
They are SU(2) scalar products
\begin{equation}\label{OLFS}
O_i = \frac{1}{N^{n-1}_c} O^{(k)}_{\ell} \cdot O^{(k)}_{SF},
\end{equation}
where  $O^{(k)}_{\ell}$ is a $k$-rank tensor in SO(3) and  $O^{(k)}_{SF}$
a $k$-rank tensor in SU(2)-spin, but invariant in SU($N_f$).
For the ground state one has $k = 0$. The excited
states also require  $k = 1$  and $k = 2$ terms.
The rank $k = 1$ tensor has as components the generators $L^i$ of SO(3). 
The components of the $k = 2$ tensor operator of SO(3) are
\begin{equation}\label{TENSOR} 
L^{(2)ij} = \frac{1}{2}\left\{L^i,L^j\right\}-\frac{1}{3}
\delta_{i,-j}\vec{L}\cdot\vec{L},
\end{equation}
which, like $L^i$, act on the orbital wave function $|\ell m \rangle$  
of the whole system of $N_c$ quarks. 
Examples will be given throughout the paper.

Presently we illustrate the formula (\ref{qcdoperatorexpansion}) with the simple case of ground state nonstrange 
baryons. Due to the operator identities \cite{Dashen:1994qi}
\begin{equation}
\{S^i,S^i\} + \{T^a,T^a\} + \{G^{ia},G^{ia}\}  =  \frac{3}{2} N_c(N_c+4), 
\end{equation}
\begin{equation}
\{T^a,T^a\}  =   \{S^i,S^i\},
\end{equation}
where the right hand side in first identity is the eigenvalue of the SU(4) Casimir operator for a spin-flavor 
symmetric state described by the partition $[N_c]$ one can reduce the mass operator to the simple form
\begin{equation}\label{GS}
 M = c_1 N_c + c_3\frac{1}{N_c}S^2 + \mathcal{O}\left(\frac{1}{N_c^3} \right),
\end{equation}
where we use a notation for $c_i$ to be consistent with the rest of the paper.
Thus the mass splitting starts at order $1/N_c$. Usually higher orders  $\mathcal{O}\left(1/N_c^3 \right)$
are neglected. Using the nucleon mass $m_N \simeq $ 940 MeV and  the $\Delta(1232)$ resonance mass one can obtain
\begin{equation}\label{gsmass}
c_1 \simeq 289 \,~ \mathrm{MeV}, ~~~c_3 \simeq 292 \,~ \mathrm{MeV}.
\end{equation}
The coefficient $c_1$ is close to the constituent mass of the quarks $u$ or $d$ and $c_3$ reproduces the 
hyperfine $\Delta-N$ splitting. 

As already mentioned, the corrections due the flavor symmetry breaking  are proportional to the parameter  
$\epsilon \sim $ 0.25  \cite{Jenkins:1995td}. The $1/N_c$ expansion, including flavor symmetry breaking,
predicted a hierarchy of spin and flavor symmetry 
relations for QCD baryons that is observed in nature. It also provided a quantitative understanding of why 
some SU(3) flavor symmetry relations in the baryon sector, 
as, for example, the Gell-Mann-Okubo mass formula or the Coleman-Glashow mass relation \cite{colemanglashow},
are satisfied to a greater precision than expected from flavor symmetry breaking suppression factors alone \cite{Jenkins:2000mi}. 
A detailed updated summary, 
including isospin breaking,
can be found in Ref. \cite{Jenkins:2001it}.

\subsection{Axial vector couplings}

In the exact SU(3) symmetry limit the baryon axial vector current operator $A^{ia}$ is a
rank one tensor operator in SU(2)-spin and in SU(3) it transforms as a flavor adjoint. Its group structure is of
the form of $G^{ia}$ of Eq. (\ref{sunf}).
An extended analysis in the $1/N_c$ expansion has included up to 3-body operators in Ref. \cite{Dashen:1994qi}.
The expansion at linear order in SU(3)-flavor breaking involves additional SU(2)-spin rank 1 operators in different 
flavor representations  \cite{Dashen:1994qi}. A comparison with the experimental data,
has been performed in Ref. \cite{DDJM96}.
The present status is summarized in Ref. \cite{Jenkins:2001it}.

\subsection{Magnetic moments}

In the exact SU(3) flavor symmetry limit the baryon magnetic moment operator 
is a rank one irreducible
tensor operator in SU(2)-spin  and an octet in SU(3)-flavor.

The $1/N_c$ expansion has been analyzed by several authors 
\cite{Dashen:1993jt,Dashen:1994qi,LMRW95,Jenkins:1994md,DDJM96,Lebed:2004fj,FloresMendieta:2009rq,Ahuatzin:2010ef,Jenkins:2011dr}.
Using the  method of Ref. \cite{Lebed:1995} to classify static observables 
the complete set of 27 linearly independent operators 
of the octet and decuplet ground state baryons organized in powers of $1/N_c$ in their matrix elements was given in
Table I of Ref. \cite{Lebed:2004fj}  for the component $i$ = 3. Operator demotions have been taken into account in the $1/N_c$
power dependence.

The "operator  demotion" was first defined in Ref. \cite{CCGL}. In a demotion one identifies
a linear combination of operators whose matrix elements are a higher order in powers of $1/N_c$ than those of the component 
operators, so that only one of the components represents an independent operator at the starting order. The result depends on the 
particular states used for evaluating the matrix elements.  The example given in Ref. \cite{CCGL} clearly clarifies 
the procedure,  which is possible when it happens that the matrix elements of the SU(6) generators contain both leading and subleading 
orders in $N_c$. A look at the Tables given in Appendix B can be convincing about this  statement. 

The most updated analysis can be found in Ref. \cite{Jenkins:2011dr}
where it was shown that the combined expansion in $1/N_c$ 
and SU(3) flavor breaking is needed to understand the hierarchy of baryon magnetic moments found in nature.
The 27 linearly independent operators were written in the basis $(U_3, Q)$ defined by the chain
SU(3) $\supset$ SU$_U$(2), in terms of the $U$-spin,   instead of the usual ($T^3,T^8$) basis related to the chain SU(3)$ \supset$ SU$_I$(2),
where $I$ is the isospin.
The reason is that the magnetic moments $M^{iQ}$ of light baryons are proportional to the quark charge matrix $T^Q$ = diag(2/3,~-1/3,~-1/3), 
where the charge operator reads 
\begin{equation}
T^{Q} = T^3 + \frac{1}{\sqrt{3}}~ T^8.
\end{equation} 
This commutes with the 3rd component of $U$ defined as 
\begin{equation}
U^3 = - \frac{1}{2}~T^3 + \frac{\sqrt{3}}{2}~ T^8,
\end{equation}
both $T^{Q}$ and $U^3$ being operators which are linear combinations of SU(3) generators corresponding to vanishing roots in Cartan's classification, 
see Eqs. (8.105d) of Ref. \cite{Stancu:1991rc}. Instead of using the standard ($I_3,Y$) coordinates
one can draw a weight diagram $(U^3, T^Q)$ where each $U$-multiplet contains baryons of identical charge and the $T^Q$ operator
changes the charge by one unit in passing from one $U$-multiplet to another.
By analogy to $T^{Q}$ one can introduce the operator 
\begin{equation} 
G^{iQ} = G^{i3} + \frac{1}{\sqrt{3}}~ G^{i8}.
\end{equation}
and define the magnetic moments $M^{iQ}$ by the linear combination  \cite{Jenkins:2011dr}
\begin{equation}
M^{iQ} =a~ G^{iQ} + b~ \frac{1}{N_c} S^i T^{Q},
\end{equation}
up to order $1/N_c$,   with coefficients $a$ and $b$ to fit experimental data. 

There are 27 magnetic moments of ground state baryons, nine for the octet including the $\Sigma^0 \rightarrow \Lambda$
transition magnetic moment and ten magnetic moments for the decuplet plus eight decuplet-octet transition  magnetic moments. 
The two new experimental results on decuplet-octet transition magnetic moments $\Lambda \Sigma^{*0}$ and $\Sigma \Sigma^{*+}$
\cite{Keller:2011nt,Keller:2011aw}
were added to the analysis made in  Ref. \cite{Jenkins:2011dr}. The conclusion was that further progress in 
understanding the hierarchy of baryon magnetic moments requires additional experimental measurements.
A significant SU(3) breaking was found and this breaking is expected to be enhanced in the magnetic moments 
relative to that of other observables, which makes studying of magnetic moments particularly useful.

\subsection{Charge radii and quadrupole moments}

Studies of baryon charge radii and quadrupole moments in the $1/N_c$ expansion have been performed 
in Refs. \cite{Buchmann:2000wf,Buchmann:2002mm} for two flavors and extended in Ref.  \cite{Buchmann:2002et}
to three flavors. The charge radius is the first moment of a Coulomb monopole transition amplitude. The calculations 
have been made in the simple single-photon exchange Ansatz which requires only two 
operators to describe both the charge radii and quadrupole moments observables. In 2003 only the charge radii of
$p$, $n$ and $\Sigma^-$ were known experimentally. For the other baryons predictions were made.
On the other hand only the $\Delta \rightarrow N$ quadrupole transition matrix element has been measured
and calculated  in the above mentioned $1/N_c$ expansion. An important feature is that the leading order 
of the diagonal quadrupole moment  is $\mathcal{O}(N^0_c)$.

The quadrupole moments and charge radii are related. In the one-gluon exchange picture one has obtained
the following relation \cite{Buchmann:2002mm}
\begin{equation}
Q_{\Delta^+p} = \frac{1}{\sqrt{2}} r^2_n \frac{N_c}{N_c + 3} \sqrt{\frac{N_c + 5}{N_c - 1}},
\end{equation}
which simplifies at $N_c$ = 3 and can be obtained within other frameworks.

\section{Excited symmetric spin-flavor states}\label{exsym}

In the $1/N_c$ expansion method 
the analysis of the masses of resonances which can be assigned to the SU(6) symmetric irreducible representation, denoted
in the following by the partition $[N_c]$, can be easily performed by analogy to the ground state. 
The most convenient framework is to use Eq. (\ref{massoperator})
and SU($2N_f$) algebra with 
operators that act on  symmetric spin-flavor states obtained as inner products of SU(2) and SU($N_f$) basis vectors.
Applications to the $[{\bf 56}, 2^+]$ and $[{\bf 56}, 4^+]$ have been considered in Refs. \cite{GSS03} and \cite{MS1} respectively.
A wave function of an orbitally symmetric state  $| \ell m \rangle $,  
spin $S, S_3$ and total angular 
momentum $J, J_3$ obtained by using Clebsch-Gordan (CG) coefficients, takes the general form
\begin{widetext}
\begin{equation}\label{statesl+}
|[N_c]\ell S; JJ_3;(\lambda\mu)YII_3 \rangle = \sum_{m,\ S_3}\left(
                            \begin{array}{cc|c}
                                \ell & S & J \\
                                m & S_3 & J_3
                            \end{array}
       ~\right) |\ell m \rangle ~  |[N_c] SS_3;(\lambda\mu),YII_3\rangle ,
\end{equation}
\end{widetext}
where $|[N_c] SS_3;(\lambda\mu)YII_3\rangle$ is a symmetric spin-flavor state under $N_c$ permutations,  
$(\lambda\mu)$  labels an SU(3) irrep  and the quantum numbers $Y, I, I_3$  stand for 
the hypercharge, isospin and its projection, labelling the basis vectors of a given $(\lambda\mu)$  irrep.  For example
 the states  (\ref{statesl+}) of  the $[{\bf 56}, 2^+]$ multiplet
are: two SU(3) octets $^28_{\frac{3}{2}}$, $^28_{\frac{5}{2}}$ and 
four decuplets
$^410_{\frac{1}{2}}$, $^410_{\frac{3}{2}}$, $^410_{\frac{5}{2}}$,
$^410_{\frac{7}{2}}$ and
for the $[{\bf 56}, 4^+]$ multiplet, 
they  are:  two SU(3) octets
$^28_{\frac{7}{2}}$, $^28_{\frac{9}{2}}$ and four decuplets
$^410_{\frac{5}{2}}$, $^410_{\frac{7}{2}}$, $^410_{\frac{9}{2}}$,
$^410_{\frac{11}{2}}$.

The operators $O_i$  can be obtained in a straightforward manner by using their definition (\ref{OLFS}).  
The operators  $B_i$ must have zero expectation values for nonstrange baryons, see, for example, Table \ref{56lplus}.

The first corrections to the leading term $O_1$ of order $N_c$, start at order $1/N_c$, like for the ground state. 
The  most dominant contributions to the mass formula given by Eq. (\ref{massoperator})
are expected from the operators  shown in Table \ref{56lplus}. Their matrix elements are easy to calculate 
(for details see for example Ref. \cite{MS1}). The angular momentum components $L^i$ act 
on the whole system so that the eigenvalue of the spin-orbit term $O_2$ becomes
\begin{equation}
\langle O_2 \rangle = \frac{1}{2N_c}[J(J+1) - \ell(\ell+1) - S(S+1)],
\end{equation}
in agreement with the results of Goity et al. \cite{GSS03}.

The SU(3) breaking operator $B_2$ can be rewritten as 
\begin{equation}\label{B2}
{B}_2 = - \frac{\sqrt{3}}{2 N_c} \vec{L}\cdot \vec{S}_s
\end{equation}
where $\vec{S}_s$ is the spin operator acting on the strange quarks. Its matrix elements can be calculated as indicated
in Ref. \cite{MS1}.

The analytic form of the first term $S_i G_{i8}$ of $B_3$ was derived from the matrix elements of the SU(6) generators 
for totally symmetric spin-flavor states \cite{Matagne:2006xx}. This is
\begin{equation}
\langle S^i G^{i8}\rangle  = \frac{1}{4 \sqrt{3}}
[3 I (I+1) - S (S+1) + \frac{3}{4} \mathcal{S} (2 - \mathcal{S})],
\end{equation}
where $\mathcal{S}$ is the strangeness.
It can be rewritten in terms of the 
number of strange quarks $N_s = - \mathcal{S}$ in order to recover the expression introduced in  Ref. \cite{Jenkins:1995td}.

\begin{table}[tbp]
\caption{Operators of Eq. (\ref{massoperator}) and coefficients $c_i$ and $d_i$ in MeV
resulting from numerical fits to data obtained for the symmetric multiplets $[{\bf 56}, 2^+]$ Ref.  \cite{GSS03} and $[{\bf 56}, 4^+]$  Ref.  \cite{MS1}.  }
\label{56lplus}
\renewcommand{\arraystretch}{1.25}
\begin{tabular}{lcc}
\hline
\hline
Operator &  $[{\bf 56}, 2^+]$   & \hspace{0.1cm} $[{\bf 56}, 4^+]$ \hspace{0.1cm} \\
\hline
\hline
$O_1 = N_c \ \1 $                &  541  $\pm$   4    &  736  $\pm$  30     \\
$O_2 =\frac{1}{N_c} L^i  S^i$    &   18  $\pm$  16     &   4  $\pm$  40   \\
$O_3 = \frac{1}{N_c}S^i S^i$     & 241  $\pm$  14   &   135  $\pm$  90   \\
\hline
$ B_1 = -{\cal S} $          &  206  $\pm$  18    &  110  $\pm$  67   \\
$ B_2 = \frac{1}{N_c} L^i G^{i8}-\frac{1}{2 \sqrt{3}} O_2$ &   104  $\pm$ 64 \\
$ B_3 = \frac{1}{N_c} S^i G^{i8}-\frac{1}{2 \sqrt{3}} O_3$ &   223  $\pm$ 68   \\
\hline
$\chi^2_{\rm dof}$   &     $\simeq 0.7$             &    $\simeq 0.26$  \\  
\hline \hline
\end{tabular}
\vspace{4cm}
\end{table}

We also found that the expectation values of $O_2$, $O_3$, $B_2$ and $B_3$ satisfy the relation
\begin{equation}\label{dependence}
\frac{\langle{B}_2\rangle}{\langle{B}_3\rangle} =\frac{\langle O_2 \rangle}{\langle O_3\rangle},
\end{equation}
for every $J$, in both the octet and the
decuplet. This can be used as a check of the analytic expressions of these operators in terms of $N_c$ given in 
Refs. \cite{GSS03} or \cite{MS1}.
 
In the numerical fit for resonances belonging to the $[{\bf 56}, 2^+]$ multiplet \cite{GSS03} ten experimentally known resonances with a status of three or four
stars  were used and predictions were made for another 14 resonances.  At higher energies,  namely the multiplet $[{\bf 56}, 4^+]$, the experimental situation
is poorer so that in Ref. \cite{MS1} only five resonances were used in the fit (with a status of one, two, three or four stars)
and 19 masses were predicted. 
\begin{table*}[pt]
\caption{The partial contribution and the total mass (MeV)
predicted by Eq. (\ref{massoperator}) as compared with the 
empirically known masses for resonances assigned to the $[{\bf 56},4^+]$ multiplet, from Ref. \cite{MS1}}
\renewcommand{\arraystretch}{1.5}
\label{4+}

\begin{tabular}{crrrrccl}\hline \hline
                    &      \multicolumn{6}{c}{1/$N_c$ expansion results}        &                     \\ 
\cline{1-6}		    
                    &      \multicolumn{4}{c}{Partial contribution (MeV)} & \hspace{.4cm} Total (MeV)  \hspace{.4cm}  & \hspace{.2cm}  Empirical \hspace{.2cm} &  Name, status \\
\cline{2-5}
                    &   \hspace{.3cm}   $c_1O_1$  & \hspace{.3cm}  $c_2O_2$ & \hspace{.3cm}$c_3O_3$ &\hspace{.3cm}  $b_1\bar B_1$   & \hspace{.4cm}   &  (MeV)   &    \\
\hline
$N_{7/2}$        & 2209 & -3 &  34 &   0  &  \hspace{.4cm} $ 2240\pm97 $ \hspace{.4cm} &\hspace*{.2cm}  \hspace{.2cm} &  \\
$\Lambda_{7/2}$  &      &     &    & 110  &  $2350\pm118 $  & & \\
$\Sigma_{7/2}$   &      &     &    & 110  &  $2350\pm118 $  &               & \\
$\Xi_{7/2}$      &      &     &    & 220  &  $2460\pm 166$  &               &  \\
\hline
$N_{9/2}  $      & 2209 & 2   & 34 &   0  &   $2245\pm95 $  & $ 2245\pm65 $ & N(2220)**** \\
$\Lambda_{9/2}$  &  &     &    & 110  &  $ 2355\pm116 $  & $ 2355\pm15 $ &  $\Lambda$(2350)***\\
$\Sigma_{9/2}$   &      &     &    & 110  &  $ 2355\pm116 $  &               &                    \\
$\Xi_{9/2}$      &      &     &    & 220  &  $2465\pm164$  &               &                    \\
\hline
$\Delta_{5/2}$   & 2209 & -9  &168 &   0  &  $ 2368\pm175$  & &  \\
$\Sigma^{}_{5/2}$&      &     &    & 110  & $2478\pm187$  & &   \\
$\Xi^{}_{5/2}$   &      &     &    & 220  &  $2588\pm220$  &               &                   \\
$\Omega_{5/2}$   &      &     &    & 330  & $2698\pm266$  &               &                    \\
\hline
$\Delta_{7/2}$   &2209  &-5   &168 &  0   &  $2372\pm153$  & $2387\pm88$ &  $\Delta$(2390)* \\
$\Sigma'_{7/2}$  &      &     &    & 110  & $2482\pm167$  &               &                  \\
$\Xi'_{7/2}$     &      &     &    & 220  & $2592\pm203$  &               &                    \\
$\Omega_{7/2}$   &      &     &    & 330  &  $2702\pm252$  &               &                    \\
\hline
$\Delta_{9/2}$   &2209  & 1   &168 &  0   &   $2378\pm144 $  & $2318\pm132  $ &  $\Delta$(2300)**\\
$\Sigma'_{9/2}$  &      &     &    & 110  &   $2488\pm159$  &               &                   \\
$\Xi'_{9/2}$     &      &     &    & 220  &  $2598\pm197$  &               &                    \\
$\Omega_{9/2}$   &      &     &    & 330  &  $2708\pm247$  &               &                    \\
\hline
$\Delta_{11/2}$  &2209  &7    &168 &  0   &  $2385\pm164$  & $ 2400\pm100$ &   $\Delta$(2420)**** \\
$\Sigma^{}_{11/2}$ &    &     &    & 110  & $2495\pm177$  &               &                     \\
$\Xi^{}_{11/2}$  &      &     &    & 220  &  $2605\pm212$  &               &                     \\
$\Omega_{11/2}$  &      &     &    & 330  &  $2715\pm260$  &               &                     \\
\hline
\hline
\end{tabular}
\end{table*}
From Table \ref{56lplus}
one can see that the  coefficient $c_1$ of the leading operator $O_1$ has far the largest value in both cases.
It is interesting to see that this coefficient is larger  for  $[{\bf 56}, 4^+]$ than for $[{\bf 56}, 2^+]$. 
It hints at a dependence of $c_1$ with energy or equivalently with  the band number  $N$. One then expects a Regge trajectory type 
behavior in terms of $N$ \cite{Matagne:2013cca} (for an illustration see Sec. \ref{versus}).

The coefficient $c_2$ of the spin-orbit operator $O_2$  has a small value, which decreases with the excitation energy.
The smallness of the spin-orbit contribution supports the quark model calculations, where the spin-orbit term is usually neglected. 
The decrease in energy is in agreement with the intuitive picture of Ref. \cite{Glozman:2002kq}
according to which, at high energies the spin dependent interactions are expected to vanish as a consequence of the chiral symmetry restoration.

The breaking of the spin-flavor symmetry is essentially given by the operator $O_3$ which 
represents the hyperfine interaction and turns out to be the most important after $O_1$.
The coefficient  $c_3$  is a measure of the splitting between octets and decuplets, as for the ground state described by the coefficients given
in  Eq. (\ref{gsmass}).
Although within numerical errors the values of $c_3$ for  $[{\bf 56}, 2^+]$ and $[{\bf 56}, 4^+]$ are compatible with each other,
the central values show a decrease with the band number $N$, or else with the excitation energy, as mentioned above. 

In general, the SU(3) flavor breaking is dominated by  $B_1 = - \mathcal{S}$.
It gives a mass
shift of about 200 MeV per unit of strangeness in $[{\bf 56}, 2^+]$. 
The operators $B_2$ and  $B_3$ can provide the $\Lambda$ - $\Sigma$ 
splitting in octets and were included in the numerical fit of Ref.  \cite{GSS03}. 
In Ref. \cite{MS1} we have ignored them in the fit because of lack of data. Then, including only   $B_1$
we have obtained a mass shift of about 110 MeV per unit of strangeness, with rather large error bars. 
From Ref. \cite{GSS03}  there is an indication that the contributions of $B_2$ and $B_3$ to the mass
sometimes roughly cancel mutually and sometimes they add to an unexpected large number, so that the 
higher $J$ states are lighter, which is unexpected. In conclusion,
more data are desired for strange excited resonances, both for $N$ = 2 and the $N$ = 4 bands.

To illustrate the discussion,  in Table \ref{4+} we reproduce the results of Ref. \cite{MS1}
for the partial contribution and the total mass predicted by 
the $1/N_c$ expansion, Eq. (\ref{massoperator}), for the $[{\bf 56}, 4^+]$ multiplet.
As a matter of fact, 
the resonance $\Sigma(2455)^{**}$ marked as "bumps" in the 2013 Review of Particle 
Physics \cite{PDG} could possibly be assigned to  $\Sigma_{5/2}(2478)$ of Table \ref{4+}.

Finally note that the operator $B_2$, through its off-diagonal matrix elements,
induces a mixing between  octet and decuplet states at fixed $J$. 
Accordingly, in Table \ref{4+} the states $\Sigma_J$ and $\Sigma^{'}_J$ are defined  
as
\begin{eqnarray}
|\Sigma_J \rangle  & = & |\Sigma_J^{(8)}\rangle \cos\theta_J^{\Sigma} 
+ |\Sigma_J^{(10)}\rangle \sin\theta_J^{\Sigma}, \\
|\Sigma_J'\rangle & = & -|\Sigma_J^{(8)}\rangle \sin\theta_J^{\Sigma}
+|\Sigma_J^{(10)} \rangle \cos\theta_J^{\Sigma}.
\end{eqnarray}
The masses of the physical states become
\begin{eqnarray}
M(\Sigma_J) & = & M(\Sigma_J^{(8)}) 
+ d_2 \langle \Sigma_J^{(8)}|{B}_2|\Sigma_J^{(10)} \rangle \tan \theta^{\Sigma}_J, \label{masssigmaj} \\
M(\Sigma_J') & = & M(\Sigma_J^{(10)}) 
- d_2 \langle \Sigma_J^{(8)}|{B}_2|\Sigma_J^{(10)} \rangle \tan \theta^{\Sigma}_J \label{masssigma'j},
\end{eqnarray}
where $M(\Sigma_J^{(8)})$ and $M(\Sigma_J^{(10)})$ are the diagonal matrix 
of the mass operator (\ref{massoperator}). The expression of the mixing angle can be found in Ref. \cite{MS1}
together with a discussion about the fitting procedure.
Similar relations hold for $\Xi$.


\section{Excited mixed symmetric spin-flavor states}\label{mixedsymmetricstates}

In fact, the first application of the large $N_c$ method was a phenomenological analysis of  strong decays 
of $\ell$ = 1 orbitally excited baryons \cite{CGKM94}. An important purpose was to show that the success of
the nonrelativistic quark model has a natural explanation in large $N_c$ QCD. It was based on the Hartree
approximation  suggested by Witten \cite{WITTEN}. 

Presently there are two procedures of applying the $1/N_c$ expansion to the study the mixed symmetric states.
We shall describe them shortly below and consider applications in the following sections.
 

\subsection{The symmetric core + excited quark procedure}\label{separation}

The first, called in the following the symmetric core + excited quark procedure,
is also inspired by the Hartree picture and is 
based on the separation of the $N_c$-quark system into a ground state symmetric core 
of $N_c - 1$ quarks and an excited quark. 
Within this procedure, the study of the matrix elements of the mass operators 
relevant at the lowest nontrivial order \cite{Goi97}
was followed by the first phenomenological analysis of electromagnetic transitions \cite{CaCa98}, 
and by an analysis of the nonstrange $\ell$ = 1 baryon masses
of the $N$ = 1 band \cite{CCGL}, later  extended to strange baryons  \cite{SGS}.

In the symmetric core + excited quark procedure each SU(2$N_f$) generator 
is split into two parts   
\begin{equation}\label{coreplusquark}
S^i = S^i_c + s^i, T^a = T^a_c + t^a, G^{ia} =   G^{ia}_c + g^{ia},
\end{equation}
where the operators carrying a lower index $c$ act on a symmetric ground state core
and $s^i$, $t^a$ and $g^{ia}$ act on the excited quark.

The procedure has the algebraical advantage that it reduces the problem of the knowledge of 
the matrix elements of the SU($2N_f$) generators  $S^i$, $T^a$ and $G^{ia}$,  acting on the whole system, 
to the knowledge of the matrix elements of $S^i_c$,  $T^a_c$ and  $G^{ia}_c$, acting on symmetric states
of partition $[N_c - 1]$,
which are simpler to find than the matrix elements of the $[N_c - 1,1]$ mixed symmetric states. In fact they were already derived for SU(4)
\cite{PY} at the time the procedure has been proposed. 

Then the operator reduction rules for the ground state \cite{Dashen:1994qi}
may be used for the core operators. However, the number of  terms  to 
be included in operators describing observables remains usually very large
as compared to the experimental data. An example is given in the following subsection.
The list of 12 linearly independent spin-singlet flavor-singlet
operators for SU(4), in powers of $1/N_c$ in their matrix
elements, shown in Table   \ref{12operators},   has been constructed in Ref. \cite{CCGL}.  

Later on the method has been formally supported by Pirjol and Schat \cite{Pirjol:2007ed}
in a large $N_c$ quark model described 
in a permutation group context and an application to $\ell$ = 1 mixed symmetric states was considered. 
Starting from an exact wave function for the whole system of $N_c$ quarks,
the authors of Ref. \cite{Pirjol:2007ed} perform a matching calculation of a general two-body quark-quark interaction
onto operators of the $1/N_c$ expansion. The separation of the system into a core + excited quark
is made on purpose by introducing Eqs. (\ref{coreplusquark}). The main result is a mass 
formula where the coefficients are defined by linear combinations
of radial overlap integrals containing the form factors of the quark-quark interaction. 
These definitions imply constraints on the dynamical coefficients because they are expressed in terms of common 
integrals. The Pauli principle is fulfilled 
provided these constraints are satisfied.
But in practice the coefficients are varied independently so that  
the Pauli principle is fulfilled only within the symmetric core and one recovers the Hartree approximation.

Moreover, one should note that the symmetric core + the excited quark procedure is simple for mixed 
symmetric states with one excited quark, i. e. those belonging to the $N$ = 1 band. 
For $N > 1 $  bands, where more than one quark is excited, the technique becomes more complicated 
as shown for mixed symmetric multiplets of the  $N$ = 2 band \cite{Matagne:2005gd}.

A simpler approach is desired. 
This is described in the next subsection. In this approach the Pauli principle is fulfilled for the entire system of
$N_c$ quarks, so that the orbital-spin-flavor wave function is totally symmetric.
This method requires and provides the matrix elements of 
SU(2N$_f$) generators between states of mixed symmetry of partition $[N_c - 1, 1]$. The procedure is valid
for any number of excited quarks which do not need to be separated from the whole system, and it can  conveniently be applied 
to any excitation  band having $N \ge $ 1.

\begin{table}
\caption{The 12 linearly independent spin-singlet flavor-singlet
operators for SU(4), in powers of $1/N_c$ in their matrix
elements.  
For $F>2$, and ignoring possible coherence in matrix
elements of $T_c^a$, one must include $\frac{1}{N_c^2} t S_c G_c$ and
$\frac{1}{N_c^2} \ell^i g^{ia} S_c^j G_c^{ja}$ in row $N_c^{-1}$. From Ref. \cite{CCGL}.}
\label{12operators}
\renewcommand{\arraystretch}{2.3}
\begin{tabular}{l|l}
\hline \hline
Order &
\multicolumn{1}{c}{Operator} \\ \hline\hline
$N_c^1$ & $N_c $ \\ \hline
$N_c^0$ & $\ell s$, \ $\frac{1}{N_c} \ell t G_c$, \ $\frac{1}{N_c}
\ell^{(2)} gG_c$ \\ \hline
$N_c^{-1}$ & ${\frac{1}{N_c} tT_c}$, \ $\frac{1}{N_c} \ell S_c$, \
$\frac{1}{N_c} \ell g T_c$, \ $\frac{1}{N_c} S_c^2$, 
\ $\frac{1}{N_c}sS_c$, \\ & $\frac{1}{N_c} \ell^{(2)} sS_c$, 
\ $\frac{1}{N_c^2} \ell^{(2)} t \{ S_c, G_c \}$, \ $\frac{1}{N_c^2}
\ell^i g^{ja} \{ S_c^j, G_c^{ia} \}$ \\ \hline \hline


\end{tabular}
\end{table}

\subsection{The totally symmetric orbital-spin-flavor wave function procedure}\label{exactwavefunction}

We remind that we deal with a system of $N_c$ quarks having $\ell$ units of
orbital excitation. Therefore the orbital ($O$) wave function  must have 
a mixed symmetry $[N_c-1,1]$, which describes the lowest excitations in
a baryon.

The color wave function being antisymmetric, the
orbital-spin-flavor wave part must be symmetric. Then the spin-flavor ($FS$)
part must have the same symmetry as the orbital part in order to obtain a 
totally symmetric
state in the orbital-spin-flavor space.

In Ref. \cite{Matagne:2006dj}, as an alternative,  we have proposed an approach
where the separation of the system into a symmetric core of $N_c$ - 1 quarks and an excited quark
is neither necessary nor desired. In that case one deals with SU($2N_f$) generators acting on the whole system of $N_c$
quarks and the number of independent operators needed in the mass formula is generally smaller 
than the number of the experimental data. The resulting mass formula is therefore more physically 
transparent and its simple form allows applications to multiplets belonging to any band with $N \ge 1$, even 
in cases where the data are more scarce. Examples will be given later on. First let us discuss the difference between
the two procedures. 

In the exact orbital-spin-flavor wave function 
both the orbital and the spin-flavor parts of the total wave function are described by the partition
$[f]$ = $[N_c - 1, 1]$. By inner product rules of the permutation group one can form a totally symmetric
orbital-spin-flavor wave function  described by the partition  $[N_c]$ as
\begin{widetext}
\begin{equation}
\label{EWF}
|[N_c] \rangle = {\frac{1}{\sqrt{d_{[N_c-1,1]}}}}
\sum_{Y} |[N_c-1,1] Y \rangle_{O}  |[N_c-1,1] Y \rangle_{FS}, 
\end{equation}
\end{widetext}
where $d_{[N_c-1,1]} = N_c - 1$ is the dimension of the representation 
$[N_c-1,1]$ of the permutation group $S_{N_c}$ and $Y$ labels a
Young tableau (or a Yamanouchi symbol). 
The sum is performed over all possible standard Young tableaux. 
In each term the first basis vector represents the orbital  
space ($O$) and the second the spin-flavor space ($FS$).
In this sum there is only one $Y$ (the normal Young tableau) 
where the $N_c$-th  particle is in 
the second row and $N_c - 2 $ terms where the $N_c$-th particle
is in the first row. 
In the  symmetric core + excited quark procedure
the latter terms are ignored. An example is given in Appendix \ref{properties}.

In our approach, the system of $N_c$ quarks is described by the wave function (\ref{EWF}). 
We therefore treat the quarks as identical, whether they are excited or not. 
Assuming that the whole system has an orbital angular momentum  $\ell$ we identify the orbital part
in Eq. (\ref{EWF}) with a spherical harmonic $|\ell m_{\ell} \rangle$
and the spin-flavor part in SU(6) with a basis vector  $| [f] (\lambda \mu ) Y I I_3; S S_3 \rangle$ 
of SU(6) defined as adequate inner products of spin  $|S S_3\rangle $  and SU(3)-flavor states  $|(\lambda \mu )Y I I_3\rangle$
which span the invariant subspace of an SU(3) irrep $(\lambda\mu)$. 
   
Following Ref. \cite{Matagne:2011fr} the most general form of such a symmetric orbital-spin-flavor wave function in SU(6) $\times$ O(3), 
having a total angular momentum $J$ and projection $J_3$ is given by
\begin{widetext}
\begin{equation}\label{WF}
  |\ell S;JJ_3;(\lambda \mu) Y I I_3\rangle  =
\sum_{m_\ell,S_3}
      \left(\begin{array}{cc|c}
	\ell    &    S   & J   \\
	m_\ell  &    S_3  & J_3 
      \end{array}\right) 
       |\ell m_\ell \rangle        
| [f]  S S_3; (\lambda \mu ) Y I I_3 \rangle ,
\end{equation}
\end{widetext}
where the first factor is the usual Clebsch-Gordan coefficient of SU(2). 
In the present case we have $[f]$ = $[N_c-1,1]$ which does not need to be specified for $|\ell m_{\ell} \rangle$.
This form is similar to that of symmetric states given in Eq. (\ref{statesl+}). For spectrum calculations or other 
observables one needs to know the matrix elements of the SU($2N_f$) generators,   $S^i$, $T^a$ and $G^{ia}$,
between the states (\ref{WF}). They are explicitly given in Appendix \ref{A}, both for  $[f]$ = $[N_c]$ and $[f]$ = $[N_c-1,1]$,
together with some of their properties.  The results for SU(4) were derived by Hecht and Pang \cite{HP} in the context  
of nuclear physics and for SU(6) the isoscalar factors were mostly obtained in Ref. \cite{Matagne:2008kb} and
completed in Ref.  \cite{Matagne:2011fr}.

To summarize, the relation between the two approaches is remotely similar to that
between Hartree and Hartree-Fock 
approaches. The symmetric core + excited quark approach is simpler, limiting the application of the Pauli
principle to a symmetric core of $N_c$ - 1 quarks. The procedure of Sec.  \ref{exactwavefunction}
is more complicated from group theory point of view, but rigorously
takes into account the Pauli principle for the entire system of $N_c$ quarks.

The symmetric core + excited quark approach has the merit of being the
firstly proposed but the separation of the system into a symmetric
core and an excited quark leads to an excessively large number of
independent operators,  making difficult the choice of dominant operators
and the understanding of their physical meaning. 
Even when the fit looks acceptable some of
the contributions to the mass cancel mutually, which may suggest 
that the decomposition of the system was not necessary. As we shall see in Sec. \ref{lowestmultiplet}
there appear peculiar situations, as for example the case of $\Lambda(1405)$
where the entire spin-spin interaction is removed by construction, because of the 
approximation made in the wave function of symmetric core + excited quark approach.
Furthermore, except for one case, the approach has not been applied to
higher mass resonances ( $N > 1 $ band) most probably because of being cumbersome 
(the core is no more in the ground state),
as it will be explained in Section \ref{positivep}.

\section{Spectrum calculations for mixed symmetric states}\label{specmixed}

Below we present a summary of results for resonances described as mixed symmetric states of either negative or positive parity.  

\subsection{The lowest negative parity $[{\bf 70},1^-]$ multiplet}\label{lowestmultiplet}

In the  baryon spectrum, the $[{\bf 70},1^-]$ multiplet 
has been most extensively studied, being the best experimentally known 
negative parity mixed symmetric multiplet.
For $N_f$ = 2 there are numerous studies as for example Refs.
\cite{PY,CGKM94,Goi97,CCGL,CaCa98}.

The above studies were in the spirit of the  Hartree approximation where  
the system of
$N_c$ quarks was split into a ground state core of $N_c-1$ quarks and an excited 
quark \cite{CCGL}, as described in Sec. \ref{separation}.
This means that each generator of SU($2N_f$) was written as a sum of two
terms, one acting on the excited quark and the other on the core, as in Eq.  (\ref{coreplusquark}).
Then, as mentioned, the  number of the coefficients $c_i$ in the 
mass formula is too large compared to the available data on resonance masses and cannot 
be uniquely determined in a numerical fit,
as it has been done for the lowest negative parity nonstrange baryons \cite{CCGL}. Accordingly, 
the choice of the most dominant operators in the mass formula (\ref{massoperator}) became  
out of control  which implies 
that important physical effects can be missed, as discussed below.

\begin{table*}[pt]
\caption{The dominant operators and the best fit coefficients for the masses of nonstrange and strange 
baryons belonging to the $[{\bf 70},1^-]$ multiplet. From Ref. \cite{SGS}.}
\renewcommand{\arraystretch}{2.}
\label{SGSbaryons}
\begin{tabular}{llrrr}
\hline
\hline
Operator & \multicolumn{4}{c}{Fitted coeff. [MeV]}\\
\hline
\hline
$O_1 = N_c \ \1 $ & $c_1 =$  & 449 & $\pm$ & 2 $\ $  \\
\hline
$O_2 = l_h \ s_h$ & $c_2 =$ & 52 & $\pm$ & 15   $\ $ \\
$O_3 = \frac{3}{N_c} \ l^{(2)}_{hk} \ g_{ha} \ G^c_{ka} $ & $c_3 =$  & 116 & $\pm$ & 44  $\ $ \\
$O_4 = \frac{4}{N_c+1} \ l_h \ t_a \ G^c_{ha}$ & $c_4 =$  & 110 & $\pm$ &  16 $\ $\\
\hline
$O_5 = \frac{1}{N_c} \ l_h \ S^c_h$ & $c_5 =$  & 74 & $\pm$ & 30 $\ $\\
$O_6 = \frac{1}{N_c} \ S^c_h \ S^c_h$ & $c_6 =$  & 480 &  $\pm$ & 15 $\ $\\
$O_7 = \frac{1}{N_c} \ s_h \ S^c_h$ & $c_7 =$ & -159 &  $\pm$ & 50 $\ $ \\
$O_8 = \frac{1}{N_c} \ l^{(2)}_{hk} s_h \ S^c_k$ & $c_8  =$  & 6  & $\pm$ &   110   $\ $\\
$O_9 = \frac{1}{N_c^2} \ l_h \ g_{ka} \{ S^c_k ,  G^c_{ha} \} $ & $c_9 =$ &  213 &  $\pm$ &  153  $\ $\\
$O_{10} = \frac{1}{N_c^2} t_a \{ S^c_h ,  G^c_{ha} \}$ & $c_{10} =$  & -168 &  $\pm$ &  56  $\ $\\
$O_{11} = \frac{1}{N_c^2} \ l_h \ g_{ha} \{ S^c_k ,  G^c_{ka} \}$ & $c_{11} =$ & -133 &  $\pm$ &  130  $\ $\\
\hline
\hline
$ B_1 = t_8 - \frac{1}{2 \sqrt{3} N_c} O_1$ & $d_1 =$  & -81 & $\pm$ & 36 $\ $\\
$ B_2 = T_8^c - \frac{N_c-1}{2 \sqrt{3} N_c } O_1 $  & $ d_2 = $  & -194 & $\pm$ & 17  $\ $\\
$ B_3 = \frac{1}{N_c} \  d_{8ab}  \ g_{ha} \ G^c_{hb}  + \frac{N_c^2 -9}{16 \sqrt{3} N_c^2 (N_c-1)} O_1 +$ &  & & $\ $\\
\hspace*{1cm} $+ \frac{1}{4 \sqrt{3} (N_c-1)} O_6 + \frac{1}{12 \sqrt{3}} O_7 $  & $ d_3 = $  & -150 & $\pm$ & 301  $\ $\\
$ B_4 = l_h \ g_{h8} - \frac{1}{2 \sqrt{3}} O_2 $ & $ d_4 = $  & -82 & $\pm$ & 57  $\ $\\
\hline \hline
\end{tabular}
\end{table*}

Several fits have been performed in Ref. \cite{CCGL} for nonstrange baryons. From our point of view the 
most interesting one is the result given in Table VII of that paper,
which is consistent with the mechanism of
the Goldstone Boson Exchange (GBE) model \cite{Glozman:1995fu,Glozman:1997ag}.
In this fit the operator $\frac{1}{N_c} \ell^{(2)} gG_c$ 
plays a crucial role and is related to a pion exchange between the excited quark and
a core quark.

To our knowledge the $N_f$ = 3 case has been considered  only in Ref.  \cite{SGS} in the symmetric
core + excited quark procedure, where
first order corrections in SU(3) symmetry breaking 
were  also included.  Both for $N_f$ = 2 and  $N_f$ = 3  cases, the conclusion was that the splitting starts 
at order $N^0_c$. 
The list of dominant operators and the best fit coefficients in the mass formula (\ref{massoperator}) is shown
in Table \ref{SGSbaryons}. The fit was made to 19 empirical quantities (17 masses and 2 mixing angles) associated
to resonances with three or more stars status and it gives $\chi^2_{\rm dof}$ = 1.29.

There are eleven operators of type $O_i$ and four of type $B_i$
included in the mass formula. 
One can see that the coefficients $c_3$ and $c_4$ are large, consistent with the 
SU(4) case where it was found that the operator $\propto \frac{1}{N_c} \ell^{(2)} gG_c$ plays a crucial role, as 
mentioned above. The contribution of the spin operators  $O_6$ and $O_7$ is large, as expected, but 
there is some mutual cancellation. Some operators of a more complex nature as $O_9$, $O_{10}$ and $O_{11}$
contribute also substantially, but the total contribution somewhat cancels out. 
One can notice the absence of the  flavor term $t \cdot T_c$, never included 
in the analyses based on the symmetric core + excited quark approach in SU(6). 

Let us remind that in the symmetric core + excited quark approach the total flavor operator was 
written  as the sum of three terms $T \cdot T = T_c \cdot T_c + 2 t \cdot T_c + 3/4$, each thought to be linearly independent.
The first term acts on the core and  its matrix elements
are identical to those of $S_c \cdot S_c$ when the SU(4) spin-flavor state is symmetric. Then its contribution cannot be distinguished from 
that of $S_c \cdot S_c$.
The last term, the constant 3/4, can be absorbed in the leading order term  but to our understanding $t \cdot T_c$  cannot be ignored.

The calculated masses are compared with the quark model results of Isgur and Karl \cite{IK78} based on an oscillator 
confinement, where the oscillator parameter was fitted to the $N$ = 1 band. In the work of Isgur and Karl the
hyperfine interaction is represented by the spin-spin and tensor parts of the Fermi-Breit Hamiltonian derived from
the one gluon exchange
\cite{rgg}. The spin-orbit part is neglected.
The resonance $\Lambda(1405)$  appeared by about 100 MeV too high, like in more recent 
studies, based on the more realistic linear confinement and the flavor dependent Goldstone boson exchange interaction
\cite{Glozman:1997ag}, which reproduces the correct level ordering of the Roper and the first negative parity nonstrange baryons,
impossible to be obtained in models based  the one gluon exchange interaction.  

The authors of Ref.  \cite{SGS} explain the lightness of  $\Lambda(1405)$ and 
$\Lambda(1520)$, seen as spin-orbit partners, by  the fact that the spin-spin terms $\frac{1}{N_c} S_c \cdot S_c$ and
$\frac{1}{N_c} s \cdot S_c$ do not contribute to their masses because the core has $S_c$ = 0.
This is the effect of the simplicity of their wave function where the part corresponding to the spin  $S_c$ = 1 is
missing, as inferred by the arguments of Sec. \ref{exactwavefunction}. However, the spin-spin interaction 
cannot be neglected, even though it is order $1/N_c$, because it is the leading term that splits $N$ and $\Delta$.
In octets and decuplets the spin-spin interaction survives, despite of the approximate wave function.
It rises their masses because there the core has a non-zero spin component.  

The spin-orbit splitting is explained 
as the combined effect of the operators $O_4$,  $O_5$, $O_9$ and $O_{11}$. 
The large error bars of the coefficients of the operators $O_8$ and $B_3$
makes these operators irrelevant in the mass formula.

\begin{table*}[pt]
\caption{Operators and their coefficients in the mass formula (\ref{massoperator}), obtained from 
three distinct numerical fits. The values of $c_i$ and $d_i$ are indicated under the heading Fit $n\ (n=1,2,3)$,
in each case \cite{Matagne:2011fr}. }
\label{ouroperators}
\renewcommand{\arraystretch}{2} 
\begin{tabular}{lrrr}
\hline
\hline
Operator \hspace{2cm} &\hspace{0.0cm} Fit 1 (MeV) & \hspace{0.cm} Fit 2 (MeV) & Fit 3 (MeV)  \\
\hline
$O_1 = N_c \ \1 $                    & $489 \pm 4$  & $492 \pm 4$  & $492\pm 4$       \\
$O_2 = \ell^i s^i$                	     & $24 \pm 6$ & $6 \pm 6$  & $6\pm5$   \\
$O_3 = \frac{1}{N_c}S^iS^i$          & $129 \pm 10$ & $123 \pm 10$ & $123\pm 10$   \\
$O_4 = \frac{1}{N_c} \left[T^aT^a - \frac{1}{12} N_c(N_c+6)\right]$  & $145 \pm 16$ & $134\pm 16$ & $135\pm 16$    \\
$O_5 =  \frac{3}{N_c} L^i T^a G^{ia}$ & $-19 \pm 7$ & $3 \pm 7$ & $4\pm 3$   \\ 
$O_6 = \frac{15}{N_c} L^{(2)ij}G^{ia}G^{ja}$    & $9 \pm 1$ & $9\pm 1$ & $9\pm 1$        \\
$O_7 = \frac{1}{N_c^2}L^iG^{ja}\{S^j,G^{ia}\}$ & $129 \pm 33$ &  $6\pm 33$  &\\ 
\hline
$B_1 = \mathcal{-S}$  & $138 \pm 8$ & $138\pm 8$ & $137\pm 8$\\
$B_2 = \frac{1}{N_c} \sum^3_{i=1}T^iT^i - O_4$ &  $-59\pm 18$ & $-40\pm 18$ & $-40\pm 18$\\ 
\hline
$\chi_{\mathrm{dof}}^2$   &  $1.7$  & $0.9$ & $0.84$   \\
\hline \hline
\end{tabular}
\end{table*}
{\squeezetable
\begin{table*}[pt]
\caption{The partial contribution and the total mass (MeV) predicted by the $1/N_c$ expansion
obtained from the Fit 1.  The last two columns give  the empirically known masses  {\protect \cite{PDG2010}} and the resonance name and status. 
From Ref. \cite{Matagne:2011fr}}\label{MASSES}
\renewcommand{\arraystretch}{2.}
\begin{tabular}{crrrrrrrrrcccl}\hline \hline
                    &      \multicolumn{9}{c}{Partial contributions (MeV)}   & \hspace{.0cm} Total (MeV)   & \hspace{.0cm}  Exp. (MeV)\hspace{0.0cm}& &\hspace{0.cm}  Name, status \hspace{.0cm} \\

\cline{2-10}
                    &   \hspace{.0cm}   $c_1O_1$  & \hspace{.0cm}  $c_2O_2$ & \hspace{.0cm}$c_3O_3$ & $c_4O_4$ &\hspace{.0cm}  $c_5O_5$ &\hspace{.0cm}  $c_6O_6$ & $c_7O_7$ &  $d_1B_1$ & $d_2B_2$&     &        \\
\hline
$N_{\frac{1}{2}}$        & 1467 & -8 &  32 & 36 & 19  & 0 & -31 & 0   & 0 &  $1499\pm 10$ & $1538\pm 18$ & & $S_{11}(1535)$****  \\
$\Lambda_{\frac{1}{2}}$  &      &    &     &    &     &   &     & 138 & 15&  $1668\pm 9$  & $1670\pm 10$  & & $S_{01}(1670)$**** \\
$\Sigma_{\frac{1}{2}}$   &      &    &     &    &     &   &     & 138 &-25&  $1628\pm 10$ &              & & \\
$\Xi_{\frac{1}{2}}$      &      &    &     &    &     &   &     & 276 & 0 &  $1791\pm 13$ &              & & \vspace{0.2cm}\\
\hline
$N_{\frac{3}{2}}$        & 1467 & 4  & 32  & 36 & -10 & 0 & 16  &  0  & 0 &  $1542\pm 10$ & $1523\pm 8$ & & $D_{13}(1520)$****  \\
$\Lambda_{\frac{3}{2}}$  &      &    &     &    &     &   &     & 138 & 15&  $1698\pm  8$ & $1690\pm 5$ & & $D_{03}(1690)$**** \\
$\Sigma_{\frac{3}{2}}$   &      &    &     &    &     &   &     & 138 &-25&  $1658\pm  9$ & $1675\pm 10$             & & $D_{13}(1670)$****\\
$\Xi_{\frac{3}{2}}$      &      &    &     &    &     &   &     & 276 & 0 &  $1821\pm 11$ & $1823\pm 5$             & & $D_{13}(1820)$***
\vspace{0.2cm} \\
\hline
$N'_{\frac{1}{2}}$       & 1467 &-20 &162  & 36 & 48  &-18& 42  & 0   & 0 &  $1648\pm 11$ & $1660\pm 20$ & & $S_{11}(1650)$****  \\
$\Lambda'_{\frac{1}{2}}$ &      &    &     &    &     &   &     & 138 &15 &  $1784\pm 16$ & $1785\pm 65$  & & $S_{01}(1800)$*** \\
$\Sigma'_{\frac{1}{2}}$  &      &    &     &    &     &   &     & 138 &-25&  $1745\pm 17$ & $1765\pm 35$             & & $S_{11}(1750)$***\\
$\Xi'_{\frac{1}{2}}$     &      &    &     &    &     &   &     & 276 & 0 &  $1907\pm 20$ &              & & \vspace{0.2cm}\\
\hline
$N'_{\frac{3}{2}}$       & 1467 & -8 & 162 & 36 & 19  & 15&-17  & 0   & 0 &  $1675\pm 10$ & $1700\pm 50$ & & $D_{13}(1700)$***  \\
$\Lambda'_{\frac{3}{2}}$ &      &    &     &    &     &   &     & 138 &15 &  $1826\pm 12$ &  & &  \\
$\Sigma'_{\frac{3}{2}}$  &      &    &     &    &     &   &     & 138 &-25&  $1787\pm 13$ &              & & \\
$\Xi'_{\frac{3}{2}}$     &      &    &     &    &     &   &     & 276 & 0 &  $1949\pm 16$ &              & & \vspace{0.2cm}\\
\hline
$N_{\frac{5}{2}}$       & 1467 & 12 & 162 & 36 &-29  & -4& 25  & 0   & 0 &  $1669\pm 10$ & $1678\pm 8$ & & $D_{15}(1675)$****  \\
$\Lambda_{\frac{5}{2}}$ &      &    &     &    &     &   &     & 138 & 15&  $1822\pm 10$ & $1820\pm 10$  & & $D_{05}(1830)$**** \\
$\Sigma_{\frac{5}{2}}$  &      &    &     &    &     &   &     & 138 &-25&  $1782\pm 11$ & $1775\pm 5$             & &$D_{15}(1775)$**** \\
$\Xi_{\frac{5}{2}}$     &      &    &     &    &     &   &     & 276 & 0 &  $1945\pm 14$ &              & & \vspace{0.2cm}\\
\hline
$\Delta_{\frac{1}{2}}$   & 1467 & 8  & 32  &181 & 38  & 0 & -24 & 0   & 0 &  $1702\pm 18$  & $1645\pm 30$ & & $S_{31}(1620)$****  \\
$\Sigma''_{\frac{1}{2}}$ &      &    &     &    &     &   &     & 138 & 34&  $1875\pm 16$  &   & &  \\
$\Xi''_{\frac{1}{2}}$    &      &    &     &    &     &   &     & 276 & 59&  $2037\pm 22$  &              & & \\
$\Omega_{\frac{1}{2}}$   &      &    &     &    &     &   &     & 413 & 74&  $2190\pm 29$  &              & & \vspace{0.2cm}\\
\hline
$\Delta_{\frac{3}{2}}$   & 1467 & -4 & 32  & 181& -19 & 0 & 12  &  0  &  0&  $1668\pm 20$  & $1720\pm 50$ & & $D_{33}(1700)$****  \\
$\Sigma''_{\frac{3}{2}}$ &      &    &     &    &     &   &     & 138 & 34&  $1841\pm 16$  &  & &  \\
$\Xi''_{\frac{3}{2}}$    &      &    &     &    &     &   &     & 276 & 59&  $2003\pm 21$  &              & & \\
$\Omega_{\frac{3}{2}}$   &      &    &     &    &     &   &     & 413 & 74&  $2156\pm 27$  &              & & \vspace{0.2cm}\\
\hline
$\Lambda''_{\frac{1}{2}}$&1467  &-24 & 32 &-108 & 0   & 0 &  -38&138  &-44&  $1421\pm 14$  & $1407\pm 4$  & & $S_{01}(1405)$**** \\
\hline
$\Lambda''_{\frac{3}{2}}$&1467  & 12 & 32 &-108 & 0   &  0&  19 & 138 &-44&  $1515\pm 14$    & $1520\pm 1$  & & $D_{03}(1520)$**** \\
\hline
$N_{1/2}-N'_{1/2}$       &0     & -8 &  0 & 0   & -10 &-55& 18  &  0  &  0&  $-55$         &   & & \\
$N_{3/2}-N'_{3/2}$       &0     & -12&  0 & 0   & -15 & 17& 28  &  0  &  0&  18            &  &  & \\
\hline
\hline
\end{tabular}
\end{table*}}

The  $[{\bf 70},1^-]$ lowest multiplet was also analyzed within the framework described in Sec. \ref{exactwavefunction},
based on the totally symmetric orbital-spin-flavor wave function, first in SU(4)  \cite{Matagne:2006dj} and next in SU(6)   \cite{Matagne:2011fr}.
The list of dominant operators and the numerical results for $c_i$ and $d_i$ obtained in  Ref.  \cite{Matagne:2011fr} are presented in Table 
\ref{ouroperators}. One can see that the number of operators used in the fit is considerably smaller than that of Table \ref{SGSbaryons}. 
The one-body spin-orbit operator $O_2$ is the same  as in Table \ref{SGSbaryons}.
The spin operator $O_3$ and the  flavor operator $O_4$ are two-body. The operator $O_4$ in SU(3)  was defined in Ref. \cite{Matagne:2011fr}
such as to be applicable to flavor singlets as well. For octets and decuplets it gives the same matrix elements 
as the isospin operator $\frac{1}{N_c} T^aT^a$ in SU(4), of order $1/N_c$. For flavor singlets the order of the matrix elements of $O_4$ is  $N^0_c$.
The operators $O_5$ and $O_6$ are two body, but $G^{ia}$ sums coherently in both and introduces a factor $N_c$ except for
the $^28$ multiplets.  The operator $O_7$ is three-body and has a more complex form, but it contains the generator  $G^{ia}$ two times
so that the order of its matrix elements becomes  $N^0_c$. However, looking at Table \ref{ouroperators} and comparing Fit 2 and  Fit 3, where $O_7$ has been removed
in the latter from the mass formula, one can see that its role is negligible. The SU(3) breaking operator $B_1$ represents the total strangeness
and $B_2$ was defined such as to account for the $\Lambda$-$\Sigma$ splitting.
The diagonal and off-diagonal matrix elements of  $O_i$ as a function of $N_c$ 
can be found in Ref. \cite{Matagne:2011fr}.  The isoscalar factors
of Tables \ref{octet_spin_one_half}, \ref{octet_spin_three_halfs}, \ref{decuplet_spin_one_half}
and \ref{singlet_spin_one_half} of Appendix \ref{A} were used to obtain their analytic expressions.

In the numerical fit we have used the 17 resonances from the Particle Data Group 2010  \cite{PDG2010}, with a status of three 
and four stars and two mixing angles. The Fit 1 is based on the experimental value $M(\Lambda(1405))$ = 1407 MeV which gives 
$\chi_{\mathrm{dof}}^2$ = 1.7.  To improve the fit we took the value 1500 MeV for the mass of 
$\Lambda(1405)$, inspired by quark model studies where  usually $M(\Lambda(1405))$ appears too high, as mentioned above. 
This is the result of Fit 2 where $\chi_{\mathrm{dof}}^2$ lowers to 0.9. 

The $\Lambda(1405)$ resonance is a long standing problem. Deeper dynamical arguments are necessary 
to understand its exceptionally low mass  (for a review see, for example, \cite{Hyodo:2011ur}).

Table \ref{MASSES} reproduces the partial 
contribution and the total mass obtained by using the coefficients of Fit 1.
One can see that in flavor singlets
the contribution of the spin operator $O_3$ is not particularly large but the flavor operator $O_4$ brings 
an essential contribution in lowering the mass of $\Lambda(1405)$ and $\Lambda(1520)$. 
The spin-orbit partners $N_J - N^{'}_J$ ( $J$ = 1/2, 3/2) receive contributions from the operators 
$O_2$, $O_5$, $O_6$ and $O_7$ via their off-diagonal matrix elements.  

The global conclusion is 
that both the spin $O_3$ and the flavor  operator $O_4$ contribute dominantly to the spin-flavor breaking. 
In particular the flavor  operator contributes to the masses of
decuplets and flavor singlets with a coefficient of the same order as that of the spin operator in octets.
Thus, in the symmetric core + excited quark approach, even if the contribution of $T_c \cdot T_c$ is
identified  to that of $S_c \cdot S_c$ there is no reason to 
ignore the isospin term $t \cdot T_c$, as a part of $T \cdot T$, as explained above.

\subsection{Highly excited negative parity states}

In the approach proposed in Ref.  \cite{Matagne:2011fr}, based on the exact wave function, as described in Sec. \ref{exactwavefunction},
the number linearly independent operators in the mass formula is considerably reduced as compared to the ground state core + excited quark
procedure.  
Thus it was possible to analyze highly excited states belonging to the N = 3 band \cite{Matagne:2012tm}
where the experimental data is still scarce.

The $N = 3$ band contains eight  SU(6) $\times$ O(3) multiplets. In the notation of Ref. \cite{Stancu:1991cz} 
these are $[{\bf 56},1^-]$,  $[{\bf 56},3^-]$, $[{\bf 70'},1^-]$,  $[{\bf 70''},1^-]$, $[{\bf 70},2^-]$, $[{\bf 70},3^-]$, $[{\bf 20},1^-]$ and $[{\bf 20},3^-]$,
where  $[{\bf 70'},1^-]$ and $[{\bf 70''},1^-]$ correspond to  radial excitations.  
This classification provides 45 non-strange states (1 state $N_{9/2^-}$, 1 state $\Delta_{9/2^-}$,  5 states $N_{7/2^-}$, 2 states $\Delta_{7/2^-}$, 
8 states $N_{5/2^-}$,  4 states 
$\Delta_{5/2^-}$, 9 states $N_{3/2^-}$, 5 states $\Delta_{3/2^-}$, 7 states $N_{1/2^-}$ and 3 states $\Delta_{1/2^-}$). 
The analysis of Ref. \cite{Matagne:2012tm} included all mixed symmetric multiplets  $[{\bf 70}, \ell^-]$ ($\ell = 1, 2$ and 3) of the band. 

%
\begin{table*}[pt]
\caption{Operators and their coefficients in the mass formula obtained from four
numerical fits of highly excited negative parity resonances of the $N$ = 3 band  \cite{Matagne:2012tm}. The values of $c_i$ and $d_i$ are indicated under the heading Fit $n\ (n=1,2,3,4)$.}
\label{highlyexcited}{\scriptsize
\renewcommand{\arraystretch}{2} 
\begin{tabular}{lrrrr}
\hline
\hline
Operator \hspace{2cm} &\hspace{0.0cm} Fit 1 (MeV) & \hspace{1.cm} Fit 2 (MeV)  & \hspace{1.cm} Fit 3 (MeV)  & \hspace{1.cm} Fit 4 (MeV)  \\
\hline
$O_1 = N_c \ \1 $                                                & $c_1 = 672 \pm 8$  & $c_1 = 673 \pm 7$      & $c_1 = 672 \pm 8$   & $c_1 = 673 \pm 7$   \\
$O_2 = \ell^i s^i$                	                         & $c_2 = 18 \pm 19$   & $c_2 = 17 \pm 18$    & $c_2 = 19 \pm 9$    & $c_2 = 20 \pm 9$ \\
$O_3 = \frac{1}{N_c}S^iS^i$                                      & $c_3 = 121 \pm 59$  & $c_3 = 115 \pm 46$   & $c_3 = 120 \pm 58$    & $c_3 = 112 \pm 42$\\
$O_4 = \frac{1}{N_c}\left[T^aT^a-\frac{1}{12}N_c(N_c+6)\right]$  & $c_4 = 202 \pm 41$  & $c_4 = 200\pm 40$   & $c_4 = 205\pm 27$   & $c_4 = 205\pm 27$  \\
$O_5 =  \frac{3}{N_c} L^{i} T^{a} G^{ia}$                  & $c_5 = 1 \pm 13$     &   $c_5 = 2 \pm 12$  &   \\ 
$O_6 =  \frac{15}{N_c} L^{(2)ij} G^{ia} G^{ja}$                  & $c_6 = 1 \pm 6$     &   &   $c_6 = 1 \pm 5$ \\ 
\hline
$B_1 = -\mathcal{S}$                                             & $d_1 = 108 \pm 93$  & $d_1 = 108 \pm 92$ & $d_1 = 109 \pm 93$  & $d_1 = 108 \pm 92$  \\     
\hline                  
$\chi_{\mathrm{dof}}^2$                                          &  $1.23$             & $0.93$   & $0.93$      & $0.75$\\
\hline \hline
\end{tabular}}

\end{table*}
%
\begin{table*}[pt]
\begin{center}
\caption{Partial contributions and the total mass (MeV) predicted by the $1/N_c$ expansion,  obtained from Fit 4 
of Table \ref{highlyexcited}. The last two columns indicate the empirically
known masses and the resonance name and status (whenever known). }\label{MASSESFIT4}
\renewcommand{\arraystretch}{2.5}{\scriptsize
\begin{tabular}{lrrrrrrrrr}\hline \hline
                    &      \multicolumn{5}{c}{Partial contributions (MeV)}  & \hspace{0.75cm} Total (MeV)   & \hspace{0.75cm}  Exp. (MeV)\hspace{0.75cm}& &\hspace{0.75cm}  Name, status \hspace{.0cm} \\

\cline{2-6}
                    &   \hspace{.35cm}   $c_1O_1$  & \hspace{.35cm}  $c_2O_2$ & \hspace{.35cm}$c_3O_3$ &\hspace{.35cm}  $c_4O_4$   &  \hspace{.35cm}  $d_1B_1$   &      \\
\hline
$^4N[{\bf 70},3^-]_{9/2}$      & 2018 & 29 & 140 & 51 &  0  &$2238\pm 46$  & $2275\pm 75$ & & $G_{19}(2250)$****  \\
$^2N[{\bf 70},3^-]_{7/2}$      & 2018 & 10 &  28 & 51 &  0  &$2107\pm 17$  & $2150\pm 50$ & & $G_{17}(2190)$**** \\
$^4N[{\bf 70},3^-]_{5/2}$      & 2018 &-23 & 140 & 51 &  0  &$2186\pm 41$  & $2180\pm 80$ & & $D_{15}(2200)$**\\
$^2N[{\bf 70},3^-]_{5/2}$      & 2018 &-39 &  28 & 51 &  0  &$2058\pm 14$  & $2060\pm 15$ & & $D_{15}(2060)$\\
$^4N[{\bf 70},3^-]_{3/2}$      & 2018 &-39 & 140 & 51 &  0  &$2170\pm 42$  & $2150\pm 60$ & & $D_{13}(2150)$\\
$^2N[{\bf 70'},1^-]_{3/2}$     & 2018 &	 3 &  28 & 51 &  0  &$2101\pm 14$  & $2081\pm 20$ & & $D_{13}(2080)$* \\
$^2N[{\bf 70'},1^-]_{1/2}$     & 2018 & -7 &  28 & 51 &  0  &$2091\pm 12$  & $2100\pm 20$ & & $S_{11}(2090)$*\\
\hline
$^2\Delta[{\bf 70},3^-]_{7/2}$ &  2018 & -10  & 28 & 256 &  0  &  $2292\pm 25$  & $2200\pm 80$ & & $G_{37}(2220)$* \\
$^2\Delta[{\bf 70},2^-]_{5/2}$ &  2018 & -7  & 28 & 256 &  0  &  $2295\pm 25$  & $2305\pm 26$ & & $D_{35}(2350)$* \\ 
\hline
$^2\Lambda[{\bf 70},3^-]_{7/2}$& 2018 &29 &  28 &-153&108  &$2030\pm 82$  & $2030\pm 82$ & & $G_{07}(2100)$**** \\ 
\hline \hline
\end{tabular}}
\end{center}
\end{table*}
Experimentally in the 1900 MeV - 2400 MeV mass region the 2010 Particle Data Group \cite{PDG2010}
provided the following resonances:  $N_{19}(2250)^{****}$,
$N_{17}(2190)^{****}$ and $\Lambda_{07}(2100)^{****}$,   $N_{15}(2220)^{**}$,   $N_{13}(2080)^{*}$,  $N_{11}(2090)^{*}$,
$\Delta_{37}(2220)^{*}$ and $\Delta_{35}(2350)^{*}$ which may interpreted as belonging to 
mixed symmetric multiples $[{\bf 70}, \ell^-]$ ($\ell = 1, 2$ and 3),
in agreement with  Ref. \cite{Stancu:1991cz}. To them we have added two new ones, $N_{15}(2060)$ and  $N_{13}(2120)$,
proposed in Refs. \cite{Anisovich:2011ye,Anisovich:2011fc},
which presently  acquired a two-star status \cite{PDG}. Note that to the latter we associated a mass of 2150 MeV as 
initially reported in Ref. \cite{Anisovich:2011fc}.

Four distinct numerical fits were performed by including the operators $O_1, ...,O_6$ and $B_1$  
of Table \ref{ouroperators},  by analogy to the $N$ = 1 band, from which the operator  $O_7$ has been neglected. 
The results of the numerical fits are exhibited in Table \ref{highlyexcited}.
One can see that the contributions of $O_5$ and $O_6$, depending on the angular momentum,  is negligible, but the coefficient of the spin-orbit operator,
although small, remains important to the fit. The spin operator $O_3$ brings a dominant contribution to $^4N$ resonances and 
the isospin operator $O_4$ even a larger contribution to the masses of $\Delta$ and $\Lambda$ resonances.  As Table \ref{MASSESFIT4} shows,
in the latter case its sign is negative and improves the agreement to the experiment.

Therefore, like in the $N$ = 1 band, we found that the isospin operator
neglected in the symmetric core + excited quark approach is very important, and definitely crucial in fitting the mass of 
the $\Lambda_{07}(2100)^{****}$ resonance.

\subsection{Positive parity mixed symmetric states}\label{positivep}

Here we present results for the masses of nonstrange and strange baryons resonances thought to belong to
the  $[{\bf 70},0^+]$ and $[{\bf 70},2^+]$ multiplets of the $N$ = 2 band.

\begin{table}[pt]
\caption{List of operators and the coefficients resulting from the fit with 
$\chi^2_{\rm dof}  \simeq 1.0$, for nonstrange and strange baryons belonging 
to the $[{\bf 70},\ell^+]$ multiplets ($\ell$ = 0 and 2).   From Ref. \cite{Matagne:2006zf}.}
\label{positiveparityoperators}
\renewcommand{\arraystretch}{1.5} 
\begin{tabular}{llrrl}
\hline
\hline
Operator & \multicolumn{4}{c}{Fitted coeff. (MeV)}\\
\hline
\hline
$O_1 = N_c \ \1 $                                   & \ \ \ $c_1 =  $  & 556 & $\pm$ & 11       \\
$O_2 = \ell_q^i s^i$                                & \ \ \ $c_2 =  $  & -43 & $\pm$ & 47    \\
$O_3 = \frac{3}{N_c}\ell^{(2)ij}_{q}g^{ia}G_c^{ja}$ & \ \ \ $c_3 =  $  & -85 & $\pm$ & 72  \\
$O_4 = \frac{4}{N_c+1} \ell^i_q t^a G_c^{ia}$         & \ \ \            &     &       &     \\
$O_5 = \frac{1}{N_c}(S_c^iS_c^i+s^iS_c^i)$          & \ \ \ $c_5 =  $  & 253 & $\pm$ & 57  \\
$O_6 = \frac{1}{N_c}t^aT_c^a$                       & \ \ \ $c_6 =  $  & -25 & $\pm$ & 86  \vspace{0.1cm}  \\ 
\hline
$B_1 = t^8-\frac{1}{2\sqrt{3}N_c}O_1$               & \ \ \ $d_1 =  $  & 365 & $\pm$ & 169 \\
$B_2 = T_c^8-\frac{N_c-1}{2\sqrt{3}N_c}O_1$         & \ \ \ $d_2 =  $  &-293 & $\pm$ & 54 \vspace{0.2cm} \\
\hline \hline
\end{tabular}
\end{table}

Although tedious in extending the symmetric core + excited quark approach to more than one excited quark
an effort has been made to apply it to the $N$ = 2 band \cite{Matagne:2005gd,Matagne:2006zf},
where the orbital wave function contains a term where two 
quarks are excited to the $p$-shell. For example, using the quark model notation $\rho$ and $\lambda$ 
for mixed symmetric three-quark states with the pair 1,2 in an antisymmetric and symmetric state respectively
one can write the orbital wave function for $\ell$ = 2 as 
\begin{widetext}
\begin{equation}\label{ell}
|{\bf N_c-1,1}, 2^+\rangle_{\rho,\lambda} =  
\sqrt{\frac{1}{3}}|[N_c-1,1]_{\rho,\lambda}(0s)^{N_c-1}(0d)\rangle 
+\sqrt{\frac{2}{3}}|[N_c-1,1]_{\rho,\lambda}(0s)^{N_c-2}(0p)^2\rangle,
\end{equation}
\end{widetext}
where the two quarks in the $p$-shell are coupled  to $\ell$ = 2.
In the first term a quark is excited in the $d$-shell so it can be treated as in 
the  $[{\bf 70},1^-]$ multiplet. The second term can be treated as an excited quark 
coupled to an excited core and one can use the fractional parentage technique developed 
in Ref. \cite{Matagne:2005gd}.
In this case the construction of the orbital part of the wave function becomes rather complicated  
which is the case for all bands with $N \geq 2$. The first application \cite{Matagne:2005gd} has been made to nonstrange
baryons using the SU(4) algebra to construct the operators $O_i$ in the mass formula (\ref{massoperator}).
The method has been extended in Ref. \ \cite{Matagne:2006zf} to include strange baryons as well. In this case 
the contribution of operators of type $B_i$ was added according to Eq. (\ref{massoperator}). 
There are many linearly independent operators which can be constructed from the excited quark and the excited core
operators. To make the method applicable we have restricted the list to those thought to be 
the most dominant. This is shown in Table \ref{positiveparityoperators}, where $\ell_q$ is the angular momentum of the
excited quark. The flavor operator $O_6$ was included. Its contribution is important, especially for flavor singlets.
Although listed and discussed in the paper, the operator $O_4$ was ignored in the fit because of
scarcity of data.  

There are two operators $B_i$ one acting on the excited quark, the other on the core. Their contribution mutually cancel to a large extent.
The $\Lambda \Sigma$ splitting obtained in the sector $^48$ is disturbingly large such that it provides for
$\Lambda$ and $\Xi$ baryons nearly equal masses.

\begin{table*}[pt]
\caption{List of dominant operators and their coefficients in the mass formula (\ref{massoperator}) obtained 
in three distinct numerical fits. From Ref. \cite{Matagne:2013cca}.}
\label{PRD87operators}
\renewcommand{\arraystretch}{1.2} 
\begin{tabular}{lrrrrr}
\hline
\hline
Operator & Fit 1 &\hspace{0.5cm} & Fit 2 & \hspace{0.5cm} & Fit 3\\
\hline
\hline
$O_1 = N_c \ \1 $               &   616 $\pm$  11     & &  616  $\pm$  11  & &  616  $\pm$  11   \\
$O_2 = \ell^i s^i$            &   150 $\pm$ 239     & &   52  $\pm$  44  & &  243  $\pm$  237            \\
$O_3 = \frac{1}{N_c}S^iS^i$     &   149 $\pm$ 30      & &  152  $\pm$  29 &  &  136  $\pm$   29    \\
$O_4 = \frac{1}{N_c}\left[T^aT^a-\frac{1}{12}N_c(N_c+6)\right]$  &  66 $\pm$ 55 & & 57 $\pm$ 51 & & 86 $\pm$ 55   \\
$O_5 =  \frac{3}{N_c} L^{i} T^{a} G^{i}$                         & -22 $\pm$ 5  & &             & &  -25 $\pm$ 52    \\ 
$O_6 =  \frac{15}{N_c} L^{(2)ij} G^{ia} G^{ja}$                  &  14 $\pm$ 5  & & 14 $\pm$ 5  & &    \\ 
\hline
$B_1 = - \mathcal{S}$           &  23 $\pm$ 38        & &  24 $\pm$ 38    & &  -22 $\pm$ 35     \\     
\hline                  
$\chi_{\mathrm{dof}}^2$         &  0.61               & &   0.52          & &   2.27   \\
\hline \hline
\end{tabular}
\end{table*}
\begin{table*}[pt]
\caption{Matrix elements of $O_i$ for flavor singlet resonances included in the analysis of Ref.\cite{Matagne:2013cca}.}
\label{SINGLET}
\renewcommand{\arraystretch}{1.2}
\begin{tabular}{lcccccc}
\hline 
\hline
   &  \hspace{ .3 cm} $O_1$ \hspace{ .3 cm}  & \hspace{ .3 cm} $O_2$  \hspace{ .3 cm} &  \hspace{ .3 cm} $O_3$  \hspace{ .3 cm} &  \hspace{ .3 cm} $O_4$  \hspace{ .3 cm} & \hspace{ .3 cm} $O_5$  & \hspace{ .3 cm} $O_6$ \\
\hline

$^21[{\bf 70},2^+]\frac{5}{2}^+$  &   $N_c$   &  $\frac{2}{3}$ & $\frac{3}{4N_c}$  & $-\frac{2 N_c + 3}{4N_c}$ & $-\frac{N_c-3}{2N_c}$ & 0\\
$^21[{\bf 70},2^+]\frac{3}{2}^+$  &   $N_c$   &  -$1$  & $\frac{3}{4N_c}$  & $-\frac{2 N_c + 3}{4N_c}$ & $\frac{3(N_c-3)}{4N_c}$ & 0 \\
$^21[{\bf 70},0^+]\frac{1}{2}^+$  &   $N_c$   &  0 & $\frac{3}{4N_c}$ & $-\frac{2 N_c + 3}{4N_c}$ & 0 & 0 \\

\hline
\hline
\end{tabular}
\end{table*}

The $[{\bf 70},\ell^+]$ baryons have been revisited \cite{Matagne:2013cca} by using the procedure
described in Sec. \ref{exactwavefunction} where the operators act on the entire system. 
In that analysis the wave function (\ref{WF}) has been used.
The list of dominant operators is given in Table \ref{PRD87operators}. Note that $O_2$ is a single particle
operator, having the same matrix elements as in Ref. \cite{Matagne:2006zf}. The matrix elements of $O_3$ and $O_4$ are
easy to calculate. 
The matrix elements of  $O_5$ and $O_6$
were obtained from the formulas (B2) and (B4) of Ref. \cite{Matagne:2011fr} and the corresponding isoscalar factors
of Tables \ref{octet_spin_one_half}, \ref{octet_spin_three_halfs}, \ref{decuplet_spin_one_half}
and \ref{singlet_spin_one_half} of Appendix \ref{A}.

The closed analytic form of the matrix elements of $O_i$, as a function of $N_c$, are not presented here
except for flavor singlets, Table \ref{SINGLET}, needed for the discussion.
Those for octets and decuplets can be found in Tables II can III  of Ref. \cite{Matagne:2013cca} respectively.

There  is a single operator 
which generates the flavor breaking, $B_1 = - \mathcal{S}$, the same for all sectors, where $\mathcal{S}$ is the strangeness.
In such a case there is no $\Lambda \Sigma$ splitting but the mass sequence with increasing number of strange quarks looks more
natural in octets and decuplets compared to the results of Ref. \cite{Matagne:2006zf} based on the symmetric core + excited quark approach.

This analysis was also motivated by the fact that the recent multichannel partial wave
analysis of Ref. \cite{Anisovich:2011fc} has revealed the existence of new positive parity resonances presently reported by 
the Particle Data Group \cite{PDG}.

Like for the $N$ = 1 and 3 bands, we found that both the  spin and flavor operators, $O_3$ and $O_4$ respectively,  
acting on the entire system, bring similar contributions to the mass, although $c_4$ is smaller than $c_3$, but 
what it matters is $c_i \langle O_i \rangle$.
The operator $O_4$, having negative 
matrix elements, see Table \ref{SINGLET}, helps in providing a good agreement of the mass of $\Lambda(1810)1/2^{+***}$ interpreted as
the flavor singlet  $^2\Lambda^{'}[{\bf 70},0^+]1/2$. Table \ref{PRD87operators} shows that the operator $O_5$ is
not important for a good fit but $O_6$ is crucial in obtaining a good $\chi_{\mathrm{dof}}^2$. 

\section{Heavy baryon masses in the combined $1/N_c$ and $1/m_Q$ expansion}\label{heavy}

The heavy quark limit was first discussed by Witten \cite{WITTEN}.
Later on the masses  of ground state baryons containing a single heavy quark $Q = c, b$ were studied in a combined  $1/m_Q$ and $1/N_c$ expansion
and SU(3) flavor symmetry breaking by Jenkins \cite{oai:arXiv.org:hep-ph/9603449,oai:arXiv.org:hep-ph/9609404}. The combined limit $m_c \rightarrow \infty $, 
$m_b \rightarrow \infty $, $N_c \rightarrow \infty$ for fixed $m_c/m_b$ and $N_c \Lambda_{QCD}/m_b$ has lead to a light quark 
$\ell$ and a heavy quark $h$ spin-flavor symmetry SU(6)$_{\ell}$ $\times$ SU(4)$_h$. For finite $m_Q$ and $N_c$ this symmetry
is violated by effects of order $1/N_c$ and $1/N_c m_Q$. A hierarchy of mass splittings was predicted together with the
masses of all of the unknown charmed baryons, as for example, $\Sigma_c^*$, $\Xi_c^{'}$ and $\Omega_c^{*}$
and of all unknown bottom baryons. The masses of the bottom baryons $\Sigma_b$,   $\Sigma_b^{*}$ and $\Xi_b$
observed ten years later were in good agreement with the theoretical predictions \cite{oai:arXiv.org:0712.0406}.

Model independent predictions for excitation energies and other observables of isoscalar heavy baryons were 
discussed in a combined heavy quark and large $N_c$ expansions \cite{oai:arXiv.org:hep-ph/0106096}. 

The mass spectrum of the $\ell$ = 1 charmed baryons was also studied in the $1/N_c$ method and the heavy 
quark effective theory and certain mass relations were derived \cite{oai:arXiv.org:hep-ph/0006267}. The simplicity 
of the approach stems from the fact that the light quark system is in the ground state  and the heavy quark
is orbitally excited. This is an improvement over previous studies made by the same authors \cite{oai:arXiv.org:hep-ph/9809576}.

\section{Mass formula in the $1/N_c$ expansion versus the quark model}\label{versus} 

It is important to see whether or not there is a compatibility between the model independent $1/N_c$ expansion and
the quark models, which successfully describe baryon spectroscopy. 
As mentioned in Sec. \ref{mixedsymmetricstates}, in the first application of the large $N_c$ method 
to a phenomenological analysis of  strong decays 
of $\ell$ = 1 orbitally excited baryons \cite{CGKM94}, a basic  purpose was to understand whether  the success of
the nonrelativistic quark model has a natural explanation in large $N_c$ QCD. 

The above application  was based on the Hartree approximation  suggested by Witten \cite{WITTEN} which inspired 
the symmetric core + excited quark procedure of Sec.  \ref{separation}. Subsequently
the validity of this procedure has been  formally supported by Pirjol and Schat  \cite{Pirjol:2007ed}
in a permutation group context by trying to match a large $N_c$ quark model Hamiltonian with the baryon mass formula
(\ref{massoperator}) of the $1/N_c$ expansion, including orbitally excited baryons, where some operators
$O_i$  contain angular momentum components. Only light baryons were considered, i. e. the SU(4) algebra.
The derivation confirmed the consistency between the order    $\mathcal{O}(1/N_c)$  of the corresponding  
operators $O_i$ shown in Table  \ref{12operators} and those resulting from large $N_c$ quark models.
Moreover, an explicit comparison of the Hamiltonian eigenvalues was made both for the one gluon-exchange (OGE) \cite{rgg} and the Goldstone-boson exchange (GBE)
\cite{Glozman:1995fu} models. 

Later on Pirjol and Schat \cite{oai:arXiv.org:1007.0964} tried to give more insight into
the spin-flavor structure of the hyperfine interaction used in  quark models.
They found that both OGE and GBE quark models are compatible with the $\ell$ = 1 nonstrange baryon data. 

Independently, a connection  between a semirelativistic quark model and the mass formula of the $1/N_c$ expansion
was established for light nonstrange baryons \cite{Semay:2007cv} and for  light nonstrange + strange baryons \cite{Semay:2007ff}, extended afterwards 
to heavy baryons \cite{oai:arXiv.org:0808.3349}, (for a detailed review see, for example, Ref. \cite{Buisseret:2008tq}).
A clear correspondence was found  between various terms of the quark model eigenvalues and those  
of the $1/N_c$ expansion mass formula.

\begin{figure*}[pt]\label{Fig1}
\begin{center}
\includegraphics[width=14cm]{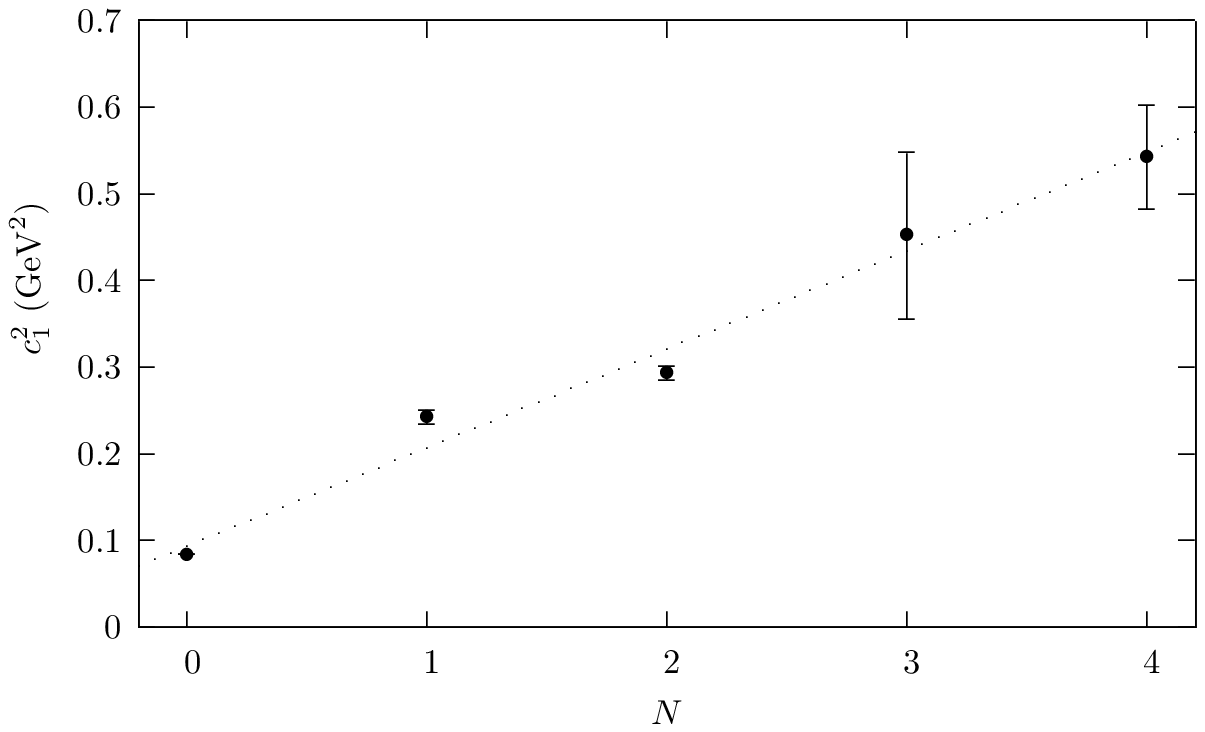} \\
\caption{Comparison between the quark model  and large $N_c$ results for $c^2_1$ (GeV$^2$) as a function of the band number $N$.
The dotted line represents  the quark model mass formula Eq. (\ref{analytic}) with the parameters (\ref{param}) taken from Ref.  \cite{Buisseret:2008tq}
and the points with error bars  
indicate large $N_c$ results:  at $N$ = 0  the value of $c_1$ was taken from Eq. (\ref{gsmass}), at  $N$ = 1 from Ref. \cite{Matagne:2011fr},   
at $N$ = 2 from Ref. \cite{GSS03} describing the multiplet $[{\bf 56},2^+]$ (see Table \ref{56lplus}), 
at $N$ = 3 from Ref. \cite{Matagne:2012tm} Fit 3 corresponding to the multiplets $[{\bf 70},\ell^-]$ ($\ell$ = 1,2,3)
and at $N$ = 4  from Ref. \cite{MS1} describing the multiplet $[{\bf 56},4^+]$ (see Table \ref{56lplus}). }
\end{center}
\end{figure*}
The spin independent  Hamiltonian used in Refs. \cite{Semay:2007cv,Semay:2007ff,oai:arXiv.org:0808.3349,Buisseret:2008tq} had 
a relativistic kinetic term and a $Y$-junction confinement interaction potential to which  
a Coulomb interaction term and a quark self-energy were added. 
Using the auxiliary field technique \cite{Silvestre-Brac:2011aua}
one can obtain an analytic expression for the mass of  light $qqq$ or heavy $qqQ$ baryons including
SU(3) breaking.  A key quantity is the band number $N$ in a harmonic oscillator picture, shown to be a good 
quantum number within the approximations considered in treating the quark model Hamiltonian. This allowed
to compare the dependence of various $c_i$ coefficients as a function of $N$ resulting from  the quark model and the $1/N_c$ expansion
results described above. For example for light baryons the quark model mass formula can be shortly written as
\begin{equation}
M_{qqq} = M_0 + n_s \Delta M_s
\end{equation}
where the first term holds for exact SU(3) flavor and the second term represents the breaking contribution. One can make the 
following identification with the mass formula (\ref{massoperator})
\begin{equation}\label{comparison}
c^2_1 = M^2_0/9, ~~~ n_s \Delta M_s = \sum_i d_i B_i
\end{equation}
where the number of strange quarks is $n_s$ = 0,1,2 or 3. The quantities $M_0$ and $\Delta M_s$ are functions of the band number $N$.
Using the analytic form of $M_0$  in terms of  quark model parameters $\sigma$, $\alpha_0$ and $f$,  defined, for example,  
in Ref. \cite{Semay:2007cv}, one can write 
\begin{equation}\label{analytic}
c^2_1 = \frac{2 \pi}{9} \sigma (N + 3) - \frac{4 \pi}{9 \sqrt{3}} \sigma \alpha_0 - \frac{f \sigma}{3},
\end{equation}
with the following choice of parameters \cite{Buisseret:2008tq}
\begin{equation}\label{param}
\sigma = 0.163 ~{\mathrm{GeV}}^2, \ \,  \alpha_0 = 0.4, \ \, f = 3.6. 
\end{equation}
This expression is plotted as a function of the band number $N$ in Fig. 1 where  it is
compared  with  large $N_c$ results. 
One can see that there is a rather good quantitative agreement between the large $N_c$ and the quark model results for
$c^2_1$.  In the quark model  $c^2_1$ contains the effect of the kinetic and of the confinement parts of the spin independent Hamiltonian 
and the nonperturbative QCD at large distances becomes dominated by confinement. 
The present agreement between the quark model results and large $N_c$ QCD brings 
further support to quark models.

Actually the quantity  $c^2_1$ is linear in the band number $N$, suggesting  a Regge-type behavior obtained from the analytic form of $M_0$,
which is
\begin{equation}\label{Regge}
M^2_0 \propto 2 \pi \sigma (N + 3)
\end{equation}
containing the quark model parameter  $\sigma$ responsible for the slope.

On the other hand a Regge-type behavior of the leading spin-flavor singlet term of the large $N_c$ mass formula has been discussed in Refs. 
\cite{Matagne:2013cca} and \cite{Goity:2007sc} where two distinct nearly parallel Regge trajectories have been found, the lower one for the 
symmetric ${\bf {56}}$-plets and  the upper one for the mixed symmetric ${\bf {70}}$-plets. It would be interesting to understand such an effect.
In addition, one could try to see if baryons and mesons lead to degenerate Regge slopes in agreement with the predictions of 
Ref. \cite{Armoni:2009zq} where massive mesons and baryons become  supersymmetric partners in the large $N_c$ limit.


\section{The quark excitation versus the meson-nucleon resonance picture}\label{MN}

The contracted  SU($2N_f$)$_c$ spin-flavor symmetry is a consequence of  large $N_c$ consistency conditions 
imposed on the meson-baryon scattering amplitudes \cite{Gervais:1983wq,Dashen:1993as}. Therefore it 
seems natural to inquire about the compatibility between  the quark excitation picture 
presented in Secs. \ref{mixedsymmetricstates} and \ref{specmixed} and the meson-nucleon resonance picture. 

According to the authors of Refs. \cite{Gervais:1983wq,Dashen:1993as}
in large $N_c$ QCD the pion-baryon couplings must satisfy a set of consistency conditions  
which require the existence of an infinite tower of degenerate baryon states with $I$ = $J$
and also determine the ratios of the pion-baryon coupling constants, which turn out to be 
identical to those given by the Skyrme model. 
This implies that the large $N_c$ QCD displays a contracted spin-flavor symmetry SU($2N_f$)$_c$ .  
The symmetry is a property of 
ground state baryons. As a matter of fact, the meson sector does not display such a symmetry.

There is no a priori justification of extending this symmetry to excited states,
which do not become stable at large $N_c$ and where, in addition,   an extension to  SU($2N_f$) $\times$ O(3) 
symmetry is necessary for introducing angular momentum components. 

In fact Witten  \cite{WITTEN}
has shown that the characteristic width of an excited baryon is of order $N^0_c$, while stable states 
are assumed in Refs. \cite{Gervais:1983wq,Dashen:1993as}.
Pirjol and Yan \cite{PY} were the first to analyze the consistency condition for excited baryons.
Their procedure is similar to that of Dashen and Manohar \cite{Dashen:1993as}.
The pions are scattered off excited baryons and one must assume that the target is stable. 
The target was described by a mixed symmetric representation of SU($2N_f$) where they claimed
that the pion-nucleon coupling goes as $N^{-1/2}_c$ to produce narrow resonances. Functional
forms of relations satisfying the consistency conditions were motivated from a simple 
nonrelativistic quark model.

The legitimacy of this procedure  has been questioned by Cohen et al. \cite{Cohen:2003fv}.
They have found that the existence of states with a width which goes as  $N^{-1}_c$
is an artifact of the simple quark model used in Ref. \cite{PY}.

To evade the difficulties of an extension of the techniques
from the ground state to excited states \cite{PY}, 
Cohen and Lebed,  in a series  of papers  \cite{COLEB1,COLEB2,Cohen:2005ct}, proposed 
to study the scattering process in large $N_c$ and compare the 
findings with the quark excitation picture, named by the authors, the
$\emph{quark-shell picture}$, and based, as we have mentioned,  on the extended symmetry 
SU($2N_f$) $\times$ O(3). For simplicity let us restrict to the SU(4) case.

The starting point was the linear relations 
of the S matrices  $S^{\pi}_{LL'RR'IJ}$ and  $S^{\eta}_{LRJ}$ of $\pi$ and $\eta$ scattering off 
a ground state baryon. They are given by the following equations 
\begin{widetext}
\begin{equation}\label{pi}
S^{\pi}_{LL'RR'IJ} = \sum_K ( - 1)^{R'-R} \sqrt{(2R+1)(2R'+1)} (2K+1)
\left\{\begin{array}{ccc}
        K& I & J \\
	R' & L' & 1
      \end{array}\right\} 
 \left\{\begin{array}{ccc}
        K& I & J \\
	R & L & 1
      \end{array}\right\}  
      s^{\pi}_{KLL'}    
\end{equation}
\end{widetext}
and
\begin{equation}\label{eta}
S^{\eta}_{LRJ} = \sum_K \delta_{KL}\delta(LRJ) s^{\eta}_{K}
\end{equation}
in terms of the reduced amplitudes $s^{\pi}_{KL'L}$ and $s^{\eta}_{K}$ respectively. 
These equations were first derived in the context  of the chiral soliton model 
\cite{HAYASHI,MAPE,MATTIS,MattisMukerjee}
where 
the mean-field breaks the rotational and isospin symmetries, so that $J$ and $I$ are not
conserved but the ${\it grand}$ ${\it spin}$  $K$ is conserved and excitations can be labelled by $K$.
These relations are exact in large $N_c$ QCD and are independent of any model assumption.
The notation is as follows. For $\pi$ scattering $R$ and $R'$ are the spin of the incoming and outgoing baryons 
respectively ($R$ = 1/2 for $N$ and $R$ = 3/2 for $\Delta$), $L$ and $L'$ are the partial wave angular momentum of the
incident and final $\pi$ respectively (the orbital angular momentum $L$ of $\eta$ remains unchanged), 
$I$ and $J$ represent the total isospin and total angular momentum
associated to a given resonance and $K$ is the 
magnitude  of the ${\it grand}$ ${\it spin}$ $\vec{K} = \vec{I} + \vec{J}$.
The $6j$ coefficients imply four triangle rules $\delta(LRJ)$, $\delta(R1I)$, $\delta(L1K)$ and 
$\delta(IJK)$.

Eqs. (\ref{pi}) and (\ref{eta}) help to relate  scattering amplitudes in various 
channels with $K$ amplitudes and look for common poles, $\emph{i. e.}$ resonances. 
These poles should correspond to degenerate towers of states. 
The quantum numbers of the channels are the quantum numbers of an $N_c$ quark system
given by a large $N_c$ quark model. Thus the quantum numbers of an $N_c$ quark system 
are the important degrees of freedom of the  $\emph{quark-shell picture}$.

According to Cohen and Lebed, if the pattern of degeneracy resulting from  Eqs. (\ref{pi}) and (\ref{eta}) 
is the same as that of the quark-shell picture it means that the two pictures are
compatible and the extension of the $1/N_c$ expansion method to excited states is justified. 
The compatibility is illustrated below for $N_f$ = 2.

The $\emph{quark-shell picture}$
requires the introduction of a Hamiltonian model
with an SU($2N_f$) $\times$ O(3) symmetry containing operators up to order $N^0_c$.

1) In the symmetric core + excited quark procedure, Sec. \ref{separation},
there are three operators up order $N^0_c$, namely  
\begin{equation}
O_1 = N_c \ \1 , ~~~  O_2 = \ell \cdot s, ~~~   O_3 =  \frac{15}{N_c} \ell^{(2)} \cdot g \cdot G_c.
\end{equation}
which generate the Hamiltonian
\begin{equation}\label{symcore}
H = c_1 N_c \ \1 + c_2 \ell \cdot s + c_3  \frac{15}{N_c} \ell^{(2)} \cdot g \cdot G_c.
\end{equation} 
The only  three distinct eigenvalues of this Hamiltonian 
can be obtained analytically. For  $\ell$ = 1 they were given in  Refs. \cite{COLEB1,Pirjol:2003ye}.
Note that the normalization of $O_3$ is different in Refs.  \cite{COLEB1} and \cite{Pirjol:2003ye}   
which is reflected in the corresponding analytic expressions of  the eigenvalues.
A similar analysis based on the Hamiltonian (\ref{symcore}) has been
extended to $\ell$ = 3 in Ref. \cite{Matagne:2011sn}.

2) In the exact basis, Sec. \ref{exactwavefunction}, there are also three operators with matrix elements up to order $\mathcal{O}(N^0_c)$.
In the notation of Ref. \cite{Matagne:2012vq} they are
\begin{equation}
O_1 = N_c \ \1 , ~~~  O_2 = \ell \cdot s, ~~~   O_6 =  \frac{15}{N_c} L^{(2)} \cdot G \cdot G,
\end{equation}
which generates the Hamiltonian
\begin{equation}\label{exact}
H = c_1 N_c \ \1 + c_2 \ell \cdot s + c_6  \frac{15}{N_c} L^{(2)} \cdot G \cdot G .
\end{equation} 
The first two terms are the same as in Eq. (\ref{symcore}) but in  $O_6$ the SO(3) tensor $L^{(2)}$ and 
the SU(4) operator $G$ act on the whole system. Interestingly, the corresponding Hamiltonian  
has analytical solutions too. These are
\begin{equation}\label{m0} 
m'_0 = c_1 N_c - c_2 - \frac{25}{4} c_6,
\end{equation}
\begin{equation}\label{m1} 
m'_1 = c_1 N_c - \frac{1}{2} c_2 + \frac{25}{8} c_6,
\end{equation}
\begin{equation}\label{m2} 
m'_2 = c_1 N_c + \frac{1}{2} c_2 - \frac{5}{8} c_6.
\end{equation}
Then, for   $\ell$ = 1 the following degenerate sets of resonances were found 
\begin{equation}\label{K0}
N_{1/2}, ~ \Delta_{3/2}, 
 ~~(s^{\eta}_0), ~~ (m'_0),  
\end{equation} 
\begin{equation}\label{Kequal1}
N_{1/2},~ \Delta_{1/2},~N_{3/2},~ \Delta_{3/2},~ \Delta_{5/2}, ~~ (s^{\pi}_{100}, s^{\pi}_{122}),
 ~~(m'_1),
\end{equation} 
\begin{equation}\label{Kequal2}
\Delta_{1/2},  ~ N_{3/2},   ~ \Delta_{3/2}, 
~ N_{5/2},  ~ \Delta_{5/2},   
~ \Delta_{7/2}, ~~ (s^{\pi}_{222}, s^{\eta}_2),
 ~~(m'_2),
\end{equation} 
where,  on the right side we  indicate  the associated   amplitudes 
$s^{\pi}_{KL'L}$ or $s^{\eta}_K$ of Eq. (\ref{pi}) or (\ref{eta}) followed by the mass of each degenerate set.
Thus the degenerate sets are  identical to those obtained from the meson-baryon picture. 
In addition the degenerate sets in the exact basis are identical  to those found 
in Refs. \cite{COLEB1} and \cite{Pirjol:2003ye}, which means that the same quantum numbers are involved.
The masses $m'_i$ of Eqs. (\ref{m0})-(\ref{m2}) shown here are naturally different from $m_i$ of 
the above references because the Hamiltonian is different in structure and it contains different 
dynamical coefficients. However it has similar large $N_c$ properties.

The conclusion is that
any resonance that do exist must fall into multiplets that become degenerate in both mass
and width (or equivalently coupling constant) at large $N_c$. The pattern of degeneracy is fully
fixed by the contracted  SU($2N_f$) symmetry. For $N_f$ = 2 each set of degenerate states is 
defined by a single quantum number $K$ = 0, 1 and 2 for  (\ref{K0}), (\ref{Kequal1}) and (\ref{Kequal2}) respectively.

For $\ell$ = 3, described within the  symmetric core plus excited quark procedure we refer to
our analysis   \cite{Matagne:2011sn}
which confirms the compatibility between the two pictures once more. 
In addition  we supported the triangular rule $\delta( K \ell 1)$ proposed in Ref \cite{COLEB2}
according to which one can associate a common $K = 2$ to both $\ell = 1$ and $\ell = 3$.  
In some sense the quark-shell picture, where $\ell$ is conserved, brings an alternative information 
to the resonance picture, which may be more relevant for experimentalists, because it implies  
an energy dependence via the $\ell$ dependence which measures the orbital excitation.

The inclusion of strange quarks complicates the
analysis. One must consider only those states within a multiplet with the same values of isospin 
and strangeness as for $N_c$ = 3. The problem was discussed qualitatively in Ref. \cite{Cohen:2005ct}.
Analyzing the  SU(3) $\times$ SU(2) content of the ${\bf 70}$ irrep of SU(6), 
twenty multiplets were found, with five distinct masses, corresponding to $K$ = 0, 1/2, 1, 3/2,
and 2. This is a  model independent result. A simple Hamiltonian expressed in terms of the  
symmetric core plus excited quark procedure containing the operators  
\begin{widetext}
\begin{equation}
O_1 = N_c \ \1 , ~~~  O_2 = \ell \cdot s, ~~~  
O_3 = \frac{3}{N_c} \ell^{(2)}\cdot g \cdot G_c, ~~~   O_4 = \ell \cdot s + \frac{4}{N_c+1} \ell\cdot t \cdot G_c,
 ~~~   O_5 = \frac{1}{N_c}(t \cdot T - \frac{\ \1}{12}) 
\end{equation}
\end{widetext}
also gives five distinct masses, which suggests that the compatibility between the quark-shell picture
and the meson-nucleon scattering picture can be achieved. Note that the additional operator $O_5$ acting only on flavor, 
usually omitted in the symmetric core + excited quark procedure \cite{CCGL},  is crucial in the compatibility issue.
The compatibility  has not yet been studied explicitly in the exact basis procedure for $N_f$ = 3. 

Finally,  we should mention that, in the nonstrange sector,
the compatibility between the two pictures was claimed on a general 
group theoretical arguments 
by Cohen and Lebed \cite{COLEB2} for completely symmetric, mixed symmetric and completely antisymmetric
flavor-spin states of $N_c$ quarks having angular momentum up to $\ell$ = 3.

A bridge between the quark models in large $N_c$ and the solitonic approach  of the Skyrme model has been 
established in Ref. \cite{Diakonov:2013qta}
within a relativistically invariant formalism to take into account $q \bar q$ pairs. In this work it was pointed out
that the advantage of the large $N_c$ limit is that the baryon physics simplifies considerably which allows one 
to take into full account important relativistic and field-theory effects which are often ignored.

\section{Baryon masses in the combined $1/N_c$ and chiral expansions}\label{combchiral}

Based on the idea that the combined $1/N_c$ expansion and chiral perturbation theory (ChPT)
can constrain the low-energy interactions of baryons with mesons a $1/N_c$ expansion of the chiral Lagrangian
has been formulated by Jenkins \cite{Jenkins:1995gc} quite early, for the lowest lying baryons. The expansion
parameters are $1/N_c$ and $m_q/\Lambda_{QCD}$ with the double limit $1/N_c \rightarrow 0$ and $m_q/\Lambda_{QCD} \rightarrow 0$
and the ratio $\frac{1}{N_c}/(m_q/\Lambda_{QCD})$ held fixed. 
The two limits cannot be taken independently from each other.
The chiral
Lagrangian correctly implements the pseudoscalar meson nonet symmetry and the contracted spin-flavor symmetry introduced in Sec. \ref{sfsymmetry}.
It describes the interaction of the spin-1/2 baryon octet and the spin-3/2 baryon decuplet with the pseudoscalar nonet.
Strong $CP$ violation was included. 

Within the same framework the combined $1/N_c$ and chiral expansions has been recently considered in Ref.
\cite{CalleCordon:2012xz} based on the important conjecture that the two
expansions do not commute \cite{Adkins:1983hy}. The dynamics underlying the noncommutativity is due 
to the behavior of the $\Delta$ resonance  \cite{Cohen:1992uy,Dashen:1993jt,Cohen:1996zz}.

We recall that  ChPT is an effective field theory that makes use of an expansion in powers of momenta $p$ \cite{Leutwyler:1994fi}.
The baryon mass splitting is taken to be $\mathcal{O}(p)$ in this expansion, named  $\xi$ expansion in Ref. \cite{CalleCordon:2012xz}.
Results for baryon masses and axial couplings
were obtained in an  expansion where $1/N_c$ = $\mathcal{O}(\xi)$ = $\mathcal{O}(p)$, thought to be the most realistic 
for studying baryons at $N_c$ = 3. Applications to lattice QCD were presented. It would be interesting to 
extend the work from $N_f$ = 2 to $N_f$ = 3. Results for the axial currents with three flavors, in a similar framework, were 
presented in Ref. \cite{FloresMendieta:2012dn}.


\section{Strong decays}\label{strongdecays}

Besides the spectrum, the strong decay of baryons represent an important field 
of application of the large $N_c$ method. Like for the spectrum, one can perform an operator analysis.
So far only a few papers were devoted to the study of strong decays within this framework.
As mentioned at the beginning of Sec. \ref{mixedsymmetricstates} the first application of the large $N_c$
method was a phenomenological analysis of strong decays of $\ell$ = 1 orbitally excited baryons
\cite{CGKM94}. This work was intended to show that the success of nonrelativistic quark models
has an explanation in large $N_c$ QCD. For this purpose it was enough to consider a restricted basis
of operators at subleading order in $1/N_c$.
This study was followed by the analysis of strong decays of the Roper resonance \cite{CC00}.

A complete analysis to $\mathcal{O}(1/N_c)$ of strong decays of nonstrange baryons belonging to the 
20-plet of SU(4) was given in Ref. \cite{Goity:2004ss} followed  by the analysis of positive 
parity nonstrange resonances  \cite{Goity:2005rg}
and  nonstrange + strange resonances   
of the ${\bf 56}$-plet of SU(6) \cite{Goity:2009wq}. Finally the study of negative parity baryon decays
was extended to SU(6) in Ref. \cite{Jayalath:2011uc}.

Note that all  the above cited studies of strong decays of negative parity mixed symmetric states
in the framework of the operator analysis rely on the Hartree approximation \cite{CGKM94}, 
or its implementation according to Sec. \ref{separation}. 

Another framework to study strong decays is based on the scattering amplitudes. This is the approach 
introduced in Sec. \ref{MN} used in Refs. \cite{Cohen:2003fv,COLEB1,COLEB2,Cohen:2005ct}.
The analytic structure of these amplitudes was used to prove the compatibility between the 
operator analysis and the meson-nucleon scattering picture. In the operator analysis
only terms of order $N^0_c$ have been used and  the name  was the $\emph{quark-shell picture}$.  
The comparison was therefore only qualitative.
Higher order terms were introduced in Ref. \cite{Pirjol:2003ye}.

Below we describe a few results obtained in the operator analysis approach 
for the strong decay widths.

\subsection{Radially excited states}\label{Radially}

For historical and pedagogical purposes we first shortly 
present the study of strong decay widths of the lowest-lying radially excited baryons of Ref. \cite{CC00},
with special attention to the Roper resonance. The large $N_c$ mass formula was written under the 
form of a G\"ursey-Radicati type. The analysis was free of any assumption regarding the interaction
potential and the quark wave functions. The decay was assumed to take place via a single quark 
interaction vertex so that the transition operator can be expressed in terms of SU(6) generators as
\begin{equation}\label{heff}
\mathcal{H}_{eff} \propto G^{ia} k^i \pi^a
\end{equation}
where $G^{ia}$ is the SU(6) generator defined by Eq. (\ref{sunf}),
$k^i$ is  the meson momentum component $i$ and  $\pi^a$ is the meson field
operator. The SU(6) operator acts on the excited quark.
The matrix elements of the operator (\ref{heff}) between the excited baryon $B_i$ and the final baryon $B_f$ + meson 
gives the transition amplitude
\begin{equation}\label{tramp}
\langle \Psi(B_f,\pi^a) | \mathcal{H}_{eff}| \Psi(B_i)\rangle = f(k) k^j \langle B_f | G^{ja} | B_i \rangle
\end{equation}
where $f(k)$  is a function that parametrizes the momentum dependence of the amplitude which encodes the 
baryon structure and therefore the binding potential. For a harmonic type confinement this function has a 
simple analytic form \cite{Koniuk:1979vy} and the above factorization takes place in general
if recoil effects of the emitting quark are ignored \cite{Sartor:1986qr}. In Ref. \cite{CC00}
a simple functional form $ f(k) = (2.8 \pm 0.2)/k $ was found to fit the data best while the  harmonic type confinement
lead to an exponential decrease with $k$ \cite{Koniuk:1979vy}.
The above analysis has been stimulated by the successful large $N_c$ study of strong decays of the $\bf 70$-plet
\cite{CGKM94} which preceded the more involved study of strong decays of Refs. \cite{Goity:2004ss,Goity:2005rg}.
A good choice of the profile function $f(k)$ as above can help in including the largest part of the momentum
dependence in the leading terms of the  large $N_c$ expansion of  the transition operator.   

Next we present the more elaborate, rather recent studies.

\subsection{The ${\bf 56}$-plet }

In Ref. \cite{Goity:2005rg} the multiplets $[{\bf 56'},0^+]$ and $[{\bf 56},2^+]$ were analyzed in SU(4). 
One interprets the Roper resonance as belonging to $[{\bf 56'},0^+]$. 

The transition operators, classified in multiplets of the O(3) $\times$ SU(2N$_f$) group   \cite{Goity:2005rg}, are reproduced in Table
\ref {GoityPRD72}. The list contains 1-body and two 2-body operators, with the order 
specified in the last column.
The name  $O^{[\ell_P,1]}$ contains  
the partial wave $\ell_P$ of the decay channel
and 1 is the isospin of the emitted  pion. 

\begin{table}[ht]
\begin{center}
\caption{Basis operators for pion decay of $[{\bf 56'},0^+]$ and $[{\bf 56},2^+]$ resonances
in SU(4). From Ref. \cite{Goity:2005rg}.}
\label{GoityPRD72}
\renewcommand{\arraystretch}{1.5}
\begin{tabular}{cccccccccc}\hline\hline
\hspace*{.cm} 
\hspace*{.cm}   &
\hspace*{.cm} Name                \hspace*{.cm}   &
\hspace*{.cm} Operator            \hspace*{.cm}   &
\hspace*{.cm} Order    \hspace*{.cm}
\\ \hline \hline
 1-body   &  $O_1^{[\ell_P,1]}$ &
$\frac{1}{N_c}\left( \xi^\ell \ G \right)^{[\ell_P,1]}$
& $\mathcal{O}(N^0_c)$ \\[2mm] 
 2-body  &   $O_2^{[\ell_P,1]}$ &
$\frac{1}{N_c^2} \left(\xi^\ell \left( [ S \ , \ G ] \right)^{[1,1]} \right)^{[\ell_P,1]}$
& $\mathcal{O}(1/N_c$)  \\[2mm]
     &   $O_3^{[\ell_P,1]}$ &
$\frac{1}{N_c^2} \left(\xi^\ell \left( \{ S \ , \ G \} \right)^{[2,1]} \right)^{[\ell_P,1]}$
&  $\mathcal{O}(1/N_c$) \\[2mm] \hline \hline
\end{tabular}
\end{center}
\end{table}
The operators  $\xi^{\ell}_m$  are components of a tensor of rank $\ell$ in SO(3), responsible for the transition between
an excited state with orbital angular momentum $\ell$ and the ground state. They were normalized to have matrix elements
of the form
\begin{equation}
\langle 0 | \xi^{\ell}_{m'}| \ell m \rangle = (-)^{\ell+m}  \delta_{m,-m'},
\end{equation} 
consistent with the Wigner-Eckart theorem provided the reduced matrix element in the right-hand side
is $\langle 0 || \xi^{\ell}|| \ell  \rangle = \sqrt{2 \ell + 1}$. 
The operators $S$ and 
$G$ are SU(4) generators. From these generators one constructs spin-flavor tensors 
$\left( {\cal G}^{[S_P, I_P]}_{[S_{3P},I_{3P}]} \right)_q$ where $S_P,S_{3P}$ are the spin 
and its projection and $I_P,I_{3P} $ are the isospin and its projection, the same as the isospin and its projection for the emitted meson. The quantity $q$ 
numbers the operators generally considered in an operator expansion study $q$ = 1,2, ..., etc.  
The adequate coupling for the partial wave $\ell_P$ of 
the meson emission defines the operators $\left( B^{[\ell_P, I_P]}_{[m_P,I_{3P} ]} \right)_q$ 
\begin{widetext}
\begin{equation}
\left( B^{[\ell_P, I_P]}_{[m_P,I_{3P} ]} \right)_q = \sum_m
\langle \ell, m ; S_P S_{3P} \mid \ell_P, m_P \rangle 
\xi^\ell_m \ \left( {\cal G}^{[S_P, I_P]}_{[S_{3P},I_{3P}]} \right)_q,
\end{equation}
\end{widetext}
which were used to construct a transition operator in the $1/N_c$ expansion as
\begin{widetext}
\begin{equation}
B^{[\ell_P, I_P]}_{[m_P, I_{3P}]} =
\left(\frac{k_P}{\Lambda}\right)^{\ell_P}\sum_q \, C_q^{[\ell_P, I_P]}(k_P)
\left( B^{[\ell_P, I_P]}_{[m_P, I_{3P}]} \right)_q,
\label{strongtrans}
\end{equation}
\end{widetext}
containing a desired number of terms each having a coefficient $ C_q^{[\ell_P, I_P]}(k_P)$
to be fit from data. The quantity 
$\left(\frac{k_P}{\Lambda}\right)^{\ell_P}$ was designed to capture the main momentum dependence 
of the dynamical coefficients $C_q^{[\ell_P, I_P]}(k_P)$ which were taken in practice as momentum independent.
The constant $\Lambda$ was chosen to be equal to 200 MeV.
Another alternative would be to introduce a momentum dependence through a profile function
as in Ref.  \cite{CC00}. The two ways are equally valid, as long as no explicit dynamics is involved.

In this notation the transition operator (\ref{heff}) of Ref. \cite{CC00} corresponds to the $O_1^{[\ell_P,1]}$
operator listed in Table \ref{GoityPRD72}.

Then the strong decay width in a nonrelativistic kinematics is defined as 
\begin{equation}
\Gamma^{[\ell_P,I_P]}
= \frac{k_P}{8 \pi^2} \frac{M_{B}}{M_B^*} \ \frac{| {\it B} (\ell_P,I_P,S,I,J^*,I^*,S^*) |^2}
{(2 J^* + 1)(2 I^*+1)},
\label{strongwidth}
\end{equation}
where ${\it B} (\ell_P,I_P,S,I,J^*,I^*,S^*)$ are the reduced matrix elements of the strong decay 
operator $B^{[\ell_P, I_P]}_{[m_P, I_{3P}]}$ defined above, with $J^*,I^*,S^*$ the quantum numbers
of the decaying resonance of mass $ M_{B^*}$ and $S,I$ the spin and isospin of the baryon ground state of mass ${M_B}$.

The reduced matrix elements ${\it B} (\ell_P,I_P,S,I,J^*,I^*,S^*)$ are defined by the generalized 
Wigner-Eckart theorem introduced in Appendix \ref{SU4} with  notations adapted to the present case.

In Ref. \cite{Goity:2005rg} the decay widths in the $p,f$ and $h$ partial waves were calculated for
a number of resonances with increasing masses starting from the Roper $N(1440)1/2^+$ till  $\Delta(2420)9/2^+$.
The decay channels were $\pi N$ and $\pi \Delta$. 
The results of \cite{Goity:2005rg} indicate that the pion decay are qualitatively well described 
at leading order described by the operator $O_1^{[\ell_P,1]}$, which
explains why the simple picture of the quark model works qualitatively well, and also justifies the 
choice of Ref. \cite{CC00}.
However the Roper resonance requires important next to the leading corrections as considered in
Ref.   \cite{CC00}, a result also consistent with the quark model studies, as e.g. \cite{Sartor:1986qr},
which give too small a width. Also the predicted suppression of the $\eta$ decay is consistent with
the experimental results obtained so far. 

In the extension to SU(6), Goity et al. \cite{Goity:2009wq}  followed 
a similar procedure to SU(4) to construct the spin-flavor transition operators.  
The SU(3) conserving  operators correspond to those of Table \ref{GoityPRD72} but written in SU(6)
notations. The SU(6) basis contains an additional SU(3) symmetry breaking (SB) operator 
\begin{equation}
 {\cal{G}}_{SB}\equiv\frac{1}{N_c}\;(d_{8ab}-\delta_{ab}/\sqrt{3})\; G_{ib}
\end{equation}
of order  ${\cal{O}}((m_s-m_{u,d})/\sqrt{N_c})$
which is necessary to carefully distinguish between 
emission of pions and $K$ mesons. The conclusions are similar to those obtained in the SU(4) case.


\subsection{The $[{\bf 70},1^-]$-plet }

As mentioned, the first analysis of the strong decays of the lowest negative parity baryons was made in
Ref. \cite{Goity:2004ss}  where the multiplet belongs to the irreducible representation 20 of SU(4).
In Ref. \cite{Jayalath:2011uc} the study of negative parity baryon decays was extended to SU(6). 
We refer the reader to these two papers for details. The construction of transition operators is similar to that
of the mass operator, using operators acting on the excited quark or on the core. This implies a larger
number of transition operators than for the  ${\bf 56}$-plet.
Both studies indicate that the 1-body operators  are dominant  in the $S$- and $D$-partial wave decay widths,
which again support the quark model picture based on the spectator model, where the pseudoscalar meson is emitted 
from the excited quark. The 2-body operators are crucial for an overall good description. 
They are thought to encode the longer range 
dynamics of the decay. However the calculated width of the $N(1535) \rightarrow \eta N$ 
and $N(1650) \rightarrow \eta N$ are too small at leading order. The  SU(3) breaking effects
turn out to be unnaturally large as the next-to-leading order analysis has shown.

An exhaustive combined analysis of the masses, strong decay widths and photo-couplings (see below)
has recently been performed in Ref. \cite{deUrreta:2013koa} for the lowest nonstrange negative parity
resonances belonging to the $[{\bf 70},1^-]$ multiplet of SU(4) $\times$ O(3) 
including an updated input for the $N_{1/2}$ baryons. The conclusion was that the composition
of the spin 1/2 and 3/2 states, which involve two mix‌ing angles, is in agreement with 
the non-relativistic quark model of Isgur and Karl obtained from the analysis of strong decays 
alone \cite{IK78}.


\section{Photoproduction amplitudes in the $1/N_c$ expansion}\label{photo}

The first analysis of the helicity amplitudes in the $1/N_c$ expansion
was devoted to  negative parity baryons  by Carlson and Carone \cite{CaCa98}. Regarding  positive parity baryons,
a particular case,  the decay $\Delta^+ \rightarrow p \gamma$ has been
studied a few years later by Jenkins et al. in Ref.  \cite{Jenkins:2002rj} where the ratio of the helicity amplitudes
$A_{3/2}/A_{1/2}$ was found to be $\sqrt{3} + \mathcal{O}(1/N_c^2)$, compatible with
experiment and  with quark models where the ratio is $\sqrt{3}$, see, for example,
Eqs. (C49) and (C50) of Ref. \cite{Sartor:1986sf}. Therefore this study showed that the ratio 
between the electric quadrupole $E2$ and the magnetic moment $M1$
is of order $1/N^2_c$. In this work the isovector electromagnetic current operator was expanded in powers of the 
photon momentum $k^j$
\begin{equation}
J^{ia}_{EM} \propto \mu^{ia} + Q^{(ij)a} k^j + ...
\end{equation}
where $i,j$ = 1,2,3 are SU(2) spin and isospin indices.
The $M1$ and $E2$ transition amplitudes are
\begin{equation}
M1 = e \sqrt{k^j} \langle N | \mu^{j3} | \Delta \rangle, 
\ \ E2 = \frac{e}{12} (k^0)^{3/2} \langle N | Q^{(20)3} | \Delta \rangle,
\end{equation}
where  spherical components of the quadrupole moment $Q^{(ij)}$ have been used in the latter equation. 
The $1/N_c$ expansion has been applied to $\mu^{j3}$ and $Q^{(20)3}$ operators.

Shortly after, this study has been extended to hyperon radiative decays by 
Lebed and Martin \cite{Lebed:2004zc} who calculated the radiative widths.

A few years later   the photoproduction amplitudes of positive parity baryons have been 
thoroughly studied by Goity and Scoccola \cite{Goity:2007ft}. This type of analysis has been 
extended to negative parity photoproduction amplitudes by Scoccola, Goity and Matagne
\cite{Scoccola:2007sn} by systematically building a complete basis of current operators 
to subleading order in $1/N_c$. The conclusion was that the one-body operators are
dominant and the subleading corrections in $1/N_c$ are important and suggest evidence for 
the need of two-body operators.

An alternative, model-independent, approach to study nonstrange resonances was based on the 
large $N_c$ consistency conditions \cite{Gervais:1983wq,Dashen:1993as}
to derive  linear relations among partial wave amplitudes for the elastic 
$\pi N \rightarrow \pi N$ and the inelastic $\pi N \rightarrow \pi \Delta$ processes
\cite{Cohen:2004qt}. The leading order relations were derived in the context of chiral
soliton models in Refs. \cite{HAYASHI,MAPE,MATTIS,MattisMukerjee} as mentioned in Sec. \ref{MN}.
Their rederivation based on group structure was obtained in Refs. \cite{COLEB1,COLEB2}.
In Ref. \cite{Cohen:2004qt} next-to-leading order were introduced and the predictions made 
were confirmed by experiment. 

The method has been extended to pion photoproduction in Ref. \cite{Cohen:2004bk}. The
corrections to order $1/N_c$ and $1/N^2_c$ give a remarkable agreement with the 
experiment. 

The approach of Ref. \cite{Cohen:2004bk} was later modified to provide a model-independent 
expansion for the electromagnetic multipole amplitudes of the pion electroproduction process
$e^- N \rightarrow e^- \pi N$ \cite{Lebed:2009aq}. The results seem to be more ambiguous.

\section{Different large $N_c$ limits}\label{differentlimits}

Soon after 't Hooft's generalization of  QCD  from $N_c$ = 3 to arbitrarily large  $N_c$  \cite{HOOFT},
it was pointed out by Corrigan and Ramond \cite{Corrigan:1979xf} that there is an ambiguity in the
generalization  of the quark content of SU(N$_c$) to  $N_c >  3$.
The argument was that the quarks can appear in other representations than the fundamental representation of SU(N$_c$).
Therefore one can construct distinct theories that agree at $N_c$ = 3 
but differ at $N_c \rightarrow \infty $. Then each distinct extrapolation
leads to a distinct $1/N_c$ expansion for the observables under study. 

So far several inequivalent large $N_c$ generalizations have been proposed. Some of them have been  discussed in Ref. \cite{Bolognesi:2006ws}.
If the quarks are in the fundamental representation one can construct a totally antisymmetric color state as defined by Eq. (\ref{CA}),
which must be combined with a symmetric orbital-spin-flavor part, as already mentioned.

As an alternative, Corrigan and Ramond \cite{Corrigan:1979xf} proposed 
a description of baryons  as formed of quarks transforming under the fundamental representation and
"larks" (antiquarks in SU(3)) transforming under the antisymmetric $N(N-1)/2$ representation, Table \ref{twoindexirreps}.

Actually there are three possible two-index representations for SU(N). They are called  tensors of rank $N^{(n,m)}$
\cite{Stancu:1991rc}, where $n+m$ = 2 in this case. They are exhibited in Table \ref{twoindexirreps}.
The superscripts refer to symmetry and subscripts to antisymmetry. 
The number of linearly independent components of each tensor gives the dimension of the corresponding irrep, denoted by $n_q$. 
Each tensor has a number of constraints given by its properties. 
This number has to be subtracted from $N^2$. Then the number of independent components is generally smaller than  $N^{n+m}$.
For example, the symmetric two-index irrep has the property
\begin{equation}
T^{ij} = T^{ji},
\end{equation}
which gives $C^2_N = N(N - 1)/2$ constraints. Then the dimension of the symmetric two-index irrep is $N^2 - N(N - 1)/2$ = $ N(N + 1)/2$.
The dimension of the antisymmetric irrep is naturally  $N(N-1)/2$. The sum of the two must be equal to $N^2$ to be consistent   
with the direct product of two fundamental representations of SU(N) which can be decomposed as
\begin{eqnarray}
\label{}
{\mbox{\begin{Young}
 \cr
\end{Young}}}\ \ 
\times
{\mbox{\begin{Young}
 \cr
\end{Young}}}\ \ 
& = & 
{\mbox{\begin{Young}
&  \cr
\end{Young}}}\ \ 
+
\raisebox{-9.0pt}{\mbox{\begin{Young}
 \cr
  \cr
\end{Young}}}~.
\end{eqnarray}
The two-index irreducible tensor $T^i_j$ must satisfy the trace condition \cite{Stancu:1991rc}
\begin{equation}
T^1_1 +  T^2_2 + ...T^N_N = 0.
\end{equation}
From here it follows that the dimension of the  representation $T^i_j$ is $N^2-1$.
Thus $T^i_j$  corresponds to the adjoint representation.
This two-index tensor can be constructed from two one-index tensors, $T^i$ and $T_j$ as
\begin{equation}
 T^i_j = T^i T_j. 
\end{equation}
We recall that the contravariant tensor $T^i$ can represent quarks
and the covariant tensor $T_j$ can describe antiquarks so that one can make the identification 
\begin{equation}
T^i = q^i, \, \, T_j = \bar q_j.
\end{equation}
 \begin{center}
\begin{table}[htb]
\caption{Two-index irreducible representations of SU(N) defined in terms of irreducible tensors, where $n_q$ (last column) is the number of linearly
independent components of each tensor.}
\label{twoindexirreps}
\renewcommand{\arraystretch}{2}
\vspace{.3cm}
 \begin{tabular}{cccc}
   \hline\hline
\vspace{0.1cm}
 Tensor   & Rank & \hspace{0.2cm} & $n_q$ \\  \hline 
$T^{ij}$   &  $(2,0)$ &  \hspace{0.2cm} &   $\frac{N(N+1)}{2} $ \\
$T_{ij}$ &  $(0,2)$ &  \hspace{0.2cm} & $\frac{N(N-1)}{2}$ \\
$T^{i}_{j}$ &  $(1,1)$ &  \hspace{0.2cm} & $N^2-1$ \\
   \hline\hline
 \end{tabular}
 \end{table}
 \end{center}

The idea of Corrigan and Ramond has been extended by Bolognesi to the two-index  
symmetric and antisymmetric representations in an effective Lagrangian approach  \cite{Bolognesi:2006ws}. 

Independently Armoni, Shifman and Veneziano \cite{Armoni:2003gp,Armoni:2003fb}
have used the two index antisymmetric representation to define a new $1/N_c$ expansion 
at a fixed number of $N_f$ flavors. For  $N_f$ = 1, in the large $N_c$ limit, their approach is equivalent
to $\mathcal{N}$ = 1 supersymmetric Yang-Mills theory. In particular, Ref.  \cite{Armoni:2003gp} predicts exactly degenerate
parity doublets. It would be interesting to find out if there are useful implications for baryons
within this context.  In addition, by using the equivalence to $\mathcal{N}$ = 1 supersymmetric Yang-Mills theory,
Armoni and Patella \cite{Armoni:2009zq} have shown that mesons and baryons become asymptotically 
superpartners, which may explain the coincidence of their Regge slopes.

Either by using 
a mean field approach \cite{WITTEN,Bolognesi:2006ws} or a diagrammatic method
\cite{Cherman:2006iy,Cherman:2009fh,Cohen:2009wm}
or within the framework of a semirelativistic constituent quark models  \cite{Buisseret:2010na}
it has been proven that in the antisymmetric case the mass of such baryons scales as $N^2_c$. 
In the symmetric case the mass scales as $N^2_c$ also \cite{Buisseret:2011aa}.
 Then it follows that both the two-index symmetric and antisymmetric 
representations lead to the same limit  for the baryon masses at $N_c \rightarrow \infty$ and that the fundamental
representation used by  't Hooft \cite{HOOFT} and the  two-index antisymmetric representation of quarks lead to the
same results at $N_c$ = 3.

The results of Ref.  
\cite{Buisseret:2010na,Buisseret:2011aa}
imply that in the Corrigan-Ramond limit, where $n_q$ = 3, the baryon mass is of order $\mathcal{O}(1)$, like for mesons, 
in agreement with \cite{Cherman:2009fh}. We should stress that in the 't Hooft limit a large number of QCD properties have a simple understanding.
However, there are cases where the 't Hooft limit is not sufficient \cite{Harada:2003em}.  In exchange, the Corrigan-Ramond limit 
has a richer structure and is convenient to study QCD at higher matter density  \cite{Frandsen:2005mb}.

One should note that in Ref. \cite{Buisseret:2010na} only the spin independent part of
the Hamiltonian was considered.
The spin contribution was analyzed in a later work \cite{Buisseret:2011aa} for ground state of light baryons
in three inequivalent large $N_c$ limits and  
it was proven that it scales as $S(S+1)/n_q$ in all cases.  Then,  Table \ref{twoindexirreps}
implies that the spin contribution to the perturbative expansion resulting from one gluon exchange in the two-index representation, 
is of order $\mathcal{O}(1/N^2_c)$, 
while in the 't Hooft's limit, the subleading order is $\mathcal{O}(1/N_c)$ in agreement with results based on the 
spin-flavor symmetry \cite{Dashen:1993jt,Dashen:1994qi}. Contrary, in the Goldstone boson exchange model \cite{Glozman:1995fu,Glozman:1997ag}
the contribution of the spin-flavor hyperfine splitting requires more attention in a perturbative expansion, 
when the coupling constants of the exchanged bosons are also considered in a large $N_c$ limit.

The large $N_c$ antisymmetric limit also implies the emergence of an SU(2N$_f$) spin-flavor symmetry and 
predicts equally successful baryon mass relations as those derived in the standard $1/N_c$ expansion \cite{Dashen:1993jt,Dashen:1994qi},
but with different $1/N_c$ suppression factors \cite{Cherman:2012eg}. Accordingly, the authors of Ref.  \cite{Cherman:2012eg}
conclude that the large $N_c$ baryons in the fundamental and antisymmetric two-index representations are about equally close to
the $N_c$ = 3 world, at least for the ground state baryon masses.
 
Studies based on the flavor adjoint representation  have been performed in Refs. \cite{Bolognesi:2006ws,Bolognesi:2007ut,Auzzi:2008hu}.

\section{Exotics}\label{exotics}

As well known, the existing quark models, inspired by QCD, can describe  the properties
of baryons as three quark systems $qqq$ and of mesons as quark-antiquark 
$q \bar q$ pairs.
These models predict the existence of new resonances, called exotics, which are formed
of more than three quarks or antiquarks ( $q^m {\bar q}^n, m + n > 3$). For a short review see,
for example, Ref. \cite{Stancu:2000zk} and references therein.
    
It was naturally to inquire about the existence of exotics in large $N_c$. 
To our knowledge the problem has been first raised by Cohen and Lebed \cite{Cohen:2003nk}
in conjunction to the presently controversial pentaquark $\theta^+$, a $q^4 \bar s $ system ($q = u$, $d$), 
of total angular momentum $J$ = 1/2, isospin $I$ = 0, strangeness $\mathcal{S} = +1$ and a mass of about 1.5 GeV, 
having a narrow width less than 15 MeV \cite{Diakonov:1997mm}, the observation of 
which was just announced by the LEPS Collaboration in 2003  \cite{Nakano:2003qx}. This was followed  by
a number of observations with either positive or negative results, reviewed, for example, in Ref. \cite{Stancu:2004ap}. 
Cohen and Lebed  argued that large $N_c$ analysis by itself cannot predict the 
mass of $\theta^+$, but it can predict the existence of degenerate partners with $\mathcal{S} = +1$, the quantum numbers of which can be 
related to poles in the $KN$ scattering amplitude. They used the  SU(3) extension \cite{MattisMukerjee} of the formalism  
presented in Sec. \ref{MN} where the baryons are described as resonances in the meson-nucleon
scattering. If, for example, one imposes the theoretical assumption that  $\theta^+$ is a state with $J$ = 1/2 the 
degenerate partners should have $I$ = 1, $J$ = 1/2,~3/2 and $I$ = 2, $J$ = 3/2,~5/2.
One does not expect the widths of these partners to be similar to that of $\theta^+$.
However Cohen and Lebed note that large $N_c$ neither implies nor precludes the 
existence of exotics. 

Shortly afterwards Jenkins and Manohar \cite{Jenkins:2004vb} have also stated that 
large $N_c$ spin-flavor symmetry does not predict that exotic baryons exist.
They have introduced the notion of exoticness $E$ \cite{Jenkins:2004tm}, as the minimal value 
for which the flavor baryon representation can be constructed from $qqq{(q \bar q)}^E$ in SU(6). They 
derived the quantum numbers of exotics both in the quark and the Skyrme model, the results being identical   
at $N_c \rightarrow \infty$, like for ordinary baryons. They proposed
an $1/N_c$ mass expansion for exotic baryons and transition operators between baryons with different values of $E$.

In the quark representation described in Sec. \ref{barexp}
the degenerate partners  predicted by Cohen and Lebed belong to the SU(3) multiplets
$\bf 27$ for $I$ = 1   and to $\bf 35$ for  $I$ = 2, respectively, while $\theta^+$ was considered as a member of an
antidecuplet ${\overline {\bf 10}}$ with spin 1/2 \cite{Diakonov:1997mm}. These three SU(3) representations have $E$ = 1.

The work of Jenkins and Manohar has been extended to one more important irreducible representation of the 
contracted spin-flavor symmetry by Pirjol and Schat \cite{Pirjol:2006ne},
who constructed a complete set of positive parity pentaquarks
with one unit of orbital angular momentum, which in the large $N_c$ limit fall into two towers with 
$K$ = 1/2 and $K$ = 3/2 of the contracted SU(4) symmetry.

After the enthusiastic wave of interest for the pentaquark $\theta^+$ and its
charmed partner  $\theta_c^0$ (a $u u d d \bar c$ system belonging to an SU(3) antisextet,
a submultiplet of the ${\overline {\bf 60}}$ irreducible representation of SU(4) \cite{Stancu:2004ap,Wu:2004wg})
during the period 2003-2005,  there followed  an overwhelming evidence that they do not exist (see the 
report by C. G. Wohl on exotic baryons in Particle Data Group \cite{PDG}).
However the common feature of most of the experimental results was that they were non-dedicated experiments until 2004.
Later on, dedicated high statistics experiments were performed (for a  review see e. g.  \cite{Liu:2014yva}). 
Recently a narrow peak structure at about 1.54 GeV in the missing mass of $K_S$ in the reaction
$\gamma + p \rightarrow p K_S K_L$ has been observed \cite{Amaryan:2011qc}. In this experiment one tries to
exploit the quantum mechanical interference between the channels $\gamma p \rightarrow  \theta^+ {\bar K}^0 
\rightarrow  p K_S K_L$
and  $\gamma p \rightarrow p \phi \rightarrow p K_S K_L$ where the latter can enhance the small amplitude 
of the $\theta^+$ channel.

More recently the pentaquark $\theta^+$ was reconsidered in a new theory 
of collective excitation  as due to a Gamow-Teller transition - like in nuclear physics - but as a transition  of the $s$
quark from the highest filled level to excited $u,d$ quark levels in a mean field \cite{Diakonov:2013qta}. In this
way  $\theta^+$ was recovered with the same mass as it was first predicted \cite{Diakonov:1997mm}.

As a consequence of Coleman's conclusion in his Erice lectures \cite{COLEMAN}
that "in the large $N$ limit, quadrilinears make meson pairs and nothing else" 
recently Weinberg  \cite{Weinberg:2013cfa} argued that exotic mesons consisting of two quarks and two antiquarks are not ruled out
in large $N_c$ QCD. He suggested that the real question is the decay rate of a tetraquark.
Weinberg's suggestion has been subsequently supported and 
analyzed by several authors \cite{Knecht:2013yqa,Lebed:2013aka,Cohen:2014via,Cohen:2014tga}.

\section{Conclusions}

The  $1/N_c$ expansion of QCD can provide  a qualitative and, to a large extent, a quantitative understanding of a large number 
of hadronic phenomena. It has been proven to be an appropriate tool for studying hadron spectroscopy in a model independent way. 
A great advantage is that it helps
to organize and relate the observables at each order in $1/N_c$.

Previous reviews have  shown that
the ground state baryons satisfy the hierarchy predicted by this expansion, whenever necessary combined with a perturbative treatment of SU(3)-flavor 
breaking. It had also successfully predicted the masses of heavy-quark baryons and can help in the discovery of the remaining 
bottom baryons. The description of axial vector couplings, magnetic moments, charge radii and quadrupole moments was also successful.

Here we have mostly been  concerned with the baryon excited states. The physics of excited states gets sorted out 
hierarchically in powers of  $1/N_c$ as well \cite{deUrreta:2013koa}.
The presently known approaches are based on an extension
of the spin-flavor symmetry to SU(2N$_f$) $\times$ O(3) symmetry. They seem to successfully explain most 
of the measured baryon masses. The  quantitative calculations allow 
to group resonances in octets, decuplets and singlets formed of excited states. Many of these are predictions,  
which may be used in the experimental discovery of unknown baryons, in particular of excited hyperons. 

The $1/N_c$ mass operator has been compared to quark models and the comparison gives strong support to quark models and 
a better understanding of the coefficients of the mass formula which encode the quark dynamics. 
The leading order term, proportional to $N_c$, can be understood as representing the contribution of the kinetic and of the 
confinement parts of a quark model Hamiltonian. This term then naturally increases with the  excitation energy, or else, with the 
band number. The deviation from spin-flavor symmetry is given by corrections in powers of  $1/N_c$ and a dominant
part is the spin term $S \cdot S$, containing a spin-spin interaction, compatible with the one used in one-gluon exchange  models. 
Especially for N$_f$ = 3 flavors, a novelty is  
that the contribution of the pure flavor term $T \cdot T$ is as important in decuplets and flavor singlets,   as it is the 
spin term in octets. The rewriting of $T \cdot T$  in terms of the spin and spin-flavor terms, by using the Casimir 
operator of SU(2N$_f$) may bring a more obvious support of models containing a Goldstone boson exchange interaction. A quantitative
analysis is highly desirable.

The present studies of strong decay widths and photoproduction amplitudes, made in the symmetric core + excited quark
approach, require sub-leading order corrections in order to fit 
experimental data. In particular, the transition operators include  terms corresponding to the pseudoscalar meson emission,
customarily used in quark model description of decays,
but also other higher order terms, unknown in quark model studies, the meaning of which could perhaps give a better insight into
transitions amplitude described by quark models. Similar studies for mixed symmetric spin-flavor states based on the totally 
antisymmetric wave function approach described in Sec. \ref{specmixed} are desirable.
They could help to extend the analysis of decays to highly excited resonances belonging to bands with $N > 1$. 

So far all studies have been devoted to a fixed SU(6) $\times$ O(3) multiplet of a given band   $N$.
On the other hand 
it has been shown that the excitation band number $N$ could be used to obtain Regge type trajectories for
the spin-independent part of the mass formula both in large $N_c$ and quark model calculations. A global fit of resonances
belonging to several multiplets of the same band $N$ would be interesting to perform. It could settle the issue 
whether symmetric and mixed symmetric multiplets can lead to the same trajectory or should lead to distinct
trajectories as shown in Sec. \ref{versus}.

The $1/N_c$ expansion method is receiving support from lattice QCD calculations showing that $N_c$ = 3 is not too
far from a larger $N_c$. 
A recent study concentrates on subleading corrections of hyperfine type \cite{Cordon:2014sda}. More accurate results are being desired.
We hope that the interplay between large $N_c$ QCD and lattice calculations will further enlighten the understanding of excited 
baryons.


\appendix

\section{The generalized Wigner-Eckart theorem and isoscalar factors of SU(6)}\label{A}

Here we follow the derivation of the isoscalar factors of SU(6) as given in Ref. \cite{Matagne:2008kb}
and completed in Ref. \cite{Matagne:2011fr}. They provide the diagonal and off-diagonal matrix elements
of the SU(6) generators needed in calculating the spectra and transition amplitudes of strong and electromagnetic
decays.

The SU(6) generators are components of an irreducible SU(6) tensor operator 
which span the invariant subspace of the adjoint representation denoted here
by the partition $[21^4]$, or  otherwise by its dimensional notation $\bf 35$.
Like for any other irreducible representation its matrix elements can be expressed 
in terms of a generalized Wigner-Eckart theorem, which factorizes each matrix element 
into products of Clebsch-Gordan coefficients and a reduced matrix element.
To write the  Wigner-Eckart theorem in its general form
we redefine the generators forming the  algebra (\ref{ALGEBRA}) as 
\begin{equation} \label{normes}
E^i =\frac{ S^i}{\sqrt{N_f}};~~~ E^a = \frac{T^a}{\sqrt{2}}; ~~~E^{ia} = \sqrt{2}
G^{ia}.
\end{equation}
where we have to take ${N_f}$ = 3 for SU(6) and  ${N_f}$ = 2 for SU(4) (Appendix \ref{SU4}).
Note that the generic name for every generator will remain $E^{ia}$ \cite{HP}.

First we discuss the SU(6) case.
An irrep of SU(6) is denoted by the partition $[f]$ and the SU(3) irreps
are labelled by $(\lambda\mu)$ following  Elliott   \cite{Elliott:1958zj}, equivalent 
to $(p,q)$ in particle physics \cite{Lichtenberg}. Then one can write the matrix element of every SU(6) generator $E^{ia}$ as 
\begin{widetext}
\begin{eqnarray}\label{GEN}
\lefteqn{\langle [f](\lambda' \mu') Y' I' I'_3 S' S'_3 | E^{ia} |
[f](\lambda \mu) Y I I_3 S S_3 \rangle =}\nonumber \\ & & \sqrt{C^{[f]}(\mathrm{SU(6)})} 
   \left(\begin{array}{cc|c}
	    S   &    S^i   & S'   \\
	    S_3  &   S^i_3   & S'_3
  \end{array}\right)
     \left(\begin{array}{cc|c}
	I   &   I^a   & I'   \\
	I_3 &   I^a_{3}   & I'_3
   \end{array}\right)  
      \sum_{\rho = 1,2}
 \left(\begin{array}{cc||c}
	(\lambda \mu)    &  (\lambda^a\mu^a)   &   (\lambda' \mu')\\
	Y I   &  Y^a I^a  &  Y' I'
      \end{array}\right)_{\rho}
\left(\begin{array}{cc||c}
	[f]    &  [21^4]   & [f]   \\
	(\lambda \mu) S  &  (\lambda^a\mu^a) S^i  &  (\lambda' \mu') S'
      \end{array}\right)_{\rho} , 
   \end{eqnarray}
\end{widetext}
where $C^{[f]}(\mathrm{SU(6)})$ is the SU(6)
Casimir operator eigenvalue associated to 
the irreducible representation $[f]$, followed by the familiar Clebsch-Gordan
coefficients of SU(2)-spin and SU(2)-isospin. The sum over $\rho$ 
contains products of isoscalar factors of SU(3) and SU(6) respectively.
The label $\rho$ is necessary whenever one has to distinguish  between 
irreps $[f']=[f]$ with multiplicities $m_{[f]}$ larger than one in the Clebsch-Gordan series \cite{Matagne:2011fr}
\begin{equation}\label{CG} 
[f] \times [21^4] = \sum_{[f']} m_{[f']} [f']. 
\end{equation}
The two values for $\rho$ both in SU(6) and SU(3) reflects the 
multiplicity problem already appearing in the direct product of SU(3) irreducible representations 
\begin{widetext}
\begin{eqnarray}\label{PROD}
\lefteqn{(\lambda \mu) \times (11)  =   (\lambda+1, \mu+1)+ (\lambda+2, \mu-1)} \nonumber \\ & + &
(\lambda \mu)_1 + (\lambda \mu)_2
+ \, (\lambda-1, \mu+2) + (\lambda-2, \mu+1)
+ (\lambda+1, \mu-2)+ (\lambda-1, \mu-1),
\end{eqnarray}
\end{widetext}
where (11) labels the SU(3) adjoint representation.
One can see  the representation $(\lambda \mu)$ which is one of the factors on the left hand side, appears twice
on the right hand side. To distinguish between the two $(\lambda \mu)$'s one introduces the index 
$\rho$, which then takes two values, both for the SU(3) and SU(6) isoscalar factors. More details can be found in
Ref. \cite{Matagne:2011fr}.

\begin{table}
\caption{Values of $\lambda$ and $\mu$ as a function of $N_c$ for all sectors of physical interest.}
\label{lambdamu}
\renewcommand{\arraystretch}{2.}
\begin{tabular}{cccccc}
\hline \hline
             &  \hspace{0.5cm}   $\lambda$   \hspace{0.5cm}  &     \hspace{0.5cm}     $\mu$  \hspace{0.5cm} \\
\hline
$^28_J$ &  $ 1 $   &$\frac{N_c - 1}{2}$ \\
$^48_J$ &  $ 1 $   &$\frac{N_c - 1}{2}$ \\
$^210_J$&  $ 3 $   &$\frac{N_c - 3}{2}$ \\
$^21_J$ &  $ 0 $   &$\frac{N_c - 3}{2}$\\
\hline \hline
\end{tabular}
\end{table}

In Eq. (\ref{GEN}) the Casimir operator eigenvalue for the 
the symmetric representation with $[f] = [N_c]$ is
\begin{equation}
C^{[N_c]}(\mathrm{SU(6)}) =\frac{5N_c(N_c+6)}{12},
\end{equation}
and for the mixed symmetric representation with  $[f] = [N_c-1,1]$  is
\begin{equation}
C^{[N_c-1,1]}(\mathrm{SU(6)}) = \frac{N_c(5 N_c+18)}{12}.
\end{equation}

The general analytic expressions of isoscalar factors of SU(3) needed for this analysis can be taken  
from the nuclear physics studies of Hecht \cite{Hecht:1965} where one has to replace $\lambda$ and $\mu$ 
by their definition in terms of $N_c$. Their properties are summarized in Appendix C.

For the reader's convenience here we reproduce 
our results for the isoscalar factors of SU(6) entering the generalized Wigner-Eckart theorem,
Eq. (\ref{GEN}).
The tables shown below give the analytic expressions of the isoscalar 
factors in terms of $N_c$ and spin. They can be used either in the  calculation of  
masses of baryons  and decay observables from the real world ($N_c$ = 3), of electromagnetic moment relations \cite{Lebed:1995}
or in the analysis of the compatibility between 
the $1/N_c$ expansion and the pion-nucleon scattering results, where one has to 
include states with $N_c \ge 3$, see Sec. \ref{MN}.

We exhibit separately our results for the symmetric $[N_c]$ representation in Table \ref{SYMM} and for the mixed symmetric
representation  $[N_c-1,1]$ in Tables \ref{octet_spin_one_half}, \ref{octet_spin_three_halfs}, \ref{decuplet_spin_one_half}
and \ref{singlet_spin_one_half}.
In each  case 
one can check that the isoscalar factors satisfy the following orthogonality relation  
\begin{widetext}
\begin{equation}
\sum_{\rho,(\lambda \mu)S,(\lambda^a\mu^a)S^i}
\left(\begin{array}{cc||c}
	[f]    &  [21^4]   & [f_1]   \\
	(\lambda \mu) S  &  (\lambda^a\mu^a) S^i  &  (\lambda_1 \mu_1) S_1
      \end{array}\right)_{\rho} 
\left(\begin{array}{cc||c}
	[f]    &  [21^4]   & [f_2]   \\
	(\lambda \mu) S  &  (\lambda^a\mu^a) S^i  &  (\lambda_2 \mu_2) S_2
      \end{array}\right)_{\rho}  
  = \delta_{f_1 f_2} \delta_{\lambda_1 \lambda_2} \delta_{\mu_1 \mu_2} \delta_{S_1 S_2}.
\end{equation}
\end{widetext}



We note that the analytic expressions obtained for  the isoscalar factors of the  symmetric representation $[N_c]$ 
were obtained in Ref. \cite{Matagne:2006xx}.

\begin{sidewaystable*}[pt]
\vspace{8cm}
\caption{Isoscalar factors of SU(6) generators defined by Eq. (\ref{GEN}, related to the product $[N_c] \times [21^4] \rightarrow [N_c]$ ).}
\renewcommand{\arraystretch}{2.5}
\begin{tabular}{l|c|c|l}
\hline
\hline

$(\lambda_1\mu_1)S_1$ \hspace{0.5cm} & \hspace{0.5cm}$(\lambda_2\mu_2)S_2$ \hspace{0.5cm} &\hspace{0.5cm}$\rho$\hspace{0.5cm} & \hspace{0.5cm}$\left(\begin{array}{cc||c}                                         [N_c]  &  [21^4]  &  [N_c] \\
                           (\lambda_1\mu_1)S_1 & (\lambda_2\mu_2)S_2 & (\lambda\mu)S
                                      \end{array}\right)_\rho$  \\
\vspace{-0.5cm} &  &   & \\
\hline
$(\lambda + 2,\mu - 1)S+1$\hspace{0.5cm} & \hspace{0.0cm}$(11)1$ & $/$ &\hspace{0.5cm}$-\sqrt{\frac{3}{2}}\sqrt{\frac{2S+3}{2S+1}}\sqrt{\frac{(N_c-2S)(N_c+2S+6)}{5N_c(N_c+6)}}$ \\
$(\lambda\mu)S$  & \hspace{0.0cm}$(11)1$ & 1 & \hspace{0.5cm}$4(N_c+3)\sqrt{\frac{2S(S+1)}{5N_c(N_c+6)[N_c(N_c+6)+12S(S+1)]}}$ \\
$(\lambda\mu)S$  &\hspace{0.0cm}$(11)1$ & 2 & \hspace{0.5cm}$-\sqrt{\frac{3}{2}}\sqrt{\frac{(N_c-2S)(N_c+4-2S)(N_c+2+2S)(N_c+6+2S)}{5N_c(N_c+6)[N_c(N_c+6)+12S(S+1)]}}$ \\
$(\lambda - 2,\mu + 1)S-1$  & \hspace{0.0cm}$(11)1$  & $/$ & \hspace{0.5cm}$-\sqrt{\frac{3}{2}}\sqrt{\frac{2S-1}{2S+1}}\sqrt{\frac{(N_c+4-2S)(N_c+2+2S)}{5N_c(N_c+6)}}$ \\
$(\lambda\mu)S$  & \hspace{0.0cm}$(00)1$ & $/$  & \hspace{0.5cm}$\sqrt{\frac{4S(S+1)}{5N_c(N_c+6)}}$ \\
$(\lambda\mu)S$  & \hspace{0.0cm}$(11)0$ & $1$  & \hspace{0.5cm}$\sqrt{\frac{N_c(N_c+6)+12S(S+1)}{10N_c(N_c+6)}}$ \\
$(\lambda\mu)S$  & \hspace{0.0cm}$(11)0$ & $2$  & \hspace{0.5cm}0 \\
\hline
\hline
\end{tabular}
\label{SYMM}
\end{sidewaystable*}



\begin{sidewaystable*}[pt]
\vspace{8cm}
\caption{Isoscalar factors of the SU(6) generators
 Eqs. (\ref{normes}) and (\ref{GEN}),
corresponding to 
the $^28$ multiplet of $N_c = 3$.}
{\scriptsize
 \renewcommand{\arraystretch}{2.5}
\begin{tabular}{l|c|c|l}
\hline
\hline
$(\lambda_1\mu_1)S_1$ \hspace{0.5cm} & \hspace{0.5cm}$(\lambda_2\mu_2)S_2$ \hspace{0.5cm} & \hspace{0.5cm}$\rho$\hspace{0.5cm} & \hspace{0.5cm}$\left(\begin{array}{cc||c}                                         [N_c-1,1]  &  [21^4]  &  [N_c-1,1] \\
                           (\lambda_1\mu_1)S_1 & (\lambda_2\mu_2)S_2 & (\lambda\mu)S
                                      \end{array}\right)_\rho$  \\
\vspace{-0.5cm} &  &   & \\
\hline
$(\lambda\mu)S+1$\hspace{0.5cm} & \hspace{0.cm}$(11)1$ & $1$ &\hspace{0.5cm}$-\frac{3\sqrt{2S(2S+3)(N_c+2S+2)}}{\sqrt{(S+1)(2S+1)\left[N_c(N_c+6)+12S(S+1)\right](5N_c+18)}}$\\
$(\lambda\mu)S+1$\hspace{0.5cm} & \hspace{0.cm}$(11)1$ & $2$ &\hspace{0.5cm}$\frac{N_c}{S+1}\sqrt{\frac{3(2S+3)(N_c-2S+4)(N_c+2S+6)}{2(2S+1)(N_c-2S)\left[N_c(N_c+6)+12S(S+1)\right](5N_c+18)}}$\\
$(\lambda\mu)S$\hspace{0.5cm} & \hspace{0.cm}$(11)1$ & $1$ &\hspace{0.5cm}$\left\{12S(S+1)+N_c[4S(S+1)-3]\right\}\sqrt{\frac{2}{S(S+1)\left[N_c(N_c+6)+12S(S+1)\right]N_c(5N_c+18)}}$\\
$(\lambda\mu)S$\hspace{0.5cm} & \hspace{0.cm}$(11)1$ & $2$ &\hspace{0.5cm}$\frac{4S^2(S+1)^2-2N_cS(S+1)-(S^2+S-1)N_c^2}{2S(S+1)}\sqrt{\frac{6(N_c-2S+4)(N_c+2S+6)}{(N_c-2S)(N_c+2S+2)\left[N_c(N_c+6)+12S(S+1)\right]N_c(5N_c+18)}}$\\
$(\lambda\mu)S-1$\hspace{0.5cm} & \hspace{0.cm}$(11)1$ & $1$ &\hspace{0.5cm}$-3\sqrt{\frac{2(S+1)(2S-1)(N_c-2S)}{S(2S+1)(N_c(N_c+6)+12S(S+1))(5N_c+18)}}$ \\
$(\lambda\mu)S-1$\hspace{0.5cm} & \hspace{0.cm}$(11)1$ & $2$ &\hspace{0.5cm}$\frac{N_c}{S}\sqrt{\frac{3(2S-1)(N_c-2S+4)(N_c+2S+6)}{2(2S+1)(N_c+2S+2)(N_c(N_c+6)+12S(S+1))(5N_c+18)}}$ \\
$(\lambda+2,\mu-1)S+1$\hspace{0.5cm} & \hspace{0.cm}$(11)1$ & $/$ &\hspace{0.5cm}$-\frac{1}{S+1}\sqrt{\frac{3S(S+2)(2S+3)(N_c-2S-2)(N_c+2S+2)(N_c+2S+6)}{2(2S+1)(N_c+2S+4)N_c(5N_c+18)}}$\\
$(\lambda+2,\mu-1)S$\hspace{0.5cm} & \hspace{0.cm}$(11)1$ & $/$ &\hspace{0.5cm}$\frac{1}{S+1}\sqrt{\frac{3(2S+3)(N_c+2S+2)(N_c+2S+6)}{2(2S+1)(N_c+2S+4)(5N_c+18)}}$ \\
$(\lambda+1,\mu-2)S+1$\hspace{0.5cm} & \hspace{0.cm}$(11)1$ & $/$ &\hspace{0.5cm}$-2\sqrt{\frac{3S(2S+3)(N_c-2S-2)}{(S+1)(2S+1)(N_c-2S)(N_c+2S+4)(5N_c+18)}}$ \\
$(\lambda+1,\mu-2)S$\hspace{0.5cm} & \hspace{0.cm}$(11)1$ & $/$ &\hspace{0.5cm}$-2\sqrt{\frac{3(N_c-2S-2)}{(S+1)(2S+1)(N_c-2S)(N_c+2S+4)(5N_c+18)}}$ \\
$(\lambda-1,\mu-1)S$\hspace{0.5cm} & \hspace{0.cm}$(11)1$ & $/$ &\hspace{0.5cm}$\sqrt{\frac{12(N_c+2S)}{S(2S+1)(N_c-2S+2)(N_c+2S+2)(5N_c+18)}}$\\
$(\lambda-1,\mu-1)S-1$\hspace{0.5cm} & \hspace{0.cm}$(11)1$ & $/$ &\hspace{0.5cm}$-2\sqrt{\frac{3(S+1)(N_c+2S)(2S-1)}{S(2S+1)(N_c-2S+2)(N_c+2S+2)(5N_c+18)}}$\\
$(\lambda-2,\mu+1)S$\hspace{0.5cm} & \hspace{0.cm}$(11)1$ & $/$ &\hspace{0.5cm}$\frac{1}{S}\sqrt{\frac{3(2S-1)(N_c-2S)(N_c-2S+4)}{2(2S+1)(N_c-2S+2)(5N_c+18)}}$\\
$(\lambda-2,\mu+1)S-1$\hspace{0.5cm} & \hspace{0.cm}$(11)1$ & $/$ &\hspace{0.5cm}$-\frac{1}{S}\sqrt{\frac{3(S-1)(S+1)(2S-1)(N_c-2S)(N_c+2S)(N_c-2S+4)}{2(2S+1)(N_c-2S+2)N_c(5N_c+18)}}$ \\
$(\lambda\mu)S$\hspace{0.5cm} & \hspace{0.cm}$(11)0$ & $1$ &\hspace{0.5cm}$\sqrt{\frac{N_c(N_c+6)+12S(S+1)}{2N_c(5N_c+18)}}$\\
$(\lambda\mu)S$\hspace{0.5cm} & \hspace{0.cm}$(11)0$ & $2$ &\hspace{0.5cm} 0\\
$(\lambda\mu)S$\hspace{0.5cm} & \hspace{0.cm}$(00)1$ & $/$ &\hspace{0.5cm} $\sqrt{\frac{4S(S+1)}{N_c(5N_c+18)}}$\\ 
\hline
\hline
\end{tabular}}
\label{octet_spin_one_half} 
\end{sidewaystable*}
\begin{sidewaystable*}[pt]
\vspace{8cm}
\caption{Isoscalar factors of the SU(6) generators,
corresponding to 
the $^48$ multiplet of $N_c = 3$.}
{\scriptsize
 \renewcommand{\arraystretch}{2.5}
\begin{tabular}{l|c|c|l}
\hline
\hline
$(\lambda_1\mu_1)S_1$ \hspace{0.5cm} & \hspace{0.5cm}$(\lambda_2\mu_2)S_2$ \hspace{0.5cm} & \hspace{0.5cm}$\rho$\hspace{0.5cm} & \hspace{0.5cm}$\left(\begin{array}{cc||c}                                         [N_c-1,1]  &  [21^4]  &  [N_c-1,1] \\
                           (\lambda_1\mu_1)S_1 & (\lambda_2\mu_2)S_2 & (\lambda-2,\mu+1)S
                                      \end{array}\right)_\rho$  \\
\vspace{-0.5cm} &  &   & \\
\hline
$(\lambda-2,\mu+1)S$\hspace{0.5cm} & \hspace{0.cm}$(11)1$ & $1$ &\hspace{0.5cm}$ \left[N_c(4S-3)+6S\right]\sqrt{\frac{2(S+1)}{S\left[N_c(N_c+6)+12(S-1)S\right]N_c(5N_c+18)}}$\\
$(\lambda-2,\mu+1)S$\hspace{0.5cm} & \hspace{0.cm}$(11)1$ & $2$ &\hspace{0.5cm}$-\frac{N_c-2S}{S}\sqrt{\frac{3(S-1)(S+1)(N_c-2S+6)(N_c+2S)(N_c+2S+4)}{2(N_c-2S+2)\left[N_c(N_c+6)+12(S-1)S\right]N_c(5N_c+18)}}$\\
$(\lambda\mu)S+1$\hspace{0.5cm} & \hspace{0.cm}$(11)1$ & $/$ &\hspace{0.5cm}$-\sqrt{\frac{3}{2}}\sqrt{\frac{2S+3}{2S+1}}\sqrt{\frac{(N_c-2S)(N_c+2S+4)}{N_c(5N_c+18)}}$\\
$(\lambda\mu)S$\hspace{0.5cm} & \hspace{0.cm}$(11)1$ & $/$ &\hspace{0.5cm}$-\frac{1}{S}\sqrt{\frac{3}{2}}\sqrt{\frac{(N_c-2S)(N_c+2S+4)}{(N_c+2S+2)(5N_c+18)}}$\\
$(\lambda\mu)S-1$\hspace{0.5cm} & \hspace{0.cm}$(11)1$ & $/$ &\hspace{0.5cm}$\frac{N_c+4S^2}{S}\sqrt{\frac{3(N_c+2S+4)}{2(2S-1)(2S+1)(N_c+2S+2)N_c(5N_c+18)}}$\\
$(\lambda-2,\mu+1)S-1$\hspace{0.5cm} & \hspace{0.cm}$(11)1$ & $1$ &\hspace{0.5cm}$\frac{3\sqrt{2(S-1)(N_c+2S)}}{\sqrt{S\left[N_c(N_c+6)+12(S-1)S\right](5N_c+18)}}$ \\
$(\lambda-2,\mu+1)S-1$\hspace{0.5cm} & \hspace{0.cm}$(11)1$ & $2$ &\hspace{0.5cm}$-\frac{N_c}{S}\sqrt{\frac{3(N_c-2S+6)(N_c+2S+4)}{2(N_c-2S+2)\left[N_c(N_c+6)+12(S-1)S\right](5N_c+18)}}$\\
$(\lambda-1,\mu-1)S$\hspace{0.5cm} & \hspace{0.cm}$(11)1$ & $/$ &\hspace{0.5cm}$-2\sqrt{\frac{3(S+1)(N_c-2S)(N_c+2S)}{S(N_c-2S+2)(N_c+2S+2)N_c(5N_c+18)}}$\\
$(\lambda-1,\mu-1)S-1$\hspace{0.5cm} & \hspace{0.cm}$(11)1$ & $/$ &\hspace{0.5cm}$2(S-1)\sqrt{\frac{3(N_c-2S)(N_c+2S)}{S(2S-1)(N_c-2S+2)(N_c+2S+2)N_c(5N_c+18)}}$\\
$(\lambda-3,\mu)S-1$\hspace{0.5cm} & \hspace{0.cm}$(11)1$ & $/$ &\hspace{0.5cm}$-2\sqrt{\frac{3(S-1)(N_c+2S-2)}{(2S-1)(N_c-2S+4)N_c(5N_c+18)}}$\\
$(\lambda-4,\mu+2)S-1$\hspace{0.5cm} & \hspace{0.cm}$(11)1$ & $/$ &\hspace{0.5cm}$-\sqrt{\frac{3}{2}}\sqrt{\frac{2S-3}{2S-1}}\sqrt{\frac{(N_c+2S)(N_c-2S+2)(N_c-2S+6)}{(N_c-2S+4)N_c(5N_c+18)}}$\\
$(\lambda-2,\mu+1)S$\hspace{0.5cm} & \hspace{0.cm}$(11)0$ & $1$ &\hspace{0.5cm}$\sqrt{\frac{N_c(N_c+6)+12(S-1)S}{2N_c(5N_c+18)}}$\\
$(\lambda-2,\mu+1)S$\hspace{0.5cm} & \hspace{0.cm}$(11)0$ & $2$ &\hspace{0.5cm}$0$\\
$(\lambda-2,\mu+1)S$\hspace{0.5cm} & \hspace{0.cm}$(00)1$ & $/$ &\hspace{0.5cm}$\sqrt{\frac{4S(S+1)}{N_c(5N_c+18)}}$\\
\hline
\hline
\end{tabular}}
\label{octet_spin_three_halfs} 
\end{sidewaystable*}

\begin{sidewaystable*}[pt]
\vspace{8cm}
\caption{Isoscalar factors of the SU(6) generators,
corresponding to 
the $^210$ multiplet of $N_c = 3$.}
{\scriptsize
 \renewcommand{\arraystretch}{2.5}
\begin{tabular}{l|c|c|l}
\hline
\hline
$(\lambda_1\mu_1)S_1$ \hspace{0.5cm} & \hspace{0.5cm}$(\lambda_2\mu_2)S_2$ \hspace{0.5cm} & \hspace{0.5cm}$\rho$\hspace{0.5cm} & \hspace{0.5cm}$\left(\begin{array}{cc||c}                                         [N_c-1,1]  &  [21^4]  &  [N_c-1,1] \\
                           (\lambda_1\mu_1)S_1 & (\lambda_2\mu_2)S_2 & (\lambda+2,\mu-1)S
                                      \end{array}\right)_\rho$  \\
\vspace{-0.5cm} &  &   & \\
\hline
$(\lambda+4,\mu-2)S+1$\hspace{0.5cm} & \hspace{0.cm}$(11)1$ & $/$ &\hspace{0.5cm}$-\sqrt{\frac{3(2S+5)(N_c+2S+4)(N_c+2S+8)(N_c-2S-2)}{2(2S+3)(N_c+2S+6)N_c(5N_c+18)}}$ \\
$(\lambda+3,\mu-3)S+1$\hspace{0.5cm} & \hspace{0.cm}$(11)1$ & $/$ &\hspace{0.5cm}$2\sqrt{\frac{3(S+2)(N_c-2S-4)}{(2S+3)(N_c+2S+6)N_c(5N_c+18)}}$\\
$(\lambda+1,\mu-2)S+1$\hspace{0.5cm} & \hspace{0.cm}$(11)1$ & $/$ &\hspace{0.5cm}$-2(S+2)\sqrt{\frac{3(N_c+2S+2)(N_c-2S-2)}{(S+1)(2S+3)(N_c-2S)(N_c+2S+4)N_c(5N_c+18)}}$\\
$(\lambda+2,\mu-1)S+1$\hspace{0.5cm} & \hspace{0.cm}$(11)1$ & $1$ &\hspace{0.5cm}$3\sqrt{\frac{2(S+2)(N_c-2S-2)}{(S+1)(N_c(N_c+6)+12(S+1)(S+2))(5N_c+18)}}$\\
$(\lambda+2,\mu-1)S+1$\hspace{0.5cm} & \hspace{0.cm}$(11)1$ & $2$ &\hspace{0.5cm}$-\frac{N_c}{S+1}\sqrt{\frac{3(N_c-2S+2)(N_c+2S+8)}{2(N_c+2S+4)(N_c(N_c+6)+12(S+1)(S+2))(5N_c+18)}}$\\
$(\lambda+2,\mu-1)S$\hspace{0.5cm} & \hspace{0.cm}$(11)1$ & $1$ &\hspace{0.5cm}$\left[N_c(4S+7)+6(S+1)\right]\sqrt{\frac{2S}{(S+1)\left[N_c(N_c+6)+12(S+1)(S+2)\right]N_c(5N_c+18)}}$\\
$(\lambda+2,\mu-1)S$\hspace{0.5cm} & \hspace{0.cm}$(11)1$ & $2$ &\hspace{0.5cm}$-\frac{N_c+2S+2}{S+1}\sqrt{\frac{3S(S+2)(N_c-2S-2)(N_c-2S+2)(N_c+2S+8)}{2(N_c+2S+4)\left[N_c(N_c+6)+12(S+1)(S+2)\right]N_c(5N_c+18)}}$\\
$(\lambda+1,\mu-2)S$\hspace{0.5cm} & \hspace{0.cm}$(11)1$ & $/$ &\hspace{0.5cm}$2\sqrt{\frac{3S(N_c+2S+2)(N_c-2S-2)}{(S+1)(N_c-2S)(N_c+2S+4)N_c(5N_c+18)}}$\\
$(\lambda\mu)S+1$\hspace{0.5cm} & \hspace{0.cm}$(11)1$ & $/$ &\hspace{0.5cm}$\frac{N_c+4(S+1)^2}{S+1}\sqrt{\frac{3(N_c-2S+2)}{2(2S+1)(2S+3)(N_c-2S)N_c(5N_c+18)}}$ \\
$(\lambda\mu)S$\hspace{0.5cm} & \hspace{0.cm}$(11)1$ & $/$ &\hspace{0.5cm} $-\frac{1}{S+1}\sqrt{\frac{3(N_c+2S+2)(N_c-2S+2)}{2(N_c-2S)(5N_c+18)}}$\\
$(\lambda\mu)S-1$\hspace{0.5cm} & \hspace{0.cm}$(11)1$ & $/$ &\hspace{0.5cm} $-\sqrt{\frac{3(2S-1)(N_c+2S+2)(N_c-2S+2)}{2(2S+1)N_c(5N_c+18)}}$\\
$(\lambda+2,\mu-1)S$\hspace{0.5cm} & \hspace{0.cm}$(11)0$ & $1$ &\hspace{0.5cm}$\sqrt{\frac{N_c(N_c+6)+12(S+1)(S+2)}{2N_c(5N_c+18)}}$\\
$(\lambda+2,\mu-1)S$\hspace{0.5cm} & \hspace{0.cm}$(11)0$ & $2$ &\hspace{0.5cm} 0\\
$(\lambda+2,\mu-1)S$\hspace{0.5cm} & \hspace{0.cm}$(00)1$ & $/$ &\hspace{0.5cm} $\sqrt{\frac{4S(S+1)}{N_c(5N_c+18)}}$\\
\hline
\hline
\end{tabular}}

\label{decuplet_spin_one_half} 
\end{sidewaystable*}

\begin{sidewaystable*}[pt]
\vspace{8cm}
\caption{Isoscalar factors of the SU(6) generators,
corresponding to 
the $^21$ multiplet of $N_c = 3$.}
{\scriptsize
 \renewcommand{\arraystretch}{2.5}
\begin{tabular}{l|c|c|l}
\hline
\hline
$(\lambda_1\mu_1)S_1$ \hspace{0.5cm} & \hspace{0.5cm}$(\lambda_2\mu_2)S_2$ \hspace{0.5cm} & \hspace{0.5cm}$\rho$\hspace{0.5cm} & \hspace{0.5cm}$\left(\begin{array}{cc||c}                                         [N_c-1,1]  &  [21^4]  &  [N_c-1,1] \\
                           (\lambda_1\mu_1)S_1 & (\lambda_2\mu_2)S_2 & (\lambda-1,\mu-1)S
                                      \end{array}\right)_\rho$  \\
\vspace{-0.5cm} &  &   & \\
\hline
$(\lambda+1,\mu-2)S+1$\hspace{0.5cm} & \hspace{0.cm}$(11)1$ & $/$ &\hspace{0.5cm}$-\sqrt{\frac{3(2S+3)(N_c-2S-2)(N_c-2S+2)(N_c+2S+4)}{2(N_c-2S)(2S+1)N_c(5N_c+18)}}$\\
$(\lambda+1,\mu-2)S$\hspace{0.5cm} & \hspace{0.cm}$(11)1$ & $/$ &\hspace{0.5cm}$-\sqrt{\frac{3(N_c-2S-2)(N_c-2S+2)(N_c+2S+4)}{2S(2S+1)(N_c-2S)N_c(5N_c+18)}}$\\
$(\lambda-1,\mu-1)S$\hspace{0.5cm} & \hspace{0.cm}$(11)1$ & $1$ &\hspace{0.5cm}$\left[N_c(4S-3)+6S\right]\sqrt{\frac{2(S+1)}{S\left[N_c^2+12(S^2-1)\right]N_c(5N_c+18)}}$\\
$(\lambda-1,\mu-1)S$\hspace{0.5cm} & \hspace{0.cm}$(11)1$ & $2$ &\hspace{0.5cm}$-\left\{N_c(N_c+6)-4\left[S(S-1)-3\right]\right\}\sqrt{\frac{3(2S-1)(S+1)(N_c-2S-2)(N_c+2S-2)}{2S(2S+1)(N_c-2S+2)(N_c+2S+2)\left[N_c^2+12(S^2-1)\right]N_c(5N_c+18)}}$\\
$(\lambda-1,\mu-1)S-1$\hspace{0.5cm} & \hspace{0.cm}$(11)1$ & $1$ &\hspace{0.5cm}$3\sqrt{\frac{2N_c(2S-1)}{S[N_c^2+12(S^2-1)](5N_c+18)}}$\\
$(\lambda-1,\mu-1)S-1$\hspace{0.5cm} & \hspace{0.cm}$(11)1$ & $2$ &\hspace{0.5cm} 0 \hspace{0.5cm} if $S=1/2$\\
$(\lambda-1,\mu-1)S-1$\hspace{0.5cm} & \hspace{0.cm}$(11)1$ & $2$ &\hspace{0.5cm} $-\left[N_c(N_c+6)-12(S^2-1)\right]\sqrt{\frac{3(N_c-2S-2)(N_c+2S-2)}{2S(2S+1)(N_c-2S+2)(N_c+2S+2)[N_c^2+12(S^2-1)]N_c(5N_c+18)}}$ \hspace{0.5cm} if $S\geq 1$ \\
$(\lambda\mu)S+1$\hspace{0.5cm} & \hspace{0.cm}$(11)1$ & $/$ &\hspace{0.5cm}$\sqrt{\frac{6(2S+3)(N_c+2S+4)}{(2S+1)(N_c-2S)N_c(5N_c+18)}}$\\
$(\lambda\mu)S$\hspace{0.5cm} & \hspace{0.cm}$(11)1$ & $/$ &\hspace{0.5cm}$\frac{1}{S}\sqrt{\frac{6(N_c+2S+4)}{(N_c-2S)(N_c+2S+2)(5N_c+18)}}$\\
$(\lambda\mu)S-1$\hspace{0.5cm} & \hspace{0.cm}$(11)1$ & $/$ &\hspace{0.5cm}$\frac{S+1}{S}\sqrt{\frac{6(2S-1)(N_c+2S+4)}{(2S+1)(N_c+2S+2)N_c(5N_c+18)}}$\\
$(\lambda-2,\mu+1)S$\hspace{0.5cm} & \hspace{0.cm}$(11)1$ & $/$ &\hspace{0.5cm} $\frac{1}{S}\sqrt{\frac{6(S+1)(N_c-2S+4)(2S-1)}{(N_c-2S+2)N_c(5N_c+18)}}$\\
$(\lambda-2,\mu+1)S-1$\hspace{0.5cm} & \hspace{0.cm}$(11)1$ & $/$ &\hspace{0.5cm} $\frac{1}{S}\sqrt{\frac{6(N_c-2S+4)(S-1)(2S-1)}{(N_c-2S+2)(N_c+2S)(5N_c+18)}}$\\
$(\lambda-3,\mu)S-1$\hspace{0.5cm} & \hspace{0.cm}$(11)1$ & $/$ &\hspace{0.5cm} 0\hspace{0.5cm}  if $S=1/2$\\
$(\lambda-3,\mu)S-1$\hspace{0.5cm} & \hspace{0.cm}$(11)1$ & $/$ &\hspace{0.5cm} $-\sqrt{\frac{3(N_c+2S-2)(N_c+2S+2)(N_c-2S+4)(S-1)}{2S(N_c+2S)N_c(5N_c+18)}}$ \hspace{0.5cm} if $S\geq 1$ \\
$(\lambda-1,\mu-1)S$\hspace{0.5cm} & \hspace{0.cm}$(11)0$ & $1$ &\hspace{0.5cm}$\sqrt{\frac{N_c^2+12(S^2-1)}{2N_c(5N_c+18)}}$\\
$(\lambda-1,\mu-1)S$\hspace{0.5cm} & \hspace{0.cm}$(11)0$ & $2$ &\hspace{0.5cm} 0 \\
$(\lambda-1,\mu-1)S$\hspace{0.5cm} & \hspace{0.cm}$(00)1$ & $/$ &\hspace{0.5cm} $\sqrt{\frac{4S(S+1)}{N_c(5N_c+18)}}$\\ 
\hline
\hline
\end{tabular}}
\label{singlet_spin_one_half} 
\end{sidewaystable*}


\section{Symmetry properties of isoscalar factors}

In Table \ref{lambdamu} we indicate the values of $\lambda$ and $\mu$ for various physical sectors
as a function of $N_c$. The flavor singlet case $^21_J$ is discussed in more details in Ref. \cite{Matagne:2008kb}.

We recall that the  isoscalar factors of SU(3) obey the following 
orthogonality relation
\begin{widetext}
\begin{equation}
\sum_{Y'' I'' Y^a I^a}
 \left(\begin{array}{cc||c}
	(\lambda'' \mu'')    &  (11)   & (\lambda' \mu')   \\
	 Y'' I'' &  Y^a I^a   &   Y I
      \end{array}\right)_{\rho}
   \left(\begin{array}{cc||c}
	 (\lambda'' \mu'')   &  (11)   & (\lambda \mu) \\
	Y'' I''  &   Y^a I^a  &   Y' I'
      \end{array}\right)_{\rho} = \delta_{\lambda' \lambda} 
      \delta_{\mu' \mu} \delta_{Y' Y}\delta_{I' I},   
 \end{equation}
 \end{widetext}
 which can be easily checked. 
For completeness also note that the isoscalar factors obey the 
following symmetry property


\begin{widetext}
\begin{equation}
 \left(\begin{array}{cc||c}  (\lambda \mu)  &  (11)  &  (\lambda' \mu') \\
                              YI            &  -Y^aI^a & Y'I'
                                      \end{array}\right)= 
(-)^{\frac{1}{3}(\mu'-\mu-\lambda'+\lambda+\frac{3}{2}Y^a)+I'-I}
\sqrt{\frac{\mathrm{dim}(\lambda'\mu')(2I+1)}{\mathrm{dim}(\lambda\mu)(2I'+1)}}
\left(\begin{array}{cc||c}   

                             (\lambda' \mu')  &  (11)  &  (\lambda \mu) \\
                              Y'I' & Y^aI^a &  YI 
                                      \end{array}\right).
\end{equation}
\end{widetext}				      
where $\mathrm{dim}(\lambda\mu) = \frac{1}{2}(\lambda+1)(\mu+1)(\lambda+\mu+2)$ 
is the dimension of the irrep $(\lambda\mu)$ of SU(3).			      

The SU(6) isoscalar factors satisfy to the following symmetry property:
\begin{widetext}
\begin{eqnarray}
\lefteqn{\left(\begin{array}{cc||c}   [f] &  [21^4]  & [f]  \\
                                             (\lambda_1 \mu_1)S_1 & (\lambda_2\mu_2)S_2 &  (\lambda\mu)S
                                      \end{array}\right) = } \nonumber \\  &  & (-1)^{1/3(\mu_1 -\mu -\lambda_1 +\lambda)}(-1)^{S_1-S}
              \sqrt{\frac{\mathrm{dim}(\lambda_1\mu_1)(2S_1+1)}{\mathrm{dim}(\lambda\mu)(2S+1)}}
 \left(\begin{array}{cc||c}   [f] &  [21^4]  & [f]  \\
                                             (\lambda \mu)S & (\lambda_2\mu_2)S_2 &  (\lambda_1\mu_1)S_1
                                      \end{array}\right).              
\end{eqnarray}
\end{widetext}


\section{Matrix elements of SU(4) generators}\label{SU4}

Here we reproduce the matrix elements of the SU(4) generators for the symmetric irrep $[N_c]$.
In Ref.  \cite{Matagne:2006xx} they were written as a particular case  of Eq. (\ref{GEN}). The form presented below is
entirely compatible with that given in Ref. \cite{HP}. Thus 
in the case of SU(4) $\supset$ SU(2) $\times$ SU(2) the analogue of Eq. (\ref{GEN}) becomes 
\begin{widetext}
\begin{equation}\label{GENsu4}
\langle [N_c] I' I'_3 S' S'_3 | E^{ia} |
[N_c] I I_3 S S_3 \rangle = \sqrt{C^{[N_c]}(\mathrm{SU(4)})}   
\left(\begin{array}{cc||c}
	[N_c]    &  [21^2]   & [N_c]   \\
	I S  &   I^a S^i  &  I'S'
      \end{array}\right) 
  \left(\begin{array}{cc|c}
	S   &    S^i   & S'   \\
	S_3  &   S^i_3   & S'_3
  \end{array}\right)
     \left(\begin{array}{cc|c}
	I   &   I^a   & I'   \\
	I_3 &   I^a_3   & I'_3
   \end{array}\right),
   \end{equation}
\end{widetext}
where
\begin{equation} 
C^{[N_c]}(\mathrm{SU(4)})=[3N_c(N_c+4)]/8
\end{equation}
is the SU(4) Casimir operator eigenvalue for the 
symmetric irrep $[N_c]$. Also, note that a symmetric state of SU(4)  has $I=S$.
We recall that the su(4) algebra is a particular case of the algebra (\ref{ALGEBRA}),
where $N_f$ = 2
\begin{eqnarray}\label{ALGEBRASU4}
[S^i,S^j] & = & i \varepsilon^{ijk} S^k,
\ \ \ [T^a,T^b]  =  i \varepsilon^{abc} T^c,\nonumber \\
\ \ \ [S^i,T^a] & = & 0, \nonumber \\
\lbrack S^i,G^{ia}\rbrack & = & i\varepsilon^{ijk}G^{ka}, \ \ \ [T^a,G^{ib}]=i\varepsilon^{abc}G^{ic}, \nonumber \\
\lbrack G^{ia},G^{jb}\rbrack & =  & \fr{i}{4} \delta^{ij} \varepsilon^{abc} T^c
+\fr{i}{2} \delta_{ab}\varepsilon_{ijk}S_k.
\end{eqnarray}
The tensor operators $E^{ia}$ are related to $S^i$, $T^a$ and $G^{ia}$ $(i=1,2,3;\ a=1,2,3)$ by
\begin{equation} \label{normes2}
E^i =\frac{S^i}{\sqrt{2}};~~~ E^a = \frac{T^a}{\sqrt{2}}; ~~~E^{ia} = \sqrt{2} G^{ia}.
\end{equation}
This is a particular case of Eqs. (\ref{normes}) where we now take  $N_f$ = 2.
In Eq. (\ref{GENsu4}) they are identified by $I^aS^i=01,10$ and 11 respectively. Now we want to obtain the SU(4) isoscalar factors as particular cases of the SU(6) results with $Y^a=0$. In SU(4) the hypercharge of a system of $N_c$ quarks takes the value $Y=N_c/3$.
By comparing (\ref{GEN}) and (\ref{GENsu4}) we obtained the relation
\begin{widetext}
\begin{equation}\label{isosu4}
\left(\begin{array}{cc||c}
	[N_c]    &  [21^2]   & [N_c]   \\
	I S  &   I^a S^i  &  I'S'
      \end{array}\right)
 = r^{I^aS^i} \sqrt{\frac{C^{[N_c]}(\mathrm{SU(6))}}{C^{[N_c]}(\mathrm{SU(4))}}}
  \sum_{\rho = 1,2}\left(\begin{array}{cc||c}
	(\lambda \mu)    &  (\lambda^a\mu^a)   &   (\lambda' \mu')\\
	\frac{N_c}{3} I   &  0 I^a  &  \frac{N_c}{3} I'
      \end{array}\right)_{\rho}
\left(\begin{array}{cc||c}
	[N_c]    &  [21^4]   & [N_c]   \\
	(\lambda \mu) S  &  (\lambda^a\mu^a) S^i  &  (\lambda' \mu') S'
      \end{array}\right)_{\rho}\, ,
\end{equation}
\end{widetext}
where
\begin{eqnarray}\label{IASI}
r^{I^aS^i}=\left\{
\begin{array}{cc} \sqrt{\frac{3}{2}} & \mathrm{for}\ I^aS^i=01 \\
1 &\mathrm{for}\ I^aS^i=10 \\
1 &\mathrm{for}\ I^aS^i=11
\end{array}\right.
,\end{eqnarray}
due to 
 (\ref{GENsu4}) and (\ref{normes2}) and taking into account that
in SU(6) one has $E^i = S^i/\sqrt{3}$ while in SU(4) one has $E^i = S^i/\sqrt{2}$. 
In Eq. (\ref{isosu4}) 
we have made the replacement
\begin{equation}\label{replacement}
\lambda = 2I,\ \mu=\frac{N_c}{2}-I;\
\lambda' = 2I',\  \mu'=\frac{N_c}{2}-I',
\end{equation}
and took
\begin{eqnarray}
(\lambda^a\mu^a)=\left\{
\begin{array}{cc}
(00) & \mathrm{for}\ I^a=0 \\
(11) & \mathrm{for}\ I^a=1 \\
\end{array}\right.
.\end{eqnarray}
In this way we have recovered the SU(4) isoscalar factors presented in  
Table A4.2 of Ref. \cite{HP} up to a phase factor. In doing these analytic calculations 
we have made use of the isoscalar factors of SU(3) obtained by Hecht in Ref. 
\cite{Hecht:1965} Table 4. These coefficients were derived in a nuclear physics
context but they can be easily rewritten in terms of $N_c$ due to 
Eqs. (\ref{replacement}).

By introducing these isoscalar factors into the matrix elements (\ref{GENsu4}) 
and the corresponding values of $I^a$ and $S^i$ according to the definition (\ref{IASI}),
we have recovered the expressions given in Eqs. (A1-A3) of  Ref. \cite{CCGL}.
\begin{widetext}
\begin{eqnarray}
  \left< [N_c] S^\prime = I^\prime; m_1^\prime, \alpha_1^\prime \left|
      G^{ia} \right| S = I; m_1, \alpha_1 [N_c]\right> & = & \frac 1
  4 \sqrt{\frac{2I+1}{2I^\prime+1}} \sqrt{(N_c+2)^2 - (I^\prime
    - I)^2 (I^\prime + I + 1)^2} \nonumber \\ & & \times \left(
    \begin{array}{cc} S & 1 \\ m_1 & i \end{array} \right| \left.
    \begin{array}{c} S^\prime \\ m_1^\prime \end{array} \right) \left(
    \begin{array}{cc} I & 1 \\ \alpha_1 & a \end{array} \right| \left.
    \begin{array}{c} I^\prime \\ \alpha_1^\prime \end{array} \right)
\label{eq:giame}
  , \\ \left< [N_c] S^\prime = I^\prime; m_1^\prime, \alpha_1^\prime \left|
      T^a \right| S = I; m_1, \alpha_1 [N_c]\right> & = & \sqrt{I
    (I+1)} \left(
    \begin{array}{cc} I & 1 \\ \alpha_1 & a \end{array} \right| \left.
    \begin{array}{c} I \\ \alpha_1^\prime \end{array} \right)
  \delta_{I^\prime I} \delta_{S^\prime S} \delta_{m_1^\prime
    m_1} , 
\\ \left< [N_c] S^\prime = I^\prime; m_1^\prime,
    \alpha_1^\prime \left| S^i \right| S = I; m_1, \alpha_1 [N_c]
  \right> & = & \sqrt{I (I+1)} \left(
    \begin{array}{cc} S & 1 \\ m_1 & i \end{array} \right| \left.
    \begin{array}{c} S \\ m_1^\prime \end{array} \right)
  \delta_{I^\prime I} \delta_{S^\prime S}
  \delta_{\alpha_1^\prime \alpha_1} .
\end{eqnarray}
\end{widetext}
We recall that  $S = I$ for a symmetric representation.  Note that the matrix elements of 
$G^{ia}$  in Ref. \cite{CCGL} refer to the symmetric representation $[N_c - 1]$ describing a core of $N_c - 1$ quarks,
while here we consider a system of  $N_c$ quarks, hence we have the term $N_c + 2$ instead of $N_c + 1$ 
under the square root in Eq. (\ref{eq:giame}). 
As an example, putting $N_c$ = 3 in Eq. (\ref{eq:giame}) one can recover the first 4 rows of Table 2 
of Ref.  \cite{CC00} by taking into account the relation between Clebsch-Gordan and 3$j$ coefficients.

\section{Isoscalar factors of the permutation group}\label{properties}

Here we shortly recall the definition of isoscalar factors of the
permutation group S$_n$. Let us denote a basis vector in the
invariant subspace of the irrep
$[f]$ of S$_n$ by $|[f] Y \rangle $, where $Y$ is the
corresponding Young tableau or Yamanouchi symbol. A basis vector obtained
from the inner product of two irreps $[f']$ and $[f'']$ is defined by
the sum over  products of basis vectors of $|[f']Y'\rangle$ and $|[f'']Y'' \rangle$ 
at fixed $[f']$ and $[f'']$
\begin{equation}
|[f] Y \rangle = \sum_{Y'Y''}
S([f']Y' [f'']Y'' | [f]Y ) |[f']Y'\rangle |[f'']Y'' \rangle,
\end{equation}
where $S([f']Y' [f'']Y'' | [f]Y )$ are  Clebsch-Gordan (CG) coefficients
of S$_n$.
Any CG coefficient can be factorized into an isoscalar factor, here called
$K$ matrix \cite{Stancu:1991rc}, and a CG coefficient of S$_{n-1}$. To apply the
factorization property it is necessary to specify the row $p$ of the 
$n$-th particle and the row $q$ of the $(n-1)$-th particle. The remaining
particles are distributed in a Young tableau denoted by $y$.
Then the isoscalar factor $K$ associated to a given CG of S$_n$ is defined as
\begin{widetext}
\begin{equation}
S([f']p'q'y' [f'']p''q''y'' | [f]pqy ) =
K([f']p'[f'']p''|[f]p) S([f'_{p'}]q'y' [f''_{p''}]q''y'' | [f_p]qy ),
\end{equation}
\end{widetext}
where  the right-hand side contains a CG coefficient
of S$_{n-1}$ containing $[f_p]$, $[f'_{p'}]$ and $[f''_{p''}]$ which are the partitions obtained from $[f]$ after
the removal of the $n$-th particle.
The $K$ matrix obeys the following orthogonality relations
\begin{widetext}
\begin{eqnarray}
\sum_{p'p''}  K([f']p'[f'']p''|[f]p) K([f']p'[f'']p''|[f_1]p_1) & = &\delta_{f f_1}
\delta_{p p_1}, \label{K1}\\
\sum_{fp}  K([f']p'[f'']p''|[f]p) K([f']p'_1[f'']p''_1|[f]p) & = &
\delta_{p'p'_1} \delta_{p'' p''_1} \label{K2}.
\end{eqnarray}
\end{widetext}
Let us consider a system of $N_c$ quarks having a total spin $S$.
The group SU(2) allows only partitions with maximum two rows, in this case
with  $N_c/2 + S$ boxes in the first row and $N_c/2 - S$
in the second row. So, one has
\begin{equation}\label{FPRIM}
[f'] = \left[\frac{N_c}{2} + S,\frac{N_c}{2} - S\right].
\end{equation}
Then one can write
a symmetric state 
of $N_c$ particles with spin $S$ as the linear combination
\begin{widetext}
\begin{eqnarray}\label{FS}
|[N_c ] 1\rangle 
&=& c^{[N_c]}_{11}(S)| [f'] 1 \rangle | [f'] 1 \rangle
  + c^{[N_c]}_{22}(S)| [f'] 2 \rangle | [f'] 2 \rangle.
\end{eqnarray} 
\end{widetext}
The isoscalar factors used to construct the spin-flavor symmetric state
(\ref{FS}) are
\begin{eqnarray}\label{SYM}
c^{[N_c]}_{11} & = & K([f']1[f']1|[N_c]1),\nonumber \\
c^{[N_c]}_{22} & = & K([f']2[f']2|[N_c]1).
\end{eqnarray}
The isoscalar factors needed to construct the state of mixed
symmetry $[N_c-1,1]$ from the same inner product are
\begin{eqnarray}\label{MS}
c^{[N_c-1,1]}_{11}  & = & K([f']1[f']1|[N_c-1,1]2),\nonumber \\
c^{[N_c-1,1]}_{22}  & = & K([f']2[f']2|[N_c-1,1]2).
\end{eqnarray}
The above coefficients and the orthogonality
relation (\ref{K2}) give 
\begin{eqnarray}
c_{11}^{[N_c-1,1]} & = & -c_{22}^{[N_c]}, \nonumber \\
c_{22}^{[N_c-1,1]} & = & c_{11}^{[N_c]}.
\end{eqnarray}
When the $N_c$-th particle is located in different rows in the flavor and
spin parts the needed coefficients are
\begin{eqnarray}
c^{[N_c-1,1]}_{12}  & = & K([f']1[f']2|[N_c-1,1]2) = 1,\nonumber \\
c^{[N_c-1,1]}_{21}  & = & K([f']2[f']1|[N_c-1,1]2) = 1,
\end{eqnarray}
which are identical because of the symmetry properties of $K$.
The identification of the so called ``elements of orthogonal basis
rotation'' of Ref. \cite{CCGL} with the above isoscalar factors  is the following.
For the symmetric states one has
\begin{equation}
 c^{[N_c]}_{11}  =  c^{\mathrm{SYM}}_{0-}, \ \
 c^{[N_c]}_{22}  =  c^{\mathrm{SYM}}_{0+},
\end{equation}
and for the mixed symmetric states there is
\begin{eqnarray}
 c^{[N_c-1,1]}_{11} & = & c^{\mathrm{MS}}_{0-}, \ \
  c^{[N_c-1,1]}_{22} =  c^{\mathrm{MS}}_{0+},\\
c^{[N_c-1,1]}_{12}  & = & c^{\mathrm{MS}}_{++},\ \
c^{[N_c-1,1]}_{21}   =  c^{\mathrm{MS}}_{--}
\end{eqnarray}
The  coefficients
$\left(c_{pp}^{[N_c]}\right)^2$ ($p = 1,2$),
can be defined in the context of SU(6) $\supset$ SU(2) $\times$  SU(3) as squares of  isoscalar factors of S$_n$.
We write the matrix elements of the generators $S_i$
in two different ways. One is to use the Wigner-Eckart theorem for SU(2)
\begin{widetext}
\begin{equation}\label{SPIN}
\langle [N_c](\lambda'\mu') Y' I' I'_3; S' S'_3 |S^i|
[N_c](\lambda \mu) Y I I_3; S S_3 \rangle =  \delta_{SS'}\delta_{\lambda \lambda'} \delta_{\mu\mu'} \delta_{YY'} \delta_{II'} \delta_{I_3I_3'}
   \sqrt{C(\mathrm{SU(2)})} \left(\begin{array}{cc|c}
	S   &    1  &  S'   \\
	S_3 &    i  &  S'_3
      \end{array}\right),
   \end{equation}
\end{widetext}
The other is  to calculate the matrix
elements of $S^i$ by using the fact that this is a one-body operator 
\begin{equation}
S^i  =  \sum_{k=0}^{N_c} s^i(k)
\end{equation}
where $s_i(k)$ is a single particle operator acting on the particle $k$.
Then for a symmetric state one can write 
\begin{equation}
\langle S^i \rangle   = N_c \langle s^i(N_c) \rangle
\end{equation} 

We define the spin state
\begin{widetext}
\begin{equation}\label{statessu(2)}
|S_1, 1/2; S S_3; p\rangle  =
\sum_{m_1,m_2}
 \left(\begin{array}{cc|c}
	S_1  &    1/2  &  S   \\
	m_1  &    m_2  &  S_3 
      \end{array}\right)
|S_1, m_1 \rangle |1/2, m_2 \rangle,
\end{equation}
\end{widetext}
in terms of an SU(2)-spin CG coefficient
with $S_1 = S - 1/2$ for $p = 1$ and  $S_1 = S + 1/2$ for $p = 2$.

For a symmetric state like (\ref{FS}) one obtains 
\begin{widetext}
\begin{eqnarray}\label{SI}
\langle S_11/2;SS'_3; p
|s_i(N_c)|S_11/2;SS_3; p \rangle & =
&  \sqrt{\frac{3}{4}}
 \sum_{m_1 m_2 m'_2} 
\left(\begin{array}{cc|c}
	S_1  &    1/2  &  S   \\
	m_1  &    m_2  &  S_3 
      \end{array}\right)
 \left(\begin{array}{cc|c}
	S_1  &    1/2  &  S   \\
	m_1  &    m'_2  &  S'_3 
      \end{array}\right) 
  \left(\begin{array}{cc|c}
	1/2  &    1  &  1/2   \\
	m_2  &    i  &  m'_2   
      \end{array}\right)  \nonumber \\
& = &  (-)^{2S}\sqrt{\frac{3}{2}~(2S+1)}
   \left(\begin{array}{cc|c}
	S   &    1  &   S   \\
	S_3  &   i  &   S'_3   
      \end{array}\right) 
    \left\{\begin{array}{ccc}
	1   &    S  &   S   \\
	S_1  &   1/2  & 1/2
      \end{array}\right\}.
\end{eqnarray}
\end{widetext}
Using all this algebra we obtain the equality
\begin{widetext}
\begin{equation}\label{B1}
\sqrt{S(S+1)}  =   (-)^{2S} N_c \sqrt{\frac{3}{2}}\sqrt{2S+1}
\left[\left(c_{22}^{[N_c]}\right)^2 \left\{\begin{array}{ccc} 1 & S & S \\ S+1/2 & 1/2 & 1/2 
\end{array}\right\}\right.  - \left.\left(c_{11}^{[N_c]}\right)^2 
\left\{\begin{array}{ccc} 1 & S & S \\ S-1/2 & 1/2 & 1/2 \end{array}\right\}\right],
\end{equation}
\end{widetext}
which is an equation for the unknown quantities 
The other equation is the normalization relation (\ref{K1})
\begin{equation}\label{normaliz}
\left(c_{11}^{[N_c]}\right)^2 +\left(c_{22}^{[N_c]}\right)^2 = 1.
\end{equation}
We found
\begin{eqnarray}
c^{[N_c]}_{11}(S) & = & \sqrt{\frac{S[N_c+2(S + 1)]}{N_c(2 S + 1)}}, \nonumber \\
c^{[N_c]}_{22}(S) & = & \sqrt{\frac{(S + 1)(N_c - 2 S)}{N_c(2 S + 1)}},
\end{eqnarray}
like in Ref. \cite{CCGL}.

In Eqs. (\ref{decupletch52})-(\ref{singletch52})  below, we illustrate the
application of isoscalar factors for mixed symmetric states of a system with $N_c = 7$ \cite{Matagne:2006ug}.   In each inner product 
the first Young diagram corresponds to spin and the second to flavor. Accordingly,
one can see that Eq. (\ref{decupletch52}) stands for $^210$, Eq. (\ref{octetch54}) 
for $^48$, Eq. (\ref{octetch52}) for $^28$ and Eq. (\ref{singletch52}) for $^21$, in the sense of Table \ref{lambdamu}.
Each inner product contains the corresponding isoscalar factors and 
the position of the $N_c$-th particle is marked with a cross. 
In the right-hand side, from the location of the cross one can read off the values of $p$ and of $p'$.
The equations are 
\begin{widetext}
\begin{eqnarray}
\label{decupletch52}
\raisebox{-9.0pt}{\mbox{\begin{Young}
 & & & & & \cr
$\times$ \cr
\end{Young}}}\
& = &
c^{[6,1]}_{21}\! \! \!
\raisebox{-9.0pt}{\mbox{
\begin{Young}
& & & \cr
& & $\times$\cr
\end{Young}}} \ \times \! \! \! \! \!
\raisebox{-9.0pt}{\mbox{
\begin{Young}
& & & & $\times$\cr
& \cr
\end{Young}}}\ ,
\\ \nonumber
\\
\label{octetch54}
\raisebox{-9.0pt}{\mbox{\begin{Young}
 & & & & & \cr
$\times$ \cr
\end{Young}}}\
& = &
c^{[6,1]}_{12}\! \! \!
\raisebox{-9.0pt}{\mbox{
\begin{Young}
& & & & $\times$\cr
& \cr
\end{Young}}} \ \times \! \! \! \! \!
\raisebox{-9.0pt}{\mbox{
\begin{Young}
& & & \cr
& & $\times$\cr
\end{Young}}}\ ,
\\ \nonumber
\\
\label{octetch52}
\raisebox{-9.0pt}{\mbox{\begin{Young}
 & & & & & \cr
$\times$ \cr
\end{Young}}}\
&=& c^{[6,1]}_{11}\! \! \!
\raisebox{-9.0pt}{\mbox{
\begin{Young}
& & & $\times$\cr
& & \cr
\end{Young}}} \ \times \! \! \! \! \!
\raisebox{-9.0pt}{\mbox{
\begin{Young}
& & & $\times$\cr
& & \cr
\end{Young}}} \nonumber \\
& & + \ c^{[6,1]}_{22}\! \! \!
\raisebox{-9.0pt}{\mbox{
\begin{Young}
& & & \cr
& & $\times$\cr
\end{Young}}} \ \times \! \! \! \! \!
\raisebox{-9.0pt}{\mbox{
\begin{Young}
& & & \cr
& & $\times$\cr
\end{Young}}}\ ,
\\ \nonumber
\\
\label{singletch52}
\raisebox{-9.0pt}{\mbox{\begin{Young}
 & & & & & \cr
$\times$ \cr
\end{Young}}}\
&=& c^{[6,1]}_{13}\! \! \!
\raisebox{-9.0pt}{\mbox{
\begin{Young}
& & & $\times$\cr
& & \cr
\end{Young}}} \ \times \! \! \! \! \!
\raisebox{-15pt}{\mbox{
\begin{Young}
& & \cr
& & \cr
$\times$ \cr
\end{Young}}}\ .
\end{eqnarray} 
\end{widetext}

The above example is a particular case of the approximate spin-flavor wave function 
used in the approach of the symmetric core + excited quark of Ref. \cite{CCGL}. One can see that the
$N_c$-th particle is always in the second row ($p$ = 2) of the spin-flavor wave function and all the terms 
with the $N_c$-th particle in the first row are missing from the exact wave function (\ref{EWF}).
Using group theoretical arguments, the relation between the exact wave function and the approximate one 
as used in  Ref. \cite{CCGL} was thoroughly discussed in Ref.  \cite{Matagne:2008fw}.
\vspace{1cm}

\centerline{\bf Acknowledgments}

We are indebted to Christian Lang for useful information on lattice QCD results.
One of us (F.S.) is most grateful to Ileana Guiasu for her constant interest in the 
subject which lead to many stimulating discussions.
This research was supported by the Fond de la Recherche Scientifique - FNRS under the
grant 4.4501.05.



\begin{thebibliography}{479}
%
\expandafter\ifx\csname natexlab\endcsname\relax\def\natexlab#1{#1}\fi
\expandafter\ifx\csname bibnamefont\endcsname\relax
  \def\bibnamefont#1{#1}\fi
\expandafter\ifx\csname bibfnamefont\endcsname\relax
  \def\bibfnamefont#1{#1}\fi
\expandafter\ifx\csname citenamefont\endcsname\relax
  \def\citenamefont#1{#1}\fi
\expandafter\ifx\csname url\endcsname\relax
  \def\url#1{\texttt{#1}}\fi
\expandafter\ifx\csname urlprefix\endcsname\relax\def\urlprefix{URL }\fi
\providecommand{\bibinfo}[2]{#2}
\providecommand{\eprint}[2][]{\url{#2}}



\bibitem[{\citenamefont{Adkins and Nappi}(1984)}]{Adkins:1983hy}
 \bibinfo{author}{\bibnamefont{Adkins}, \bibfnamefont{G.~S.}}, and \bibinfo{author}{\bibfnamefont{C.~R.}~\bibnamefont{Nappi}}, \bibinfo{year}{1984},
  \bibinfo{journal}{Nucl.\ Phys.\ B {\bf 233}}, \bibinfo{pages}{109}.

\bibitem[{\citenamefont{Ahuatzin} \emph{et~al.}(2014)\citenamefont{Ahuatzin, Flores-Mendieta, Hernandez-Ruiz}}]{Ahuatzin:2010ef}
 \bibinfo{author}{\bibnamefont{Ahuatzin}, \bibfnamefont{G.}}, 
\bibinfo{author}{\bibfnamefont{R.}~\bibnamefont{Flores-Mendieta}}, 
\bibinfo{author}{\bibfnamefont{M.~A.}~\bibnamefont{Hernandez-Ruiz}}, and
\bibinfo{author}{\bibfnamefont{C.~P.}~\bibnamefont{Hofmann}},
\bibinfo{year}{2014},
\bibinfo{journal}{Phys.\ Rev.\ D {\bf 89}}, 
\bibinfo{pages}{034012}.

\bibitem[{\citenamefont{Alexandrou} \emph{et~al.}(2014)
\citenamefont{Alexandrou, Korzec, Koutsou, Leontiou}}]{Alexandrou:2013fsu}
  Alexandrou, C., T.~Korzec, G.~Koutsou and T.~Leontiou, 2014,
  Phys.\ Rev.\ D {\bf 89}, 034502.

\bibitem[{\citenamefont{Amaryan} \emph{et~al.}(2012)\citenamefont{Amaryan, Gavalian, Nepali, Polyakov, Azimov, Briscoe, Dodge, Hyde}}]{Amaryan:2011qc}
  Amaryan, M.~J., G.~Gavalian, C.~Nepali, M.~V.~Polyakov, Y.~.Azimov, W.~J.~Briscoe, G.~E.~Dodge and C.~E.~Hyde {\it et al.}, 2012,
  Phys.\ Rev.\ C {\bf 85}, 035209.



  \bibitem[{\citenamefont{Anisovich} \emph{et.~al.}(2011)\citenamefont{Anisovich, Klempt, Nikonov, Sarantsev, Thoma}}]{Anisovich:2011ye}
  Anisovich, A.~V., E.~Klempt, V.~A.~Nikonov, A.~V.~Sarantsev and U.~Thoma, 2011,
  Eur.\ Phys.\ J.\ A {\bf 47}, 153.

  \bibitem[{\citenamefont{Anisovich} \emph{et.~al.}(2012)\citenamefont{Anisovich, Beck, Klempt, Nikonov, Sarantsev Thoma}}]{Anisovich:2011fc}
  Anisovich, A.~V., R.~Beck, E.~Klempt, V.~A.~Nikonov, A.~V.~Sarantsev and U.~Thoma, 2012,
  Eur.\ Phys.\ J.\ A {\bf 48}, 15.

\bibitem[{\citenamefont{Armoni} \emph{et.~al.}(2003{\natexlab{a}})\citenamefont{Armoni, Shifman, Veneziano}}]{Armoni:2003gp}
  Armoni, A., M.~Shifman and G.~Veneziano, 2003,
  Nucl.\ Phys.\ B {\bf 667}, 170.



\bibitem[{\citenamefont{Armoni} \emph{et.~al.}(2003{\natexlab{b}})\citenamefont{Armoni, Shifman, Veneziano}}]{Armoni:2003fb}
  Armoni, A., M.~Shifman and G.~Veneziano, 2003,
  Phys.\ Rev.\ Lett.\  {\bf 91}, 191601.


\bibitem[{\citenamefont{Armoni and Patella}(2009)}]{Armoni:2009zq}
  Armoni, A., and A.~Patella, 2009,
  JHEP {\bf 0907}, 073.


\bibitem[{\citenamefont{Auzzi} \emph{et.~al.}(2008)\citenamefont{Auzzi, Bolognesi, Shifman}}]{Auzzi:2008hu}
  Auzzi, R., S.~Bolognesi and M.~Shifman, 2008,
  Phys.\ Rev.\ D {\bf 77}, 125029.


\bibitem[{\citenamefont{Aziza Baccouche} \emph{et.~al.}(2001)\citenamefont{Aziza Baccouche, Chow, Cohen, Gelman}}]{oai:arXiv.org:hep-ph/0106096}
  Aziza Baccouche, Z., C.~-K.~Chow, T.~D.~Cohen and B.~A.~Gelman, 2001,
  Phys.\ Lett.\ B {\bf 514}, 346.

\bibitem[{\citenamefont{Bhaduri}(1988)}]{BHADURI}
Bhaduri, R.~K., 1988,  {\it Models of the nucleon: from quarks to solitons}, Lecture notes and supplements in physics, Addison Wesley, 22.

\bibitem[{\citenamefont{Beringer} \emph{et.~al}(2012)}]{PDG} 
Beringer, J., \emph{et.~al.}, 2012, (Particle Data Group),  Phys.\ Rev.\ D {\bf  86}, 010001.

\bibitem[{\citenamefont{Bolognesi and Shifman}(2007)}]{Bolognesi:2007ut}
  Bolognesi, S., and M.~Shifman, 2007,
  Phys.\ Rev.\ D {\bf 75}, 065020.

\bibitem[{\citenamefont{Bolognesi}(2007)}]{Bolognesi:2006ws}
  Bolognesi, S., 2007,
  Phys.\ Rev.\ D {\bf 75}, 065030.

\bibitem[{\citenamefont{Buchmann and Lebed}(2000)}]{Buchmann:2000wf}
  Buchmann, A.~J., and R.~F.~Lebed, 2000,
  Phys.\ Rev.\ D {\bf 62}, 096005.

\bibitem[{\citenamefont{Buchmann \emph{et.~al.}}(2002)\citenamefont{Buchmann, Hester, Lebed}}]{Buchmann:2002mm}
  Buchmann, A.~J., J.~A.~Hester and R.~F.~Lebed, 2002,
  Phys.\ Rev.\ D {\bf 66}, 056002.



\bibitem[{\citenamefont{Buchmann and Lebed}(2003)}]{Buchmann:2002et}
  Buchmann, A.~J., and R.~F.~Lebed, 2003,
  Phys.\ Rev.\ D {\bf 67}, 016002.


\bibitem[{\citenamefont{Buisseret \emph{et.~al.}}(2008)\citenamefont{Buisseret, Semay, Stancu and Matagne}}]{Buisseret:2008tq}
  Buisseret, F., C.~Semay, Fl.~Stancu and N.~Matagne, 2008,
  Bled Workshops in Physics, DMFA, Zalonistvo, Ljubljana, Vol. 9, No. 1, p. 9,
  arXiv:0810.2905 [hep-ph].

\bibitem[{\citenamefont{Buisseret and Semay}(2010)}]{Buisseret:2010na}
  Buisseret, F., and C.~Semay, 2010,
  Phys.\ Rev.\ D {\bf 82}, 056008.

\bibitem[{\citenamefont{Buisseret \emph{et.~al.}}(2012)\citenamefont{Buisseret, Matagne, Semay}}]{Buisseret:2011aa}
  Buisseret, F., N.~Matagne and C.~Semay, 2012,
  Phys.\ Rev.\ D {\bf 85}, 036010.

\bibitem[{\citenamefont{Carlson \emph{et.~al.}}(1998)\citenamefont{Carlson, Carone, Goity and Lebed}}]{CCGL}
Carlson, C.~E., C.~D. Carone, J.~L. Goity and R.~F. Lebed, 1998,
 Phys. Lett. B {\bf 438}, 327; (1999),
 Phys. Rev. D {\bf 59}, 114008.


\bibitem[{\citenamefont{Carlson and Carone}(1998)}]{CaCa98}
Carlson, C.~E., and C.~D.~Carone, 1998,
 Phys.\ Lett.\ B {\bf  441}, 363; 1998,
 Phys.\ Rev.\  D {\bf  58}, 053005.
 


\bibitem[{\citenamefont{Carlson and Carone}(2000)}]{CC00}
Carlson, C.~E., and C. D. Carone, 2000,
Phys. Lett. {\bf B484}, 260.



\bibitem[{\citenamefont{Carone} \emph{et.~al.}}(1994{\natexlab{a}})]{CGO94}
Carone, C.~D., H.~Georgi and S.~Osofsky, 1994,
Phys.~Lett. B {\bf 322}, 227.




\bibitem[{\citenamefont{Carone} \emph{et.~al.}}(1994{\natexlab{b}})]{CGKM94} 
Carone, C.~D., H.~Georgi, L.~Kaplan and D.~Morin, 1994,
 Phys. Rev. D {\bf 50}, 5793. 

\bibitem[{\citenamefont{Cherman and Cohen}(2006)}]{Cherman:2006iy}
  Cherman, A., and T.~D.~Cohen, 2006,
  JHEP {\bf 0612}, 035.


\bibitem[{\citenamefont{Cherman \emph{et.~al.}}(2009)}]{Cherman:2009fh}
  Cherman, A., T.~D.~Cohen, and R.~F.~Lebed, 2009,
  Phys.\ Rev.\ D {\bf 80}, 036002.


\bibitem[{\citenamefont{Cherman \emph{et.~al.}}(2012)}]{Cherman:2012eg}
  Cherman, A., T.~D.~Cohen, and R.~F.~Lebed, 2012,
  Phys.\ Rev.\ D {\bf 86}, 016002.

\bibitem[{\citenamefont{Cohen}(1996)}]{Cohen:1996zz}
  Cohen, T.~D., 1996,
  Rev.\ Mod.\ Phys.\  {\bf 68}, 599.

\bibitem[{\citenamefont{Cohen and Broniowski}(1992)}]{Cohen:1992uy}
  Cohen, T.~D., and, W.~Broniowski, 1992
  Phys.\ Lett.\ B {\bf 292}, 5.

\bibitem[{\citenamefont{Cohen \emph{et.~al.}}(2004)}]{Cohen:2003fv}
  Cohen,, T.~D., D.~C.~Dakin, A.~Nellore, and R.~F.~Lebed, 2004,
  Phys.\ Rev.\ D {\bf 69}, 056001.


\bibitem[{\citenamefont{Cohen and Lebed}(2003{\natexlab{a}})}]{COLEB1}
  Cohen, T.~D., and R.~F.~Lebed, 2003,
  Phys.\ Rev.\ Lett.\  {\bf 91}, 012001; 2003,
  Phys. Rev. D {\bf 67}, 096008.




\bibitem[{\citenamefont{Cohen and Lebed}(2003{\natexlab{b}})}]{COLEB2}
Cohen, T.~D., and R. F. Lebed, 2003, Phys. Rev. D {\bf 68}, 056003.

\bibitem[{\citenamefont{Cohen and Lebed}(2004)}]{Cohen:2003nk}
  Cohen, T.~D., and R.~F.~Lebed, 2004,
  Phys.\ Lett.\ B {\bf 578}, 150.



\bibitem[{\citenamefont{Cohen and Lebed}(2005)}]{Cohen:2005ct}
  Cohen, T.~D., and R.~F.~Lebed, 2005,
  Phys.\ Rev.\ D {\bf 72}, 056001.

\bibitem[{\citenamefont{Cohen and Lebed}(2014{\natexlab{a}})}]{Cohen:2014via}
  Cohen, T.~D., and R.~F.~Lebed, 2014,
  Phys.\ Rev.\ D {\bf 89}, 054018


\bibitem[{\citenamefont{Cohen and Lebed}(2014{\natexlab{b}})}]{Cohen:2014tga}
  Cohen, T.~D., and R.~F.~Lebed, 2014,
  Phys.\ Rev.\ D {\bf 90}, 016001.




\bibitem[{\citenamefont{Cohen \emph{et.~al.}}(2004)}]{Cohen:2004qt}
  Cohen, T.~D., D.~C.~Dakin, A.~Nellore, and R.~F.~Lebed, 2004,
  Phys.\ Rev.\ D {\bf 70}, 056004.


\bibitem[{\citenamefont{Cohen \emph{et.~al.}}(2005)}]{Cohen:2004bk}
  Cohen, T.~D., D.~C.~Dakin, R.~F.~Lebed, and D.~R.~Martin, 2005,
  Phys.\ Rev.\ D {\bf 71}, 076010.

\bibitem[{\citenamefont{Cohen \emph{et.~al.}}(2010)}]{Cohen:2009wm}
  Cohen, T.~D., D.~L.~Shafer, and R.~F.~Lebed, 2010,
  Phys.\ Rev.\ D {\bf 81}, 036006.



\bibitem[{\citenamefont{Coleman}(1985)}]{COLEMAN}
Coleman, S., 1985, {\it Aspects of Symmetry}, Chapter 8, Cambridge University Press, Cambridge.

\bibitem[{\citenamefont{Coleman and Glashow}(1961)}]{colemanglashow}
Coleman, S., and S. L. Glashow, 1961,
Phys.\ Rev.\ Lett.\  {\bf 6}, 423.

\bibitem[{\citenamefont{Cordon and Goity}(2013)}]{CalleCordon:2012xz}
  Cordon, A.~C., and J.~L.~Goity, 2013,
  Phys.\ Rev.\ D {\bf 87}, 016019.


\bibitem[{\citenamefont{Cordon \emph{et.~al.}}(2014)}]{Cordon:2014sda}
  Cordon,  A.~C., T.~DeGrand and J.~L.~Goity, 2014,
  Phys.\ Rev.\ D {\bf 90}, 014505.


\bibitem[{\citenamefont{Corrigan and Ramond}(1979)}]{Corrigan:1979xf}
  Corrigan, E., and P.~Ramond, 1979,
  Phys.\ Lett.\ B {\bf 87}, 73.


\bibitem[{\citenamefont{Dai \emph{et.~al.}}(1996)}]{DDJM96}
Dai, J., R.~Dashen, E.~Jenkins, and A.~V.~Manohar, 1996,
Phys.~Rev.  D {\bf 53}, 273.


\bibitem[{\citenamefont{Dashen and Manohar}(1993)}]{Dashen:1993as}
  Dashen, R.~F., and A.~V.~Manohar, 1993,
  Phys.\ Lett.\ B {\bf 315}, 425; 1993,
  Phys.\ Lett.\ B {\bf 315}, 438.


\bibitem[{\citenamefont{Dashen \emph{et.~al.}}(1994)}]{Dashen:1993jt}
  Dashen, R.~F., E.~E.~Jenkins, and A.~V.~Manohar, 1994,
  Phys.\ Rev.\ D {\bf 49}, 4713; 1995,
   Erratum-ibid. \ D {\bf 51}, 2489.

\bibitem[{\citenamefont{Dashen \emph{et.~al.}}(1995)}]{Dashen:1994qi}
  Dashen, R.~F., E.~E.~Jenkins, and A.~V.~Manohar, 1995,
  Phys.\ Rev.\ D {\bf 51}, 3697.


\bibitem[{\citenamefont{De Rujula \emph{et.~al.}}(1975)}]{rgg}
  De Rujula, A., H.~Georgi, and S.~L.~Glashow, 1975,
  Phys.\ Rev.\ D {\bf 12}, 147.


\bibitem[{\citenamefont{Diakonov \emph{et.~al.}}(1997)}]{Diakonov:1997mm}
  Diakonov, D., V.~Petrov, and M.~V.~Polyakov, 1997,
  Z.\ Phys.\ A {\bf 359}, 305.


\bibitem[{\citenamefont{Diakonov \emph{et.~al.}}(2013)}]{Diakonov:2013qta}
  Diakonov, D., V.~Petrov, and A.~A.~Vladimirov, 2013,
  Phys.\ Rev.\ D {\bf 88}, 074030.


\bibitem[{\citenamefont{Edwards \emph{et.~al.}}(2013)}]{Edwards:2012fx}
  Edwards, R.~G., N.~Mathur, D.~G.~Richards, and S.~J.~Wallace, 2013,
  Phys.\ Rev.\ D {\bf 87}, 054506.

\bibitem[{\citenamefont{Elliot}(1958)}]{Elliott:1958zj}
  Elliott, J.~P., 1958,
  Proc.\ Roy.\ Soc.\ Lond.\ A {\bf 245}, 128; 1958, ibid. \ A {\bf 245}, 562.


\bibitem[{\citenamefont{Frandsen \emph{et.~al.}}(2006)}]{Frandsen:2005mb}
  Frandsen, M.~T, C.~Kouvaris and F.~Sannino, 2006,
  Phys.\ Rev.\ D {\bf 74}, 117503.



\bibitem[{\citenamefont{Flores-Mendieta}(2009)}]{FloresMendieta:2009rq}
  Flores-Mendieta, R., 2009,
  Phys.\ Rev.\ D {\bf 80}, 094014.


\bibitem[{\citenamefont{Flores-Mendieta \emph{et.~al.}}(2012)}]{FloresMendieta:2012dn}
  Flores-Mendieta, R., M.~A.~Hernandez-Ruiz and C.~P.~Hofmann, 2012,
  Phys.\ Rev.\ D {\bf 86}, 094041.


\bibitem[{\citenamefont{Gervais and Sakita}(1984)}]{Gervais:1983wq}
  Gervais, J.~L., and B.~Sakita, 1984,
  Phys.\ Rev.\ Lett.\  {\bf 52}, 87; 1984,
  Phys.\ Rev.\ D {\bf 30}, 1795.

\bibitem[{\citenamefont{Glozman and Riska}(1996)}]{Glozman:1995fu}
  Glozman, L.~Y., and D.~O.~Riska, 1996,
  Phys.\ Rept.\  {\bf 268}, 263.

\bibitem[{\citenamefont{Glozman \emph{et.~al.}}(1998)}]{Glozman:1997ag}
  Glozman, L.~Y., W.~Plessas, K.~Varga, and R.~F.~Wagenbrunn, 1998,
  Phys.\ Rev.\ D {\bf 58}, 094030.


\bibitem[{\citenamefont{Glozman}(2002)}]{Glozman:2002kq}
  Glozman, L.~Y., 2002,
  Phys.\ Lett.\ B {\bf 541}, 115.
 







\bibitem[{\citenamefont{Goity}(1997)}]{Goi97}
Goity, J.~L., 1997, Phys. Lett. B {\bf 414}, 140. 

\bibitem[{\citenamefont{Goity \emph{et.~al.}}(2003)}]{GSS03}
Goity, J.~L.,  C.~L.~Schat, and N.~N.~Scoccola, 2003,
 Phys. Lett. B {\bf 564}, 83.

\bibitem[{\citenamefont{Goity \emph{et.~al.}}(2005)}]{TRENTO2004}
Goity,  J.~L., R. L. Lebed, A. Pich, C. L. Schat, and N. N. Scoccola (Eds.), 2005,
Proceedings of the workshop ''Large $N_c$ QCD 2004'', Trento, Iatly, July 5-11 2004, World Scientific.



\bibitem[{\citenamefont{Goity and Matagne}(2007)}]{Goity:2007sc}
  Goity,   J.~L., and N.~Matagne, 2007,
  Phys.\ Lett.\  B {\bf 655}, 223.

\bibitem[{\citenamefont{Goity \emph{et.~al.}}(2005)}]{Goity:2004ss}
  Goity,   J.~L., C.~Schat, and N.~Scoccola, 2005,
  Phys.\ Rev.\ D {\bf 71}, 034016.

\bibitem[{\citenamefont{Goity and Scoccola}(2005)}]{Goity:2005rg}
  Goity, J.~L., and N.~N.~Scoccola, 2005,
  Phys.\ Rev.\ D {\bf 72}, 034024.

\bibitem[{\citenamefont{Goity and Scoccola}(2007)}]{Goity:2007ft}
  Goity, J.~L., and N.~N.~Scoccola, 2007,
  Phys.\ Rev.\ Lett.\  {\bf 99}, 062002.

\bibitem[{\citenamefont{Goity \emph{et.~al.}}(2009)}]{Goity:2009wq}
  Goity, J.~L., C.~Jayalath, and N.~N.~Scoccola, 2009,
  Phys.\ Rev.\ D {\bf 80}, 074027.

\bibitem[{\citenamefont{Harada \emph{et.~al.}}(2004)}]{Harada:2003em}
  Harada, M., F.~Sannino, and J.~Schechter, 2004,
  Phys.\ Rev.\ D {\bf 69}, 034005.


\bibitem[{\citenamefont{Hayashi \emph{et.~al.}}(1984)}]{HAYASHI}
 Hayashi, A., G. Eckart, G. Holzwart, and H. Walliser, 1984, Phys. Lett.  {\bf  147B}, 5.

\bibitem[{\citenamefont{Hecht}(1965)}]{Hecht:1965}
Hecht, K.~T., 1965, Nucl. Phys. {\bf 62}, 1.

\bibitem[{\citenamefont{Hecht and Pang}(1969)}]{HP}
Hecht, K.~T., and S.~C.~Pang, 1969,
J.\ Math.\ Phys.\ (N.Y.) {\bf 10}, 1571. 


\bibitem[{\citenamefont{Hermann}(1966)}]{HERMANN}
Hermann, R., 1966,``Lie groups for physicists,''
The mathematical physics monograph series, Princeton University, Benjamin, New York, Ch. 11.


\bibitem[{\citenamefont{Hyodo and Jido}(2012)}]{Hyodo:2011ur}
  Hyodo, T., and D.~Jido, 2012,
  Prog.\ Part.\ Nucl.\ Phys.\  {\bf 67}, 55.


\bibitem[{\citenamefont{In\"on\"u and E.~P.~ Wigner}(1953)}]{INONUWIGNER53}
In\"on\"u, E., and E.~P.~ Wigner, 1953, Proc. Natl. Acad. Sci. U.S.  {\bf 39}, 510.


\bibitem[{\citenamefont{Isgur and Karl}(1978)}]{IK78}
  Isgur, N., and G.~Karl, 1978,
  Phys.\ Rev.\ D {\bf 18}, 4187.



\bibitem[{\citenamefont{Jayalath \emph{et.~al.}}(2011)}]{Jayalath:2011uc}
  Jayalath, C., J.~L.~Goity, E.~Gonzalez de Urreta and N.~N.~Scoccola, 2011,
  Phys.\ Rev.\ D {\bf 84}, 074012.

\bibitem[{\citenamefont{Jenkins}(1993)}]{Jenk1}
Jenkins, E. E., 1993, Phys.~Lett. B {\bf 315}, 431; 1993, ibid. B {\bf 315}, 441; 1993, ibid. B {\bf 315}, 447.

\bibitem[{\citenamefont{Jenkins}(1996)}]{Jenkins:1995gc}
  Jenkins, E.~E., 1996,
  Phys.\ Rev.\ D {\bf 53}, 2625.

\bibitem[{\citenamefont{Jenkins}(1996)}]{oai:arXiv.org:hep-ph/9603449}
  Jenkins, E.~E., 1996,
  Phys.\ Rev.\ D {\bf 54}, 4515.

\bibitem[{\citenamefont{Jenkins}(1997)}]{oai:arXiv.org:hep-ph/9609404}
  Jenkins, E.~E., 1997,
  Phys.\ Rev.\ D {\bf 55}, 10.



\bibitem[{\citenamefont{Jenkins}(1998)}]{Jenkins:1998wy}
  Jenkins, E.~E.,1998,
  Ann.\ Rev.\ Nucl.\ Part.\ Sci.\  {\bf 48} (1998) 81.

\bibitem[{\citenamefont{Jenkins}(2001)}]{Jenkins:2001it}
  Jenkins, E. E., 2001,
  hep-ph/0111338.


\bibitem[{\citenamefont{Jenkins}(2008)}]{oai:arXiv.org:0712.0406}
  Jenkins, E. E., 2008,
  Phys.\ Rev.\ D {\bf 77}, 034012.


\bibitem[{\citenamefont{Jenkins}(2012)}]{Jenkins:2011dr}
  Jenkins, E. E., 2012,
  Phys.\ Rev.\ D {\bf 85}, 065007.

\bibitem[{\citenamefont{Jenkins \emph{et.~al.}}(2002)}]{Jenkins:2002rj}
  Jenkins, E. E., X.~D.~Ji, and A.~V.~Manohar, 2002,
  Phys.\ Rev.\ Lett.\  {\bf 89}, 242001.


\bibitem[{\citenamefont{Jenkins and Lebed}(1995)}]{Jenkins:1995td}
  Jenkins, E. E., and R.~F.~Lebed, 1995,
  Phys.\ Rev.\ D {\bf 52}, 282.


\bibitem[{\citenamefont{Jenkins and Lebed}(2000)}]{Jenkins:2000mi}
  Jenkins E. E., and R.~F.~Lebed, 2000,
  Phys.\ Rev.\ D {\bf 62}, 077901.

\bibitem[{\citenamefont{Jenkins and Manohar}(1994)}]{Jenkins:1994md}
  Jenkins, E. E., and A.~V.~Manohar, 1994,
  Phys.\ Lett.\ B {\bf 335}, 452.



\bibitem[{\citenamefont{Jenkins and Manohar}(2004{\natexlab{a}})}]{Jenkins:2004vb}
  Jenkins, E. E., and A.~V.~Manohar, 2004,
  JHEP {\bf 0406}, 039.

\bibitem[{\citenamefont{Jenkins and Manohar}(2004{\natexlab{b}})}]{Jenkins:2004tm}
  Jenkins, E. E., and A.~V.~Manohar, 2004,
  Phys.\ Rev.\ Lett.\  {\bf 93}, 022001.








\bibitem[{\citenamefont{Jenkins \emph{et.~al.}}(2010)}]{Jenkins:2009wv}
 Jenkins, E. E., A.~V.~Manohar, J.~W.~Negele, and A.~Walker-Loud, 2010,
  Phys.\ Rev.\ D {\bf 81}, 014502.




\bibitem[{\citenamefont{Keller \emph{et.~al.}}(2011)}]{Keller:2011nt}
  Keller, D., {\it et al.}, 2011,  [CLAS Collaboration],
  Phys.\ Rev.\ D {\bf 83}, 072004.

\bibitem[{\citenamefont{Keller \emph{et.~al.}}(2012)}]{Keller:2011aw}
  Keller, D., {\it et al.}, 2012  [CLAS Collaboration],
  Phys.\ Rev.\ D {\bf 85}, 052004.




\bibitem[{\citenamefont{Knecht and Peris}(2013)}]{Knecht:2013yqa}
  Knecht, M., and S.~Peris, 2013,
  Phys.\ Rev.\ D {\bf 88}, 036016.

\bibitem[{\citenamefont{Koniuk and Isgur}(1980)}]{Koniuk:1979vy}
  Koniuk, R., and N.~Isgur, 1980,
  Phys.\ Rev.\ D {\bf 21}, 1868
   [Erratum-ibid., 1981, D {\bf 23}, 818].

\bibitem[{\citenamefont{Lang and Verduci}(1980)}]{Lang:2013eca}
  Lang, C.~B., and V.~Verduci, 2013,
  Acta Phys.\ Polon.\ Supp.\  {\bf 6}, 3,  967.


\bibitem[{\citenamefont{Lebed}(1995)}]{Lebed:1995}
Lebed, R.~F., 1995,
Phys.\ Rev.\ D {\bf 51}, 5039.

\bibitem[{\citenamefont{Lebed}(2013)}]{Lebed:2013aka}
  Lebed, R.~F., 2013,
  Phys.\ Rev.\ D {\bf 88}, 057901.

\bibitem[{\citenamefont{Lebed and Martin}(2004{\natexlab{a}})}]{Lebed:2004fj}
  Lebed, R.~F., and D.~R.~Martin, 2004,
  Phys.\ Rev.\ D {\bf 70}, 016008.

\bibitem[{\citenamefont{Lebed and Martin}(2004{\natexlab{b}})}]{Lebed:2004zc}
  Lebed, R.~F., and D.~R.~Martin, 2004,
  Phys.\ Rev.\ D {\bf 70}, 057901.


\bibitem[{\citenamefont{Lebed and Martin}(2009)}]{Lebed:2009aq}
  Lebed, R.~F., and L.~Yu, 2009,
  Phys.\ Rev.\ D {\bf 80}, 076006.

\bibitem[{\citenamefont{Lee \emph{et.~al.}}(1999)}]{oai:arXiv.org:hep-ph/9809576}
  Lee, J.~-P., C.~Liu and H.~S.~Song, 1999,
  Phys.\ Rev.\ D {\bf 59}, 034002.


\bibitem[{\citenamefont{Lee \emph{et.~al.}}(2000)}]{oai:arXiv.org:hep-ph/0006267}
  Lee, J.~-P., C.~Liu, and H.~S.~Song, 2000,
  Phys.\ Rev.\ D {\bf 62}, 096001.



\bibitem[{\citenamefont{Leutwyler}(1995)}]{Leutwyler:1994fi}
  Leutwyler, H., 1995,
  Varenna 1995, Selected topics in nonperturbative QCD
  [hep-ph/9406283].


\bibitem[{\citenamefont{Lichtenberg}(1970)}]{Lichtenberg}
  Lichtenberg, D.~B., 1970,
  ``Unitary symmetry and elementary particles,''
  Academic Press, New York.


\bibitem[{\citenamefont{Liu \emph{et.~al.}}(2014)}]{Liu:2014yva}
  Liu, T., Y.~Mao, and B.~-Q.~Ma, 2014,
  Int.\ J.\ Mod.\ Phys.\ A {\bf 29}, 13,  1430020.



\bibitem[{\citenamefont{Lucini \emph{et.~al.}}(2013)}]{Lucini:2012gg}
  Lucini, B., and M.~Panero, 2013,
  Phys.\ Rept.\  {\bf 526}, 93.


\bibitem[{\citenamefont{Luty and March-Russell}(1994)}]{Luty:1993fu}
  Luty, M.~A., and J.~March-Russell, 1994,
  Nucl.\ Phys.\ B {\bf 426}, 71.



\bibitem[{\citenamefont{Luty \emph{et.~al.}}(1995)}]{LMRW95}
Luty, M.~A., J.~March-Russell and  M.~White, 1995,
Phys.~Rev. D {\bf 51}, 2332.


\bibitem[{\citenamefont{Mackey}(1995)}]{Mackey}
Mackey,G.~W., 1968, ''Induced representations of groups and quantum mechanics'', Benjamin, New York.


\bibitem[{\citenamefont{Manohar}(1998)}]{Manohar:1998xv}
  Manohar, A.~V., 1998,
  ``Large N QCD,'', Les Houches Lectures Session LXVIII: {\it Probing the Standard Model of Particle Interactions}, 
eds. F.~David and R.~Gupta,
  hep-ph/9802419.


\bibitem[{\citenamefont{Matagne}(2006)}]{Matagne:2006ug}
  Matagne, N., 2006,
  ``Baryon resonances in large $N_c$ QCD,'' PhD Thesis, University of Li\`ege,
  hep-ph/0701061.

\bibitem[{\citenamefont{Matagne and Stancu}(2005{\natexlab{a}})}]{MS1}
 Matagne, N.,  and Fl.~Stancu, 2005, Phys. Rev. D {\bf 71}, 014010.

\bibitem[{\citenamefont{Matagne and Stancu}(2005{\natexlab{b}})}]{Matagne:2005gd}
  Matagne, N.,  and Fl.~Stancu, 2005,
  Phys.\ Lett.\  B {\bf 631}, 7.

\bibitem[{\citenamefont{Matagne and Stancu}(2006{\natexlab{a}})}]{Matagne:2006xx}
  Matagne, N., and Fl.~Stancu, 2006,
  Phys.\ Rev.\ D {\bf 73}, 114025.

\bibitem[{\citenamefont{Matagne and Stancu}(2006{\natexlab{b}})}]{Matagne:2006zf}
  Matagne, N., and Fl.~Stancu, 2006,
  Phys.\ Rev.\ D {\bf 74}, 034014.

\bibitem[{\citenamefont{Matagne and Stancu}(2008{\natexlab{a}})}]{Matagne:2006dj}
  Matagne, N., and Fl.~Stancu, 2008,
  Nucl.\ Phys.\ A {\bf 811}, 291.

\bibitem[{\citenamefont{Matagne and Stancu}(2008{\natexlab{b}})}]{Matagne:2008fw}
  Matagne, N., and Fl.~Stancu, 2008,
  Phys.\ Rev.\ D {\bf 77}, 054026.

\bibitem[{\citenamefont{Matagne and Stancu}(2009)}]{Matagne:2008kb}
  Matagne, N., and Fl.~Stancu, 2009,
  Nucl.\ Phys.\  A {\bf 826}, 161.

\bibitem[{\citenamefont{Matagne and Stancu}(2010)}]{Matagne:2010qt}
  Matagne, N., and Fl.~Stancu, 2010,
  Phys.\ Rev.\ D {\bf 81}, 116002.

\bibitem[{\citenamefont{Matagne and Stancu}(2011{\natexlab{a}})}]{Matagne:2011fr}
  Matagne, N., and Fl.~Stancu, 2011,
  Phys.\ Rev.\ D {\bf 83}, 056007.

\bibitem[{\citenamefont{Matagne and Stancu}(2011{\natexlab{b}})}]{Matagne:2011sn}
  Matagne, N., and Fl.~Stancu, 2011,
  Phys.\ Rev.\ D {\bf 84}, 056013.

\bibitem[{\citenamefont{Matagne and Stancu}(2012{\natexlab{a}})}]{Matagne:2012tm}
  Matagne, N., and Fl.~Stancu, 2012,
  Phys.\ Rev.\ D {\bf 85}, 116003.

\bibitem[{\citenamefont{Matagne and Stancu}(2012{\natexlab{b}})}]{Matagne:2012vq}
  Matagne, N., and Fl.~Stancu, 2012,
  Phys.\ Rev.\ D {\bf 86}, 076007.

\bibitem[{\citenamefont{Matagne and Stancu}(2013)}]{Matagne:2013cca}
  Matagne, N., and Fl.~Stancu, 2013,
  Phys.\ Rev.\ D {\bf 87}, 076012.

\bibitem[{\citenamefont{Mattis}(1986)}]{MATTIS}
Mattis, M.~P., 1986, Phys. Rev.  Lett. {\bf 56}, 1103; 1989, Phys. Rev. D {\bf 39}, 994; 1989, Phys. Rev.  Lett. {\bf 63}, 1455.

\bibitem[{\citenamefont{Mattis and Mukerjee}(1988)}]{MattisMukerjee}
Mattis, M.~P., and M. Mukerjee, 1988, Phys. Rev. Lett.  {\bf 61}, 1344.


\bibitem[{\citenamefont{Mattis and Peskins}(1985)}]{MAPE}
Mattis, M.~P., and M. E. Peskin, 1985, Phys. Rev. D {\bf 32}, 58;


\bibitem[{\citenamefont{Mohler}(2012)}]{Mohler:2012nh}
  Mohler, D., 2012,
  PoS LATTICE {\bf 2012}, 003
  [arXiv:1211.6163 [hep-lat]].

\bibitem[{\citenamefont{Nakamura \emph{et.~al.}}(2010)}]{PDG2010}
Nakamura, K., {\it et al.}, (2010), (Particle Data Group), J. Phys. G {\bf 37}, 075021.
 

\bibitem[{\citenamefont{Nakano \emph{et.~al.}}(2003)}]{Nakano:2003qx}
  Nakano, T., {\it et al.}  [LEPS Collaboration], 2003,
  Phys.\ Rev.\ Lett.\  {\bf 91}, 012002.



\bibitem[{\citenamefont{Pirjol and Yan}(1998)}]{PY}
 Pirjol, D., and T.~M. Yan, 1998,
 Phys. Rev. D {\bf 57}, 1449; 1998, ibid. D {\bf 57}, 5434.

\bibitem[{\citenamefont{Pirjol and Schat}(2003)}]{Pirjol:2003ye}
 Pirjol, D., and C.~Schat, 2003, Phys. Rev. {\bf  D67}, 096009.

\bibitem[{\citenamefont{Pirjol and Schat}(2007)}]{Pirjol:2006ne}
  Pirjol, D., and C.~Schat, 2007,
  Phys.\ Rev.\ D {\bf 75}, 076004.


\bibitem[{\citenamefont{Pirjol and Schat}(2008)}]{Pirjol:2007ed}
  Pirjol, D., and C.~Schat, 2008,
  Phys.\ Rev.\ D {\bf 78}, 034026.

\bibitem[{\citenamefont{Pirjol and Schat}(2010)}]{oai:arXiv.org:1007.0964}
  Pirjol, D., and C.~Schat, 2010,
  Phys.\ Rev.\ D {\bf 82}, 114005.



\bibitem[{\citenamefont{Florence}(2011)}]{FLORENCE2011}
Proceedings of the workshop ''Large-N gauge theories'', \\
4-04-2011 to 17-06-2011, Florence, Italy, \\
http://ggi-www.fi.infn.it



\bibitem[{\citenamefont{Sartor and Stancu}(1986{\natexlab{a}})}]{Sartor:1986sf}
  Sartor, R., and Fl.~Stancu, 1986,
  Phys.\ Rev.\ D {\bf 33}, 727.




\bibitem[{\citenamefont{Sartor and Stancu}(1986{\natexlab{b}})}]{Sartor:1986qr}
Sartor, R., and Fl.~Stancu, 1986
  Phys.\ Rev.\ D {\bf 34}, 3405.



\bibitem[{\citenamefont{Schat \emph{et.~al.}}(2002)}]{SGS}
Schat, C.~L., J.~L. Goity, and N.~N. Scoccola, 2002,  Phys. Rev. Lett.
{\bf 88}, 102002;
Goity, J.~L., C.~L.~Schat, and N.~N.~Scoccola, 2002
  Phys.\ Rev.\ D {\bf  66}, 114014.




\bibitem[{\citenamefont{Scoccola \emph{et.~al.}}(2008)}]{Scoccola:2007sn}
  Scoccola, N.~N., J.~L.~Goity, and N.~Matagne, 2008,
  Phys.\ Lett.\ B {\bf 663}, 222.


\bibitem[{\citenamefont{Segal \emph{et.~al.}}(1951)}]{SEGAL51}
Segal, I.~E., Duke Math J.,  1951, {\bf 18}, 221.


\bibitem[{\citenamefont{Semay \emph{et.~al.}}(2007{\natexlab{a}})}]{Semay:2007cv}
  Semay, C., F.~Buisseret, N.~Matagne, and Fl.~Stancu, 2007,
  Phys.\ Rev.\  D {\bf 75}, 096001.


\bibitem[{\citenamefont{Semay \emph{et.~al.}}(2007{\natexlab{b}})}]{Semay:2007ff}
  Semay, C., F.~Buisseret, and Fl.~Stancu, 2007,
  Phys.\ Rev.\ D {\bf 76}, 116005.


\bibitem[{\citenamefont{Semay \emph{et.~al.}}(2008)}]{oai:arXiv.org:0808.3349}
  Semay, C., F.~Buisseret, and Fl.~Stancu, 2008
  Phys.\ Rev.\ D {\bf 78}, 076003.




\bibitem[{\citenamefont{Silvestre-Brac \emph{et.~al.}}(2012)}]{Silvestre-Brac:2011aua}
  Silvestre-Brac, B., C.~Semay, and F.~Buisseret, 2012
  J.\ Phys.\ Math.\  {\bf 4}, P120601
  [arXiv:1101.5222 [quant-ph]].


\bibitem[{\citenamefont{Stancu}(1996)}]{Stancu:1991rc}
  Stancu, Fl., 1996,
  ``Group theory in subnuclear physics,''
  Oxford Stud.\ Nucl.\ Phys.\  {\bf 19}, 1.


\bibitem[{\citenamefont{Stancu}(2001)}]{Stancu:2000zk}
  Stancu, Fl., 2001,
  Few Body Syst.\ Suppl.\  {\bf 13}, 225.


\bibitem[{\citenamefont{Stancu}(2005)}]{Stancu:2004ap}
  Stancu, Fl., 2005,
  Int.\ J.\ Mod.\ Phys.\ A {\bf 20}, 209
  [hep-ph/0408042].


\bibitem[{\citenamefont{Stancu and Stassart}(1991)}]{Stancu:1991cz}
  Stancu, Fl., and P.~Stassart, 1991,
  Phys.\ Lett.\  B {\bf 269}, 243.
 



\bibitem[{\citenamefont{Teper}(1999)}]{Teper:1998te}
  Teper, M.~J., 1999,
  Phys.\ Rev.\ D {\bf 59}, 014512.


\bibitem[{\citenamefont{'t Hooft}(1999)}]{HOOFT}
 't Hooft, G., 1974, Nucl. Phys. {\bf 72}, 461.

\bibitem[{\citenamefont{de Urreta \emph{et.~al.}}(2014)}]{deUrreta:2013koa}
  de Urreta, E.~G., J.~L.~Goity, and N.~N.~Scoccola, 2014,
  Phys.\ Rev.\ D {\bf 89}, 034024.


\bibitem[{\citenamefont{Weinberg}(2013)}]{Weinberg:2013cfa}
  Weinberg, S., 2013,
  Phys.\ Rev.\ Lett.\  {\bf 110}, 261601.


\bibitem[{\citenamefont{Wilson}(1974)}]{WILSON}
 Wilson, K.~G., 1974,  Phys.\ Rev.\ D {\bf 10}, 2445.



\bibitem[{\citenamefont{Witten}(1979{\natexlab{a}})}]{WITTEN0}
 Witten, E., 1979, Nucl. Phys. {\bf B149}, 285.

\bibitem[{\citenamefont{Witten}(1979{\natexlab{b}})}]{WITTEN}
 Witten, E., 1979, Nucl. Phys. {\bf B160}, 57.


\bibitem[{\citenamefont{Wu and Ma}(2004)}]{Wu:2004wg}
 Wu, B., and B.~Q.~Ma, 2004,
  Phys.\ Rev.\ D {\bf 70}, 034025.



\end{thebibliography}
\end{document}